\documentclass[final,3p,times]{elsarticle}

\usepackage{amssymb}
\usepackage{amsmath}
\usepackage{booktabs}
\usepackage{multirow}
\usepackage{tikz}
\usepackage{bm}
\usepackage{algorithm}
\usepackage{algpseudocode}
\usepackage{xcolor}

\biboptions{sort&compress}
\begin{document}

\begin{frontmatter}



  \title{
Exploiting repeated matrix block structures for more efficient CFD on modern
  supercomputers}


  \author[cttc]{Josep Plana-Riu} 
  \author[cttc]{F.Xavier Trias} 
  \author[cttc,padova]{Àdel Alsalti-Badellou} 
  \author[cttc,surf]{Xavier Álvarez-Farré}
  \author[cttc]{Guillem Colomer}
  \author[cttc]{Assensi Oliva}

  \affiliation[cttc]{organization={Heat and Mass Transfer Technological Centre, Technical University of Catalonia},
            addressline={Carrer de Colom 11}, 
            city={Terrassa},
            postcode={08222}, 
            country={Spain}}
  \affiliation[padova]{organization={
Department of Information Engineering, University of Padova},
            addressline={Via Giovanni Gradenigo 6/B}, 
            city={Padova},
            postcode={35131}, 
            country={Italy}}
   \affiliation[surf]{organization={High Performance Computing and Visualiztion
   Team, SURF},
            addressline={Science Park 140}, 
            city={Amsterdam},
            postcode={1098 XG}, 
            country={the Netherlands}}           
\begin{abstract}

  Computational Fluid Dynamics (CFD) simulations are often constrained by the memory-bound nature of sparse matrix-vector operations, which eventually limits performance on modern high-performance computing (HPC) systems. This work introduces a novel approach to 
accelerate the numerical solution of the incompressible Navier-Stokes equations by leveraging repeated matrix block structures. The method transforms the conventional sparse matrix-vector product (\texttt{SpMV}) into a generalized sparse matrix-matrix product (\texttt{SpMM}), enabling simultaneous processing of multiple right-hand sides. This shifts the computation towards a more compute-bound regime by reusing matrix coefficients.

Additionally, an inline mesh-refinement strategy is proposed: simulations initially run on a coarse mesh to establish a statistically steady flow, then refine to the target mesh. This reduces the wall-clock time to reach transition, leading to faster convergence with equivalent computational cost.

  The methodology is evaluated using theoretical performance bounds and validated through three test cases: a turbulent channel flow, Rayleigh-Bénard convection, and an industrial airfoil simulation. Results demonstrate substantial speed-ups—from modest improvements in basic configurations to over 50\% in the mesh-refinement setup—highlighting the benefits of integrating \texttt{SpMM} across all CFD operators, including divergence, gradient, and Laplacian.
\end{abstract}


\end{frontmatter}



\section{Introduction}
\label{introduction}

Numerical solutions of the Navier-Stokes equations typically involve a few
algebraic operations, or in a stencil-based implementation the equivalent
counterparts: the dot product (\texttt{dot}), the linear combination of vectors
(\texttt{axpy}), the elementwise product of vectors (\texttt{axty}), and the
sparse matrix-vector product (\texttt{SpMV}), regardless of the method used
(finite-volume method (FVM), finite-element method (FEM), discontinuous
Galerkin (DG), finite difference method (FDM), etc.). Even though the expense of each
and every operation is implementation dependent, \texttt{SpMV} usually becomes
one of the bottlenecks of these simulations, given the big amount of data that
needs to be transferred. Despite having sufficient computational
power, the system cannot supply data quickly enough to utilize efficiently the
available resources. 

\begin{figure}[h]
  \centering
  \includegraphics[width=0.5\textwidth]{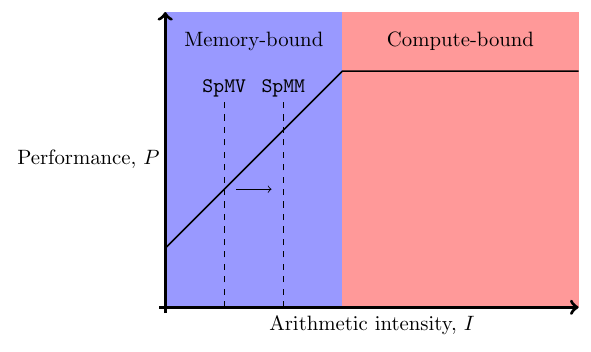}
  \caption{Simplified version of a roofline model in which the memory-bound
  (blue) and compute-bound (red) regions are depicted. The goal of the present
  paper is represented in pushing the arithmetic intensity $I$ towards the
  compute-bound zone.}
  \label{fig:roofline_dummy}
\end{figure}

How efficiently the data provided can be used is shown by the roofline model
\citep{williams_roofline_2009}, which shows the performance bounds for a given arithmetic
intensity, $I$, which is defined as the ratio between the number of
floating-point operations and the data to transfer. Given that the system
cannot supply data quickly enough, it can be stated that these operations are
memory-bound, as shown in Figure \ref{fig:roofline_dummy}.

Therefore, addressing the memory-boundness of the implementation is
critical to reaching the full potential of modern
high-performance computing (HPC) systems. According to this roofline model, by
increasing the $I$ of the computations, the performance of the method defined as the available
floating-point operations per second will become closer to the peak performance of the
device used, which would define a compute-bound implementation.

Given the definition of $I$, it
is straightforward to consider two ways of increasing the arithmetic intensity: increasing the
amount of floating-point operations or reducing the amount of data to transfer.
The former is limited by the operation performed; e.g., in
a \texttt{dot}, the number of operations will be proportional to $n$, being $n$
the length of the vector. This is why an increase in the
arithmetic intensity is more related to reducing the amount of data to
transfer.

In order to deal with these issues related to the strong scaling limits of the
different codes and implementations, the literature has considered multiple
strategies to increase the {performance} of \texttt{SpMV}. Considering
the structure of the matrices, \citet{greathouse_efficient_2014} aimed to improve the
performance of the operation in GPU-accelerated nodes by properly mapping the
loads of the sparse matrix, which led eventually to remarkable speed-ups
compared to the original CSR-based algorithms. {Similarly, sparse matrix formats
such as ELLPACK, tailored for GPUs, have been developed for improved performance. In this line,
the SELL-C-$\sigma$ format \cite{Kreutzer2014} has been proposed to improve Single Instruction Multiple Data (SIMD) utilization across various architectures, including CPUs and many-core processors. Moreover, specialized formats such as HYB \cite{Bell2009} are often employed to regularize memory access patterns and handle the irregular sparsity found in unstructured meshes. These strategies align with modern trends in hardware acceleration, such as the use of tensor cores of GPUs \cite{Markidis2018} for high-throughput matrix operations and the adoption of mixed-precision computing \cite{Anzt2019}, to further reduce memory pressure while maintaining numerical accuracy.}

{While these strategies improve the performance of \texttt{SpMV}, the memory wall is still not circumvented. Even though their work did not directly improve the arithmetic intensity of the operation, \citet{makarashvili_performance_2017} set the first stone for improvements in this direction. By exploiting }
the ergodicity of the statistically steady state, i.e., the time average of the
results can be exchanged by an ensemble average of different uncorrelated
solutions and obtained relevant speed-ups starting from developed turbulent flows and
applying incompressible perturbations to the whole domain. These simulations
were run on different devices, i.e., each case had its dedicated computing
units, so no arithmetic intensity improvements were obtained.
According to \citet{nastac_lyapunov_2017}, the perturbations will grow following
the Lyapunov exponent, which ensures that after this transition time, $T_T$,
all different flow states will be uncorrelated. Later on,
\citet{tosi_use_2022} developed a statistical analysis of this method, which reduced
the transition times and improved the method's efficiency.

{Extending on this work,} \citet{krasnopolsky_approach_2018} exploited the possibility of
running the different flow states, i.e., the different simultaneous simulations
that have been perturbed, in the same device. By doing so, these $m$ flow states are run
simultaneously for a shorter simulation such that $T_T$ is fully simulated
while the averaging time, $T_A$, is downscaled with the number of flow states.
This is run in the same device so that the \texttt{SpMV} operations can be
transformed into a single generalized sparse matrix-vector product
(\texttt{GSpMV}). This was performed by rearranging the $m$ vectors into
a dense matrix with $m$ columns and then performing the products simultaneously.
This reduced the amount of data to transfer, i.e., the sparse matrix was read
only once while preserving the amount of floating-point operations, which led
to increased arithmetic intensity in the solution of the Poisson equation.

The method presented by \citet{krasnopolsky_approach_2018} can be improved by
(i) incrementing the performance given by the \texttt{GSpMV}, and (ii) reducing
the time spent to reach $T_T$ in the simulation. With regards to this
performance optimization, some recent studies \citep{alsalti-baldellou_lighter_2024, alsalti-baldellou_multigrid_2024} applied the
same principle of the $m$-column dense matrix in simulations with domains with
symmetries or repeated patterns by making use of, when properly ordered, the
block structure pattern of the matrix, and applied the method not only in the
Poisson equation but in all the \texttt{SpMV} products throughout the
simulation, in what it was called sparse matrix-matrix product (\texttt{SpMM}),
which not only improved remarkably the arithmetic intensity but reduced the
memory footprint of those simulations. This was implemented within the
in-house framework of HPC$^2$ \citep{alvarez_hpc2fully-portable_2018} in which an algebraic framework is
implemented, which allows the portability of the code, making it suitable for
architectures ranging from CPU clusters to modern accelerators. This proves the
possibility of extending the method to the whole simulation, expecting better
results as the speed-up is found throughout the whole algorithm.

This implementation was tested both for CPUs and GPUs, leading to remarkable
speed-ups in the application of the \texttt{SpMM} not only in the solution of
the Poisson equation, even though it is there where it is the most useful, but
in the gradient, divergence and Laplacian operators that naturally arise from
discretizing the equations, thus leading to bigger speed-ups globally, as all
three operators present remarkable speed-ups \citep{alsalti-baldellou_lighter_2024}.

With regards to optimizing the time spent to reach $T_T$,
\citet{krasnopolsky_optimal_2018} proposed an analogous methodology to
\citet{makarashvili_performance_2017}: running a single simulation until the flow is
developed, to then perturb the developed flow and let it develop until
uncorrelated following \citep{nastac_lyapunov_2017}, which would set the starting point
of the time averaging process, which improved the speed-ups obtained in the
baseline study \citep{krasnopolsky_approach_2018}. Nonetheless, further improvements are
studied in the present work by running a coarser mesh until developing the flow
and then refining, developing the flow again, and proceeding with the averaging
process.

In this context, the present work presents a strategy to 
{accelerate the numerical solution of
the incompressible Navier-Stokes equations by replacing the \texttt{SpMV} with} 
the more compute-intensive \texttt{SpMM} not only in the solution of the
Poisson equation but in the whole simulation following the lines of
\citep{alsalti-baldellou_exploiting_2023,alsalti-baldellou_lighter_2024}, as remarkable speed-ups were observed when
applying this novel method to the Laplacian, gradient, and divergence
operators given the effect from Figure \ref{fig:roofline_dummy}. Nonetheless, instead of exploiting spatial symmetries or
repeated patterns spatially, the repeated patterns will be temporal, as the
framework presented by \citet{krasnopolsky_approach_2018} runs different flow states
simultaneously. {On the other hand, an inline mesh-refinement strategy
for ensemble averaging simulations is presented. This inline mesh-refinement strategy will reduce the wall-clock time
to reach transition, leading to faster convergence with equivalent computational cost.} {Therefore, the present
work presents a performance optimization strategy orthogonal to those related to the structure of the sparse matrix itself \cite{greathouse_efficient_2014, Bell2009, Kreutzer2014}, as it focuses on the data management of the operations rather than on the structure of the sparse matrix.}

This paper is organized as follows. First, in section \ref{sec:ai} the different methods to
increase the arithmetic intensity of \texttt{SpMV} are presented, with the derivation of
the different bounds to the performance improvements. Later, section \ref{sec:cfd} presents
the discretization techniques applied to the Navier-Stokes equations both in space and time,
and presents the conditions under the repeated matrix block structures appear in CFD simulations.
In section \ref{sec:exp}, the numerical experiments to validate
the methodology are presented for three different cases: a turbulent channel flow, a Rayleigh-Bénard
convection and the numerical simulation of a 30P30N airfoil.
In section \ref{sec:disc} a discussion on the applicability and benefits of
the method is presented. Eventually, section \ref{sec:conc} presents the
conclusions of the work.

\section{Increasing arithmetic intensity of sparse matrix-vector products} \label{sec:ai}

As previously depicted, the numerical simulation of most partial differential
equations (PDE) eventually relies on
the implementation of kernels for the basic algebra operations, \texttt{dot,
axty, axpy}, and \texttt{SpMV}, all of which have a low arithmetic intensity.

According to the roofline model \citep{williams_roofline_2009}, this indicates that the
performance obtained by the code is not limited by the peak performance of the
compute unit. Instead, it is limited by its memory
bandwidth, which leads to a memory-bound code. Hence, should this
arithmetic intensity be improved, the performance of the codes would increase
without any additional computational cost, with the benefit coming from a more efficient
data management.

Among the four basic operations, improving the performance of the sparse
matrix-vector operation is the most feasible option. Let us consider a sparse
matrix whose structure reads $A
= I_s\otimes\tilde{A}\in\mathbb{R}^{sn\times sm}$, being $I_s$ the identity matrix of size $s$, which
essentially is $\tilde{A}$ repeated $s$ times forming a block-diagonal
structure. Thus, given $s$ vectors $\tilde{\mathbf{x}}\in\mathbb{R}^m$, which can be
arranged in such a way that $\mathbf{x}
= (\tilde{\mathbf{x}}_1~\tilde{\mathbf{x}}_2~\dots~\tilde{\mathbf{x}}_s)^T$, it
is straightforward to observe that the outcome of $A\mathbf{x}$ would be the
same as the outcome of the sparse matrix-matrix product

\begin{equation}
  \tilde{A}(\tilde{\mathbf{x}}_1~\tilde{\mathbf{x}}_2~\dots~\tilde{\mathbf{x}}_s).
\end{equation}

It is clear to see that the amount of operations that the \texttt{SpMM} has to
do compared to the \texttt{SpMV} is the same, as the matrix $A$ is block
diagonal, and the vectors are the same, just rearranged. Moreover, as the matrix
is the same for all \texttt{SpMV}s, it is only transferred once instead of
$s$-times, which will eventually increase the arithmetic intensity of the operation.

This increased arithmetic intensity, according to the roofline model (see Fig. \ref{fig:roofline_dummy}), will
increase the performance of the code without any additional resources. This is
explained by the fact that the maximum performance that a memory-bound operation can
extract is given by the product of the arithmetic intensity and the memory
bandwidth.
Thus, increasing the arithmetic intensity increases linearly the
potential performance, generating speed-up in the operation.

The speed-up from transforming \texttt{SpMV} to \texttt{SpMM} is bounded with the ratio of arithmetic intensities from
a \texttt{SpMM} and a \texttt{SpMV}, {which algorithm implementation is detailed in Algorithm \ref{alg:spmm}}. 

\begin{algorithm}
  \caption{Sparse matrix-matrix product (\texttt{SpMM}) implementation for a sparse matrix in CSR format assuming an array of structures (AoS) ordering.}
  \label{alg:spmm}
  \hspace*{\algorithmicindent} \textbf{Input}: $A\in\mathbb{R}^{n_r/d\times n_c/d}$, $\mathbf{x}\in\mathbb{R}^{n_c}$, $\mathbf{c}\in\mathbb{R}^d$ \\
  \hspace*{\algorithmicindent} \textbf{Output}: $\mathbf{y}\in\mathbb{R}^{n_c}$
  \begin{algorithmic}[1]
    \For{$i\gets 1$ to $n_c/d$}
    \State $\mathbf{s}\gets \mathrm{zeros}(d)$
    \For{$j \gets A.\mathrm{ptr}[i]$ to $A.\mathrm{ptr}[i+1]$}
    \For{$k\gets 1$ to $d$}
    \State $s[k] \gets s[k] + A.\mathrm{val}[j][k] \cdot \mathbf{x}[d\cdot A.\mathrm{idx}[j]+k]$
    \EndFor
    \EndFor
    \For{$k\gets 1$ to $d$}
    \State $\mathbf{y}[d\cdot i + k] \gets \mathbf{c}[k]\cdot \mathbf{s}[k]$
    \EndFor
    \EndFor
  \end{algorithmic}
\end{algorithm}

    According to \citep{alsalti-baldellou_lighter_2024}, the
arithmetic intensity of a \texttt{SpMM} with $m$ rhs and a sparse matrix
with $n_c$ columns, $n_r$ rows and $\text{nnz}(A)$ non-zeros, can be defined for a sparse matrix
in CSR format as

\begin{equation}
  I_{\texttt{SpMM}}(m)
  = \frac{(2\text{nnz}(A)+1)m}{8\text{nnz}(A)+4\text{nnz}(A)+4(n_r+1)+m(8n_r+8n_c+8)}.
  \label{eq:ai}
\end{equation}

\noindent Both the upper- and
lower-bounds of $P_{m,\text{\texttt{SpMM}}}$ are computed as in
\citep{alsalti-baldellou_lighter_2024}, where the upper-bound is the ratio of operational
intensities of an \texttt{SpMM} and \texttt{SpMV},
$I_{\texttt{SpMM}}(m)/I_{\texttt{SpMV}}$, and the lower-bound is computed
assuming zero temporal locality when accessing the input vector coefficients by
replacing $8n_cm$ with $8\text{nnz}(A)m$,

\begin{subequations}
  \begin{align}
    P_{m,\text{\texttt{SpMM}}}^{ub}
    &= \frac{m[(3(\text{nnz}(A)+n_r+1)+2n_c]}{3\text{nnz}(A)+(1+2m)(n_r+1)+2mn_c}, \label{eq:pmspmm_ub}\\
    P_{m,\text{\texttt{SpMM}}}^{lb}
    &= \frac{m[5\text{nnz}(A)+3(n_r+1)]}{(3+2m)\text{nnz}(A)+(1+2m)(n_r+1)}. \label{eq:pmspmm_lb}
  \end{align}
  \label{eq:pmspmm}
\end{subequations}

\begin{figure}[h]
  \centering
  \includegraphics[height=0.3\textwidth]{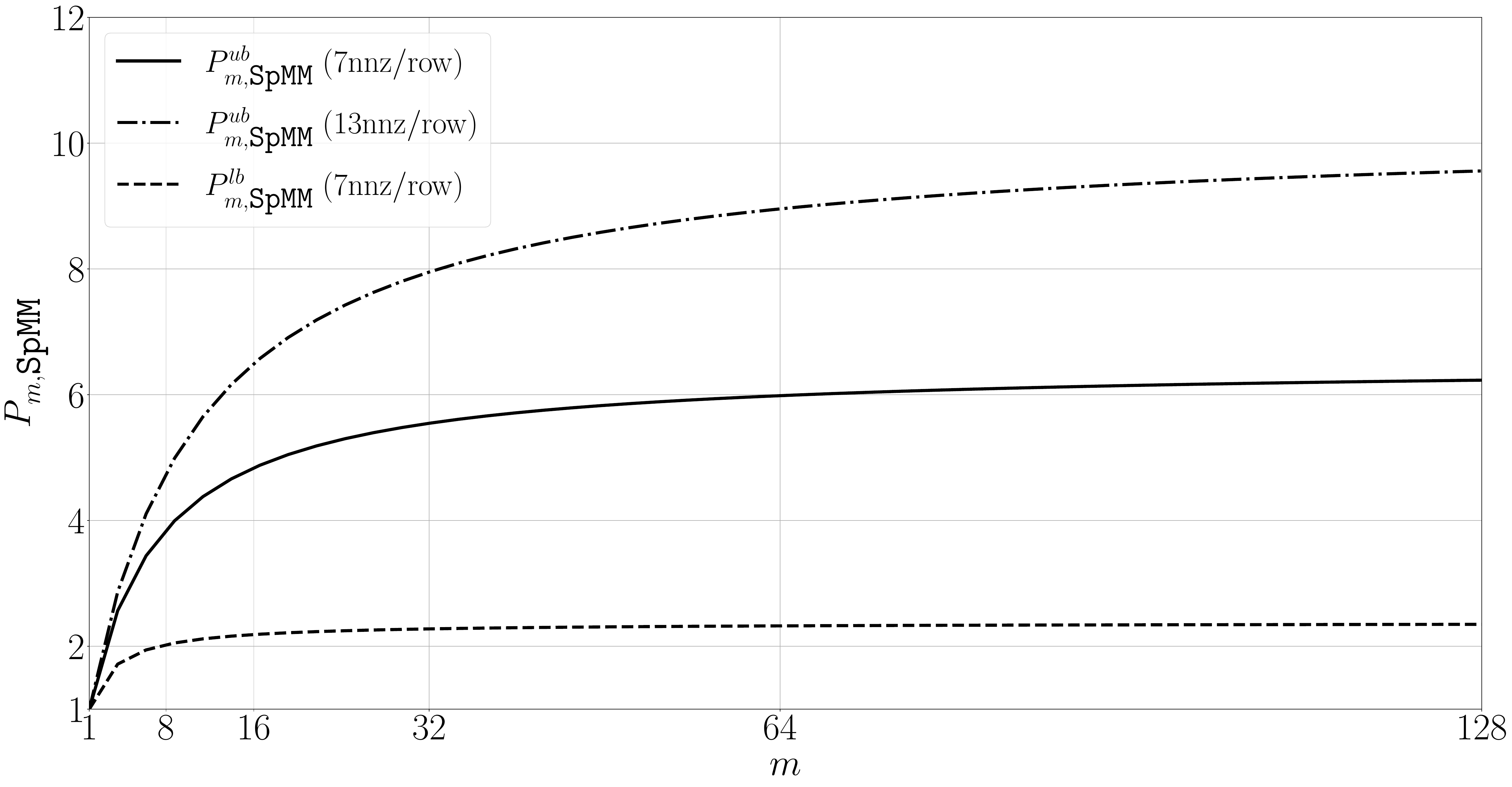}
  \includegraphics[height=0.3\textwidth]{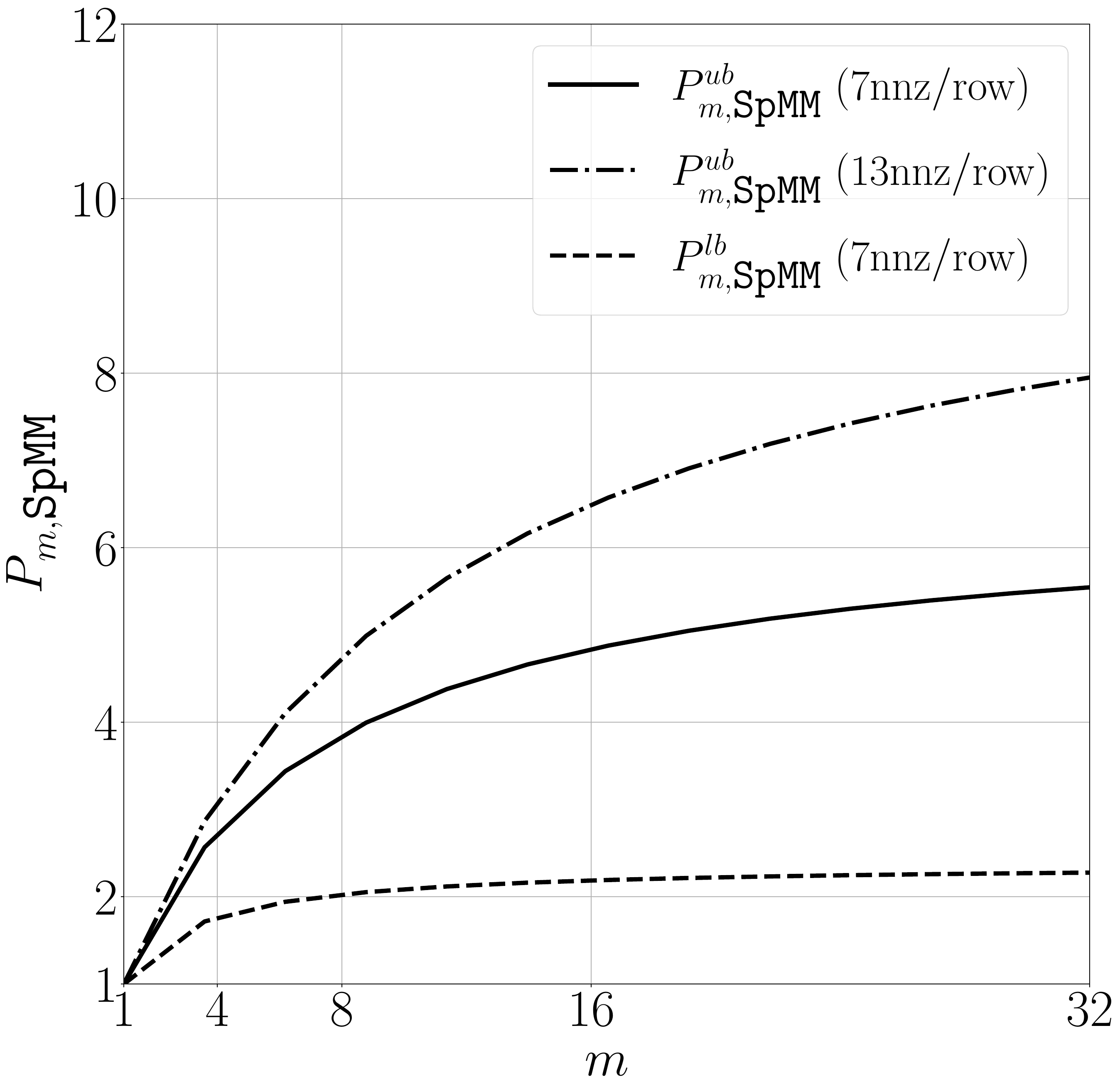}
  \caption{Speed-up bounds associated to a \texttt{SpMM} with a sparse square matrix with $n_c=n_r=10^6$
  with 7 and 13 non-zeros per row for $m=[1,128]$ (left) and a zoom up to 32
  rhs (right).}
  \label{fig:theo_speedupspmm}
\end{figure}

Figure \ref{fig:theo_speedupspmm} provides an example with 7 and 13 non-zeros
per row in which it can be seen than denser matrices, generally related to
higher-order discretization schemes, provide improved potential speed-ups in the
\texttt{SpMM} as the amount of data not to be transferred thanks to using
a more compute-bound operation provides better speed-ups than the sparser
matrices. The lower-bound presents marginal differences with different number of non-zeros. Thus,
only the lower-bound for 13 non-zeros per row is shown.

This kind of block structure naturally appears in simulations in which the
geometry presents symmetrical patterns or repeated geometries
\citep{alsalti-baldellou_exploiting_2023, alsalti-baldellou_lighter_2024} if the mesh ordering is properly set. This
means that the amount of memory required to execute the case is notably reduced
while the performance of the implementation increases with no additional
resources, which ends up leading to faster and lighter simulations. Moreover,
this structure can be forced to appear in other situations, without the
additional benefit of reducing the memory footprint of the simulation
\cite{krasnopolsky_approach_2018}. First of
all, a multiple parameter simulation, in which a set of parameters wants to be
studied, e.g. an airfoil in which a reduced range of angles of attack wants to be
simulated. On the other hand, if a turbulent flow presents a statistically
steady-state, the ergodicity of this phase, i.e. time-averaging and ensemble
averaging are equivalent, will allow running multiple ensembles simultaneously,
which allows to
artificially build the block structure to exploit the increased arithmetic
intensity of \texttt{SpMM} compared to \texttt{SpMV}.

It is worth noticing that most of the speed-up will be obtained in the solution
of the Poisson equation that naturally arises in the numerical solution of the 
incompressible Navier-Stokes equations. Hence, an estimation of the speed-up
for this operation itself is also provided. Namely, by definition, the speed-up in 
the Poisson equation solution given $m$ RHS will be

\begin{equation}
  P_{m,\text{Poisson}} = \frac{mT_{\text{Poisson},1}}{T_{\text{Poisson},m}},
  \label{eq:speedup_poisson}
\end{equation}

where $T_{\text{Poisson},m}$ indicates the wall-clock time of the solution of
the system of equations. This wall-clock time may be split into two different
contributions. First of all, the contribution that will benefit of the
increased arithmetic intensity, which will be the \texttt{SpMM},
$T_{\text{\texttt{SpMM}},m}$. Then, $T_{\text{Poisson},m}^{\text{\texttt{SpMM}}}$ denotes the 
wall-clock time to perform all the \texttt{SpMM}s in a Poisson solution. The other terms, which it may be assumed
to grow linearly with $m$, are denoted by $T_{\text{others},m} = m T_{\text{others},1}$. Let $\xi$ be
the fraction of the wall-clock time for the 1-rhs case devoted to the other
operations. Thus, the wall-clock time for the 1-rhs Poisson equation may be
expressed as
\begin{equation}
  T_{\text{Poisson},1} = \frac{1}{1-\xi}T_{\text{Poisson},1}^{\text{\texttt{SpMM}}} = \frac{1}{1-\xi}T_{\text{Poisson}}^{\text{\texttt{SpMV}}}.
\end{equation}
Hence, the wall-clock time for the $m$-rhs case will read as
\begin{equation}
  T_{\text{Poisson},m} = T_{\text{\texttt{SpMM}},m} + m\xi T_{\text{Poisson},1}
  = T_{\text{\texttt{SpMM}},m} + \frac{m\xi}{1-\xi}T_\text{Poisson}^{\text{\texttt{SpMV}}}.
\end{equation}

By applying Eq. \eqref{eq:speedup_poisson} and introducing the weight of the
Poisson equation in the solution of the system of equations, $\chi = 1-\xi$, the
speed-up of the Poisson equation solution will be estimated with

\begin{equation}
  P_{m,\text{Poisson}} = \frac{P_{m,\text{\texttt{SpMM}}}}{\chi
  + (1-\chi)P_{m,\text{\texttt{SpMM}}}}.
  \label{eq:poipm}
\end{equation}

\section{Application to CFD simulations} \label{sec:cfd}

In the previous section, a strategy to take benefit from repeated matrix block structures
in sparse matrices present in \texttt{SpMV} operations has been presented, where the standard
\texttt{SpMV} is replaced by \texttt{SpMM}, a more compute-intensive method as its arithmetic
intensity is increased by reducing the amount of data transferred. This strategy can be applied to virtually any PDE
where these repeated matrix block structures are present. As an application example, the numerical solution of the incompressible
Navier-Stokes equations using this method is shown. These read, in a dimensionless form, as follows:

\begin{equation}
  \frac{\partial\mathbf{u}}{\partial t} + (\mathbf{u}\cdot\nabla)\mathbf{u}
= \frac{1}{\text{Re}}\nabla^2\mathbf{u} - \nabla p, \hspace{1in} \nabla\cdot\mathbf{u}=0,
  \label{eq:ns_cont}
\end{equation}

\noindent where $\mathbf{u}(\mathbf{x},t)$ and $p(\mathbf{x},t)$ are the velocity and kinematic pressure fields, respectively, and $\text{Re}=\frac{UL}{\nu}$
is the Reynolds number, where $U$ and $L$ are the reference velocity and length
respectively, and $\nu$ is the kinematic viscosity. In order to solve these PDEs, they have to be discretized both in space and time. The former will determine
the coefficients for the discrete operators, which correspond to sparse matrices, which will eventually determine properties such
as the number of non-zeros per row, this being an important parameter as seen in the previous section. The latter will eventually define
the overall algorithm to solve the equations.

If the incompressible Navier-Stokes equations are discretized in space, it leads to the semi-discrete formulation of the equations,
following the notation from \citep{Trias2014},

\begin{subequations}
  \begin{align}
    M\mathbf{u}_s &= \mathbf{0}_c, \\
    \Omega\frac{d\mathbf{u}_c}{dt} + C(\mathbf{u}_s)\mathbf{u}_c
    + D\mathbf{u}_c + \Omega G_c\mathbf{p}_c &= \mathbf{0}_c,
  \end{align}
  \label{eq:ns}
\end{subequations}

where $C$ is the cell-centered convective operator, $D$ is the cell-centered diffusive operator,
$G_c$ is the discrete cell-to-cell gradient operator, and $\Omega$ is a diagonal matrix with
the volume of the cells; $\mathbf{u}_c, \mathbf{p}_c$ are the velocity and
kinematic pressure defined at the cell
center, respectively, and $\mathbf{u}_s = \Gamma_{c\rightarrow s}\mathbf{u}_c$ is the velocity
defined at the cell faces, where $\Gamma_{c\rightarrow s}$ is the cell-to-face
interpolator.

\subsection{Parallel-in-time ensemble averaging solutions} \label{sec:theory}

Numerical simulations of turbulent flows, if the boundary conditions are
not time-dependent, can generally be split into two different phases:
a transition time, which ranges from the initial conditions set by the user to
the statistically steady-state situation, and an averaging time, in which
statistics and data are collected by performing a time average of those
relevant magnitudes. This is done by considering the statistically steady state
period of the simulation ergodic, i.e., doing an ensemble
average of different simulations is equivalent to a time average of a single
simulation \citep{makarashvili_performance_2017}.

The ergodicity principle was applied as it would be more expensive to develop a single
long simulation and perform a time average of those results than performing
different independent simulations and then performing an ensemble average.
Hence, these different and independent simulations can be expressed in the same
way as in Eq. \eqref{eq:ns-nrhs}, in which the different cases in
$\mathbf{U}_c,\mathbf{P}_c$ are identical both geometrically and in terms of
boundary conditions, with only a small perturbation given to the initial field
to trigger the statistical difference between all flow states.

Hence, let this overall time averaging period, $T_A$, be divided
into $m$ flow states, i.e., $m$ statistically independent simulations. Then, each
of the $m$ flow states should last the transition time, $T_T$, and
$T_A/m$, so the whole averaging period is preserved. Hence, each
flow state will be run until $T_T + T_A/m$, ensuring that $T_A/m$ is greater
than a single time unit \citep{krasnopolsky_approach_2018}. According
to the results from \citet{makarashvili_performance_2017}, for the ergodic ensemble
averaging to yield the expected results, the different ensembles should be
statistically independent, i.e., uncorrelated.

{
This situation will lead to all the arising simulations having identical divergence,
gradient and Laplacian operators. By doing so, the construction of the convective
and diffusive operators will be equivalent for all of the $m$ cases.}

In the framework with $m$ simulations, the semi-discrete equations can be
rewritten as
\begin{subequations}
  \begin{align}
    \tilde{M}\mathbf{U}_s &= \tilde{\mathbf{0}}_c, \\
    \tilde{\Omega}\frac{d\mathbf{U}_c}{dt} + \tilde{M}\tilde{\mathbf{U}}_s\tilde{\Gamma}\mathbf{U}_c
    + \tilde{D}\mathbf{U}_c + \tilde{\Omega}\tilde{G}_c\mathbf{P}_c &=
    \tilde{\mathbf{0}}_c.
    \label{eq:ns-nrhs}
  \end{align}
\end{subequations}

Under this framework, $\tilde{M}=I_m\otimes M$,
$\tilde{\Omega}=I_m\otimes\Omega$, $\tilde{D} = I_m\otimes D$, $\tilde{G}_c
= I_m\otimes G_c$, $\tilde{\Gamma} = I_m\otimes(\Gamma_{c\rightarrow s})$;
$\mathbf{U}_c= (\mathbf{u}_{c,1}~\mathbf{u}_{c,2}~\dots\mathbf{u}_{c,m})^T$, $\mathbf{P}_c
= (\mathbf{p}_{c,1}~\mathbf{p}_{c,2}~\dots\mathbf{p}_{c,m})^T$. 

With regards to the convective term, even though it can be applied regardless of the discretization schemes used,
it has been rewritten following the symmetry-preserving discretization that reads $C=MU_s\Gamma_{c\rightarrow s}$, 
where $U_s=diag(\mathbf{u}_s)$. By doing so, in a multiple parameter set-up, $\tilde{\mathbf{U}}_s = diag(\mathbf{u}_{s,1}
~\mathbf{u}_{s,2}~\dots~\mathbf{u}_{s,m})$ is the block diagonal matrix containing the face-velocity components for the $m$ 
simulations, which essentially changes the equation as now the operator with repeated structures is not the convective term 
overall, but the divergence operator, which is applied to $\tilde{\mathbf{U}_s}\tilde{\Gamma}\mathbf{U}_c$, which may be interpreted as
a \texttt{axty} of the face-velocity components and the face-interpolated cell velocities.

These definitions will end up producing the (virtual) block structure that allows transforming
the \texttt{SpMV} to \texttt{SpMM} that will end up increasing the arithmetic
intensity. Hence, the equations will be solved in an equivalent form to Eq.
\eqref{eq:ns}, yet for $\{\mathbf{U}_c,\mathbf{P}_c\}$ as all the $m$ cases are
solved simultaneously.

The speed-up obtained in an iteration by using the improved \texttt{SpMM}
compared to using $m$ \texttt{SpMV} can be defined as
\begin{equation}
  P_{m,ite} = \frac{mt_{ite,1}}{t_{ite,m}},
\end{equation}

\noindent where $P_{m,ite}$ is the iteration speed-up by using $s$ parameters,
$t_{ite,1}$ is the iteration time of a single simulation with one parameter,
and $t_{ite,m}$ is the iteration time of the multiple parameter simulation.
Hence, if the iteration times are split as $t_{ite}^L$ and $t_{ite}^S$, being
the times in which operations that are assumed to grow linearly with $m$, e.g.
\texttt{dot,axpy,axty}, are performed, and the time spent doing operations
involving sparse matrices, respectively. Hence, 
$t_{ite,m} = t_{ite,m}^L + t_{ite,m}^S = mt_{ite,1}^L + t_{ite,m}^S$.
Introducing $\Theta=t_{ite,1}^S/t_{ite,1}$, and simplifying the equations,
\begin{equation}
  P_{m,ite} = \frac{1}{1-\Theta+\Theta P_{m,\text{\texttt{SpMM}}}^{-1}},
  \label{eq:pmite}
\end{equation}

\noindent where $P_{m,\text{\texttt{SpMM}}} = \frac{mt_{\text{\texttt{SpMM}},1}^S}{t_{\text{\texttt{SpMM}},m}}$ and its
value is bounded by Eqs. \eqref{eq:pmspmm_ub},\eqref{eq:pmspmm_lb}.

Note that from a computational standpoint, the use of more right-hand sides in
the simulations will imply a greater memory footprint compared to running an
individual simulation. This increment in terms of memory footprint will only be
due to the fields used in the run, including all the additional fields used
apart from the simulated magnitudes. This could indeed be a downside when using
the method in memory-limited systems in which having a large number of
doubles-per-cell may be problematic. Nonetheless, opposite to
\citet{krasnopolsky_approach_2018}, the usage of the method in all operators and not
just the solution of the Poisson equation will imply that all operators are
stored as for the single right-hand side case, thus leaving a smaller memory
footprint compared to \citep{krasnopolsky_approach_2018}. Moreover, as the whole
computational domain would still be used, the benefits obtained by
\citet{alsalti-baldellou_lighter_2024} in terms of lighter simulations will not be found in this
approach.

{\subsection{Obtention of uncorrelated flow states}}

The whole methodology relies on the fact that the different flow states that are going to be
simulated simultaneously are uncorrelated.

According to the works of \citet{nikitin_disturbance_2009}, the growth of disturbances,
$\varepsilon$, in
turbulent flows, is generally a wall effect. It is defined as,
\begin{equation}
  \varepsilon^2(t) = \frac{1}{V}\int_V{|\Delta\mathbf{u}|^2dV},
\end{equation}
where $V$ is the domain volume and $\Delta\mathbf{u} = (\Delta u_x,~\Delta
u_y,~\Delta u_z)^T$ is the vector of disturbances. This growth is exponential in time until 
approximately $t^+=10^3$, being $t^+$ the elapsed time
normalized with the viscous time scale $t_\tau = l_\tau/u_\tau$, with
$l_\tau=\nu/u_\tau^2$, in which these disturbances
remain constant at approximately $\varepsilon\approx10^{-1}$. This exponential
growth corresponds to the classical behavior of chaotic systems where the
perturbations grow with the so-called Lyapunov exponent \citep{Keefe1992}.
According to the results from \citet{badii_correlation_1988}, the correlation among different
flow states will decay exponentially with the generalized Lyapunov exponent for
1D cases, which indicates that for $t\rightarrow\infty$,there will not be any
correlation among all flow states. Later on, this behavior was extended to 2D
analysis by \citet{mendes_decay_2019}.

Even after introducing pertirbations, the average solution is expected to behave similarly \citep{cheskidov_global_2006,Bortolan2024}. 
This is ensured by a weak global attractor, i.e. a statistical steady state, for the three-dimensional Navier-Stokes
equations that constrain the possible behaviors of the solutions, even
though these might not be unique \citep{Bortolan2024}. However, suppose these
solutions contained in the weak attractor are strongly continuous, according to
\citet{cheskidov_global_2006}. In that case, there will be a strong global attractor,
and the solutions will behave predictably and stably over time instead of just their
average values as in the presence of a weak global attractor. Hence, as suggested by
\citet{Keefe1992}, the existence of this attractor underlies predictable solutions to the
Navier-Stokes equations in periodic flows.

{\subsection{Inline mesh refinement for faster convergence}}

Hence, as previously stated, every flow state will be integrated for
$T_T+\frac{T_A}{m}$, in which the transition time is solved completely for
every flow state while the averaging time is divided among all the flow states.
In this case, the transition time is defined as the time in which the
different flow states become uncorrelated. In order to quantify how relevant
the transition time is compared to the averaging time,
\citep{krasnopolsky_approach_2018} defines the times ratio as the ratio between the
averaging time and the transition time, $\beta=T_A/T_T$, and estimates the
whole simulation speed-up as

\begin{equation}
  P_m = \frac{1+\beta}{m+\beta}\frac{5m}{5m-3\theta(m-1)},
  \label{eq:pm_krasn}
\end{equation}

where $\theta$ is the weight of the Poisson equation in the whole iteration.
The speed-up calculation assumes the time spent in all the iterations of the whole time
integration is the same. Nonetheless, the actual factor computed by
\citet{krasnopolsky_approach_2018} assumes the cost per iteration is maintained
constant throughout the simulation.

\begin{figure}[h]
  \centering
  \includegraphics[width=0.8\textwidth]{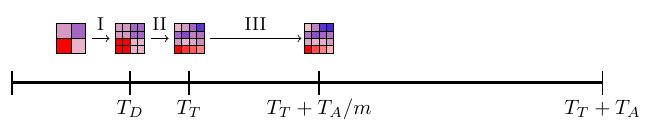}
  \caption{Proposed ensemble averaging strategy in which the case is run until
  $T_D$ in a coarser setup, and then in the intended setup for
  $t\in\left(T_D,T_T+T_A/m\right)$. I: mapping, II: developing, III: averaging.}
  \label{fig:time-split}
\end{figure}

Assume the simulation is started with a less costly procedure than
the expected simulation, e.g., with a coarser mesh. After the flow has
developed and the simulations are expected to be independent in the coarse
mesh, $T_D$; the mesh is refined to the objective mesh, and the flow is developed 
until $T_T$, in the time period named $T_{DT}$. Afterwards, the procedure is identical to the one presented in
\citet{krasnopolsky_approach_2018}. Thus, the followed strategy for the mesh refinement is represented schematically in Fig.
\ref{fig:time-split}.  {The mesh refinement strategy used in this work maps the solution from the coarse to the fine nodes consistently, and it is designed to work both for structured and unstructured grids. In both cases, the strategy follows classical adaptive mesh refinement techniques. A number of open-source libraries implementing such a procedure exist, including p4est \cite{burstedde_p4est_2011}, AMReX \cite{Zhang2019}, or Chombo \cite{Martin2025}.} Note that this in-flight mesh refinement will indeed have
a wall-clock time overhead compared to running the fine mesh since the
beginning of the simulation and thus it should be considered when analyzing the
results with and without mesh refinement.

Hence, it can be assumed that $\overline{\Delta t}_{DT,m} \approx \overline{\Delta
t}_{A,m}=\overline{\Delta t}^F_m$. Moreover,
$\bar{t}_{DT,m}\approx\bar{t}_{A,m}=\bar{t}^F_m$, being $\bar{t}$ the
average wall-clock time per iteration, while the superscript $F$ indicates it
corresponds to the fine mesh. On the other hand, $\overline{\Delta t}^C_m$ and
$\bar{t}^C_m$ indicate the average timestep and wall-clock time for the coarse
mesh, respectively. Hence,
for an arbitrary case with $m$ flow states, let $\tau_m$ be the wall clock time
of the whole simulation with $m$ flow states,

\begin{equation}
  \tau_m = \frac{\bar{t}_m^C}{\overline{\Delta t}_m^C}T_D
  + \frac{\bar{t}_m^F}{\overline{\Delta t}_m^F}\left(T_{DT}+\frac{1}{m}T_A\right)
\end{equation}

Let $T_{D}=\gamma T_{DT}$, and $f_m^X = \bar{t}_m^X/\overline{\Delta t}_m^X$, where
$X$ stands for both $F,C$. Thus,

\begin{equation}
  \tau_m
  = \frac{T_T}{1+\gamma}\left[f_m^C\gamma+f_m^F\left(1+\frac{1+\gamma}{m}\beta\right)\right]
\end{equation}

By definition, the overall speed-up becomes $P_m = \tau_1/\tau_m$, which leads
to

\begin{equation}
  P_m =\frac{1+\delta_1\gamma
  + (1+\gamma)\beta}{m(\delta_m\gamma+1)+(1+\gamma)\beta}\frac{mf_1^F}{f_m^F},
  \label{eq:speedup1}
\end{equation}

\noindent where $f_m^C = \delta_mf_m^F$. The formulation from Eq. \eqref{eq:speedup1} is
equivalent to the single mesh speed-up from \citet{krasnopolsky_approach_2018} if
$\gamma=0$, which sets $T_D=0$. Eq. \eqref{eq:speedup1} can be rewritten as

\begin{equation}
  P_m = \frac{1
  + \tilde{\beta}}{\Phi m+\tilde{\beta}}\frac{mf_1^F}{f_m^F},
  \label{eq:pm_full_v1}
\end{equation}

\noindent where $\tilde{\beta} = \frac{1+\gamma}{1+\delta_1\gamma}\beta$, and $\Phi
= \frac{1+\delta_m\gamma}{1+\delta_1\gamma}$. In this case, it may be assumed
that $\overline{\Delta t}_m^F \approx \overline{\Delta t}_1^F$, which simplifies Eq.
\eqref{eq:pm_full_v1} to

\begin{equation}
  P_m = \frac{1
  + \tilde{\beta}}{\Phi m+\tilde{\beta}}P_{m,ite}^F.
  \label{eq:pm_full_v2}
\end{equation}

Eventually, Eq. \eqref{eq:pm_full_v2} is characterized by three parameters:
$P_{m,ite}^F$, $\Phi$ , and $\delta_m$ within $\tilde{\beta}$. Note that $\delta_m^{-1}$ indicates how faster the coarse mesh is in advancing
in time compared to the fine mesh, considering both the bigger timestep that
preserves the stability of the integration and the faster the iterations due to
a reduced problem size. Hence, this factor will be strongly related to the
refinement factor of the mesh, $r_f$. Hence, it can be assumed that values of
$\delta_m^{-1}$ will be around $r_f^4$. Let $\Pi=\delta_1^{-1}$. Then,

\begin{equation}
  \tilde{\beta} = \frac{\Pi(1+\gamma)}{\gamma+\Pi}\beta.
  \label{eq:bbtilde}
\end{equation}

\begin{figure}[h]
  \centering
  \includegraphics[width=0.4\textwidth]{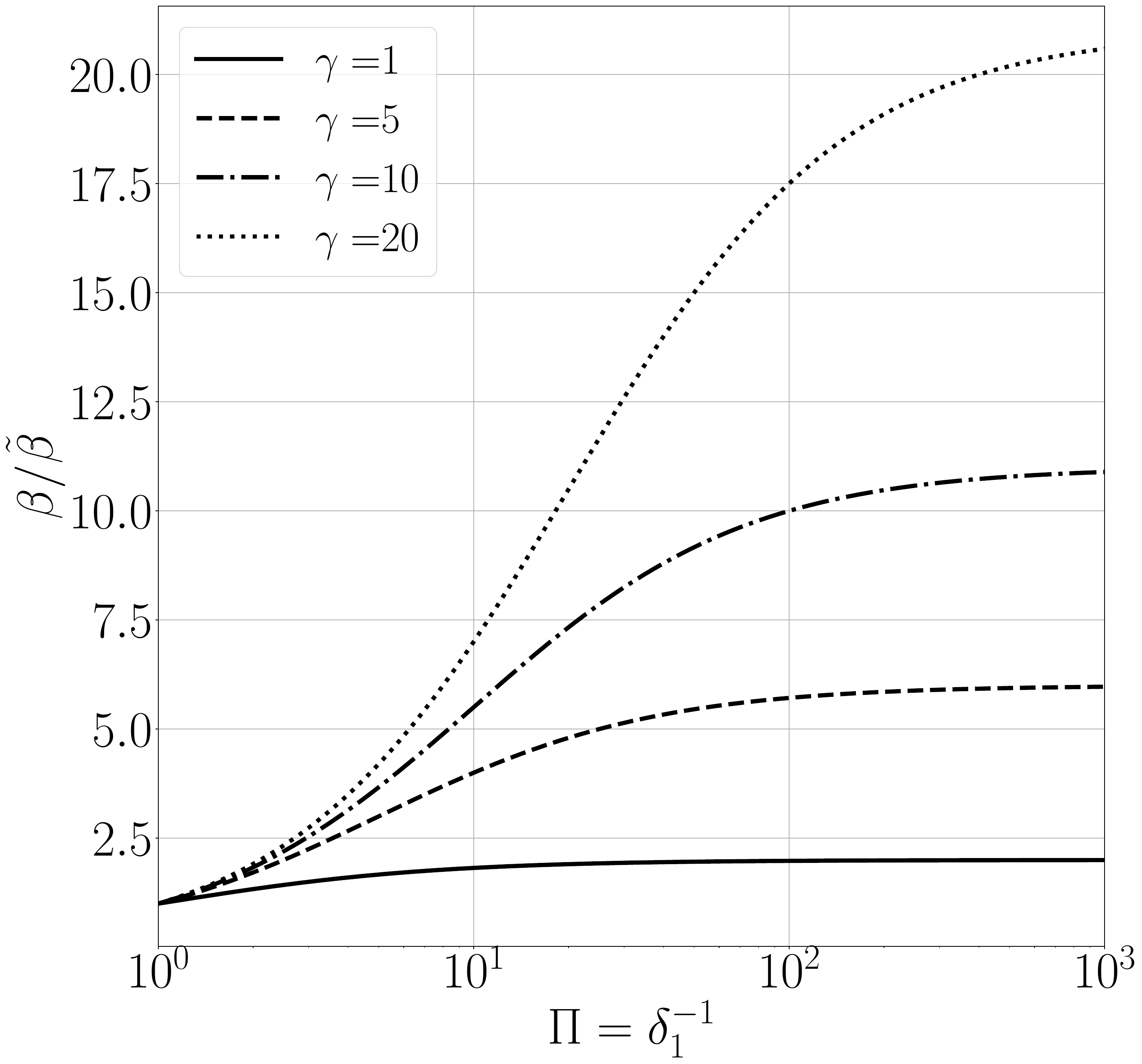}
  \caption{$\tilde{\beta}/{\beta}$ as a function of $\Pi$ and
  $\gamma$.}
  \label{fig:bbtilde}
\end{figure}

Fig. \ref{fig:bbtilde} shows the behaviour of $\tilde{\beta}/\beta$ as
a function of $\gamma$ and $\Pi$, which clearly shows that the presence of
a refinement of the mesh after $T_D$ will ultimately increase the effective times ratio,
which will imply bigger speed-ups in the whole simulation.

On the other hand, the selection of $\gamma$ generates a trade-off for the
user, as a smaller value will ensure bigger convergence to a statistically
steady state before the start of the averaging process, leading to smaller
speed-ups compared to using a greater value of $\gamma$, yet with the
\textit{a priori} uncertainty if the case would be properly converged.
Introducing the definition from Eq. \eqref{eq:pmite} as well as from all the relevant parameters defined at Table \ref{tab:params}, the speed-up obtained in the whole simulation
considering the refinements can be estimated as

\begin{equation}
  P_m = \frac{1+\tilde{\beta}}{\Phi m+\tilde{\beta}}\frac{1}{1-\Theta+\Theta
  P_{m,\text{\texttt{SpMM}}}^{-1}}.
  \label{eq:fullpm}
\end{equation}

\begin{table}[h]
  \centering
  \caption{Relevant parameters for the parallel-in-time speed-ups.}
  \begin{tabular}{cccc} \toprule
    \multicolumn{2}{c}{\textbf{Speed-ups and times}} & \multicolumn{2}{c}{\textbf{Parallel-in-time
    parameters}} \\ \toprule
    Parameter & Definition & Parameter & Definition \\ \toprule
    Poisson speed-up, $P_{m,\text{Poisson}}$ & Eq. \eqref{eq:poipm} & Times ratio, $\beta$ & $T_A/T_T$ \\
    Iteration speed-up, $P_{m,ite}$ & Eq. \eqref{eq:pmite} & Transition ratio, $\gamma$ & $T_D/T_{DT}$ \\
    Simulation speed-up, $P_m$ & Eq. \eqref{eq:fullpm} & Simulation pace, $f_m$ & $\bar{t}_m/\overline{\Delta t}_m$ \\
    Averaging time, $T_A$ & Fig. \ref{fig:time-split} & Mesh pace ratio, $\delta_m$ & $f_m^C/f_m^F$ \\
    Transition time, $T_T$ & Fig. \ref{fig:time-split} & Advancing ratio, $\Pi$ & $\delta_1^{-1}$\\
    Coarse transition time, $T_D$ & Fig. \ref{fig:time-split}  & Right-hand side multiplier, $\Phi$ & $(1+\delta_m\gamma)/(1+\delta_1\gamma)$  \\
    Fine transition time, $T_{DT}$ & Fig. \ref{fig:time-split} & Effective times ratio, $\tilde{\beta}$ & Eq. \eqref{eq:bbtilde} \\ \bottomrule
  \end{tabular}
  \label{tab:params}
\end{table}

In order to make \textit{a priori} estimations on the overall speed-up of the
methods, $\Phi=1$ for all $m$. This hypothesis indicates that the effect of the in-flight mesh
refinement does not depend on $m$.  Nonetheless, this assumption will be validated
when analyzing the numerical experiments. With these results and using Eq. \eqref{eq:fullpm}, the speed-ups given $\{m,
\tilde{\beta}\}$, can be bounded so that the performance obtained by the
implementation should lay in between those values. 

To exemplify the effects in terms of speed-up, a weight of the 
\texttt{SpMV} in the whole simulation of 65\%, i.e. $\Theta = 0.65$, will be assumed. Moreover, the advancing ratio $\Pi$ will be 
set to 16, which is a reasonable value for a Cartesian mesh where the number of divisions per direction has been doubled, as it 
assumes a factor of 8 ($2^3$) in the number of cells per direction, and a factor of 2 in the timestep. Moreover, $\gamma=5$ will 
be used, which indicates that the amount of time units spent in the coarse mesh is five times the time spent transition in the 
fine mesh. Therefore, with these parameters, the estimated upper-bound simulation speed-ups for
a simulation in which spatial discretization considers only the first neighbors
are presented in Table \ref{tab:ub_7nnz}. Note that the conditions presented
for the density of the matrix (7 non-zeros per row) present the worst-case
scenario in the implementation, as a denser matrix will provide better results. However, it can be observed that figures ranging
from 14\% in the most ill-conditioned case (i.e. $\beta=1$) to 78\% in the most favourable case ($\beta=40$) are present. 
It is remarkable to notice that for bigger values of $\beta$, the overall speed-up is not affected by using more rhs.

In order to test with a more favorable condition,
a denser matrix ($\text{nnz}(A)/n=27$), which would appear in a discretization
where not only the first neighbors normal to the faces are considered but also the
diagonal neighbors, has also been tested and presented in Table \ref{tab:ub_27nnz}. In this case, the actual weight of the \texttt{SpMM}
should increase from 65\%. Nonetheless, for the sake of comparing speed-ups under the same circumstances, it has been left untouched.
Note that even for times ratio
values of 1, a 22\% speed-up can be found for two flow states. At the same
time, a 19\% can be obtained with four. Moreover, for bigger times ratio values up to 40,
it can be observed that some cases exceed speed-ups of 100\%, using the same
amount of computational power. Note that previous works ensured that
the speed-up estimates peaked for $m$ around $\mathcal{O}(1)$, which led to
$T_A/m$ greater than a time unit in most cases. Nonetheless, with the results
in Table \ref{tab:ub_27nnz} which would be possible with high-order spatial
discretizations, these speed-ups
become the greatest for greater number of right-hand sides, thus implying that
the restriction $T_A/m>1$ has to be considered for cases with bigger $m$.

\begin{table}[h]
  \centering
  \caption{Estimated upper bound speed-ups considering $\Theta=0.65$, $\Pi=16$,
  $\gamma=5$, given a matrix with $\text{nnz}(A)/n=7$, for times ratio
  $\beta$ values ranging from 1 to 40 and 1 to 16 simultaneous flow states.}
\begin{tabular}{@{}cccccccc@{}}
  \cmidrule(l){3-8}
  \multirow{5}{*}{\rotatebox{90}{$m$}} & \multicolumn{1}{c|}{16} & 0.52 & 0.78 & 1.18  & 1.46  & 1.66  & 1.78   \\
                                            & \multicolumn{1}{c|}{8}  & 0.81 & 1.07 & 1.40  & 1.58  & 1.69  & 1.75   \\
                                            & \multicolumn{1}{c|}{4}  & 1.06 & 1.26 & 1.44  & 1.53  & 1.58  & 1.60   \\
                                            & \multicolumn{1}{c|}{2}  & 1.14 & 1.22 & 1.29  & 1.31  & 1.33  & 1.34   \\
                                            & \multicolumn{1}{c|}{1}  & 1.00 & 1.00 & 1.00  & 1.00  & 1.00 & 1.00   \\ \cmidrule(l){3-8} 
                                            &                         & 1    & 2    & 5     & 10    & 20    & 40     \\
                                            &                         & \multicolumn{6}{c}{$\beta$}    
\end{tabular}
  \label{tab:ub_7nnz}
\end{table}

\begin{table}[h]
  \centering
  \caption{Estimated upper bound speed-ups considering $\Theta=0.65$, $\Pi=16$,
  $\gamma=5$, given a matrix with $\text{nnz}(A)/n=27$, for times ratio
  $\beta$ values ranging from 1 to 40 and 1 to 16 simultaneous flow states.}
\begin{tabular}{@{}cccccccc@{}}
  \cmidrule(l){3-8}
  \multirow{5}{*}{\rotatebox{90}{$m$}} & \multicolumn{1}{c|}{16} & 0.63 & 0.93 & 1.42  & 1.75  & 1.99  & 2.14   \\
                                            & \multicolumn{1}{c|}{8} & 0.94 & 1.25 & 1.64 & 1.85 & 1.98 & 2.05   \\
                                            & \multicolumn{1}{c|}{4} & 1.19 & 1.41 & 1.62 & 1.72 & 1.77 & 1.80   \\
                                            & \multicolumn{1}{c|}{2} & 1.22 & 1.30 & 1.37 & 1.40 & 1.42 & 1.42   \\
                                            & \multicolumn{1}{c|}{1} & 1.00 & 1.00 & 1.00  & 1.00  & 1.00 & 1.00   \\ \cmidrule(l){3-8} 
                                            &                         & 1    & 2    & 5     & 10    & 20    & 40     \\
                                            &                         & \multicolumn{6}{c}{$\beta$}    
\end{tabular}
  \label{tab:ub_27nnz}
\end{table}

\subsection{Multiple parameter simulations}

Parametric studies such as the numerical computation of the $C_L-\alpha$ curve of an
airfoil, or the effect of the thermal conductivity in a wall, usually requires a number of simulations to be run independently, i.e.
the user launches multiple simulations with a given angle of attack $\alpha$, or with different thermal conductivities $\lambda$, respectively; in
order to compute the pressure and viscous forces over the airfoil which end up
generating the lift and drag forces to the airfoil and thus the characteristics
of this airfoil could be extracted.

In general, let a flow simulation in which an arbitrary parameter $\xi$
varies in such a way that there are $m$ options, $\bm{\xi}
= \{\xi_1,\xi_2,\dots,\xi_m\}$, with an identical mesh and geometry.
{Therefore, the presence of multiple parameters will lead as well to the same
virtual matrix block structure as in the case of multiple flow states. However, while the 
inline mesh refinement strategy cannot be applied in this case, the use of the multiple parameters
will lead to the same benefits in terms of performance as in the case of multiple flow states.}

\section{Numerical experiments} \label{sec:exp}

In order to validate the method presented hereabove, the incompressible
Navier-Stokes equations are discretized in space following
a finite-volume symmetry-preserving discretization \citep{Verstappen2003,Trias2014} so that
the properties of the divergence, gradient, and Laplacian operators are
preserved in such a way that no additional numerical dissipation is added in
the case of using a staggered grid. The semi-discretized equations are
integrated in time using Heun's third-order Runge-Kutta scheme and applied to
the Navier-Stokes equations as in \citep{sanderse_accuracy_2012}. Note that this
implementation will introduce some numerical dissipation, as shown by
\citet{capuano_explicit_2017}. The time-step size will be calculated on the run using the
\texttt{AlgEigCD} framework developed by \citet{trias_efficient_2024} in which the time step is computed
following by bounding the eigenvalues of the system using a \texttt{SpMV}-based
system so that the matrices are not reconstructed every time-step as it is
based in the incidence matrices as well as the mass flows and diffusivities at
the faces. The use of this adaptive time-step technique will ensure stability in the
integration even after the in-flight mesh refinement is performed.

The method has been implemented in the in-house code Termofluids Algebraic
(TFA) in which the calculations are carried out using the HPC$^2$ framework
\citep{alvarez_hpc2fully-portable_2018}, which allows the portability of the code to multiple
architectures as well as integrating it with external libraries such as HYPRE
\citep{falgout_design_2006} or Chronos \citep{isotton_chronos_2021}, while still having its own solvers and
kernels. The runs with CPU's have been performed in MareNostrum
5 General Purpose Partition (MN5-GPP) supercomputer. In
  this case the numerical solution of the Poisson equation was 
tackled with a conjugate gradient solver preconditioned with the adaptive algebraic multigrid (aAMG) \cite{paludetto_magri_novel_2019}
provided by Chronos. The configuration used was tailored for CFD problems and consisted of a PMIS coarsening
\cite{de_sterck_reducing_2006}, an extended+I interpolation \cite{de_sterck_distancetwo_2008}, and one pre- and post-smoothing steps
using a lightweight factored sparse approximate inverse with five nonzeros per row.

  The techniques previously presented have been applied to a turbulent planar
  channel flow, a Rayleigh-Bénard natural convection case as two examples of
  academic flows that are relevant given the reproducibility of the results in
  previous publications and benchmarks, thus leading to a verification of the
  method, where a parallel-in-time approach is applied. 
  Finally, the method has been applied as well to an industrially
  relevant case such as an airfoil with both
  flaps and slats depleted, in a multiple parameter framework, in order to prove the applicability of the method
  not only to academic cases but also to industrial flows.

\subsection{Channel flow}

First, the methodology has been tested in a turbulent planar channel flow with
$\text{Re}_\tau=180$. The method has been tested with a $160^3$ grid with
a hyperbolic tangent refinement in the $y$ direction with $\gamma=1.5$, which
ensures a theoretical $y^+$ at the wall of 0.65. Moreover, 
the test has been run using both of the proposed frameworks in this paper: with
and without mesh refinement. In the case in which the refinement was applied,
the refinement factor used was $r_f=2$, leading to a $80^3$ mesh before the
mapping is applied.

The simulation was run for 352 half-channel-height time-units, $\delta/u_\tau$, with a 32-time-unit transition period, 
determined from probe data observations in
the core of flow, showing steady-state behaviour had been reached. In the case
of using the coarse mesh to launch faster the simulation, the coarse mesh has
been run for 30 time units before interpolating to the fine mesh. Note that
these 32 time units exceed notably the expected time according to
\cite{nikitin_disturbance_2009}, which sets the time to have independent fields to
$t^+=10^3$, which in the present case would correspond to 5.6 time units, given the ratio between the viscous time-units and the half-channel-height
time units ($t^+/t^*=\delta\text{Re}_\tau u_\tau$). This assumes
a fully developed turbulent flow, whereas our case starts from a perturbed Poiseuille profile. Moreover,
a safety factor of 0.95 has been set to the time step calculated using the
\texttt{AlgEigCD} method in order to ensure the simulation is stable. Regarding the space discretization, 
the chosen has been a second-order symmetry-preserving discretization \cite{Verstappen2003}.

The simulation has been run in 64 cores of a node of MN5-GPP supercomputer in
order to set a reasonable CPU-load of 64k cells per CPU core, so that the
cores are loaded enough to perform efficiently. The parallelization strategy used has
been the combination of MPI+OpenMP.

\begin{figure}[h]
  \centering
  \includegraphics[height=0.3\textwidth]{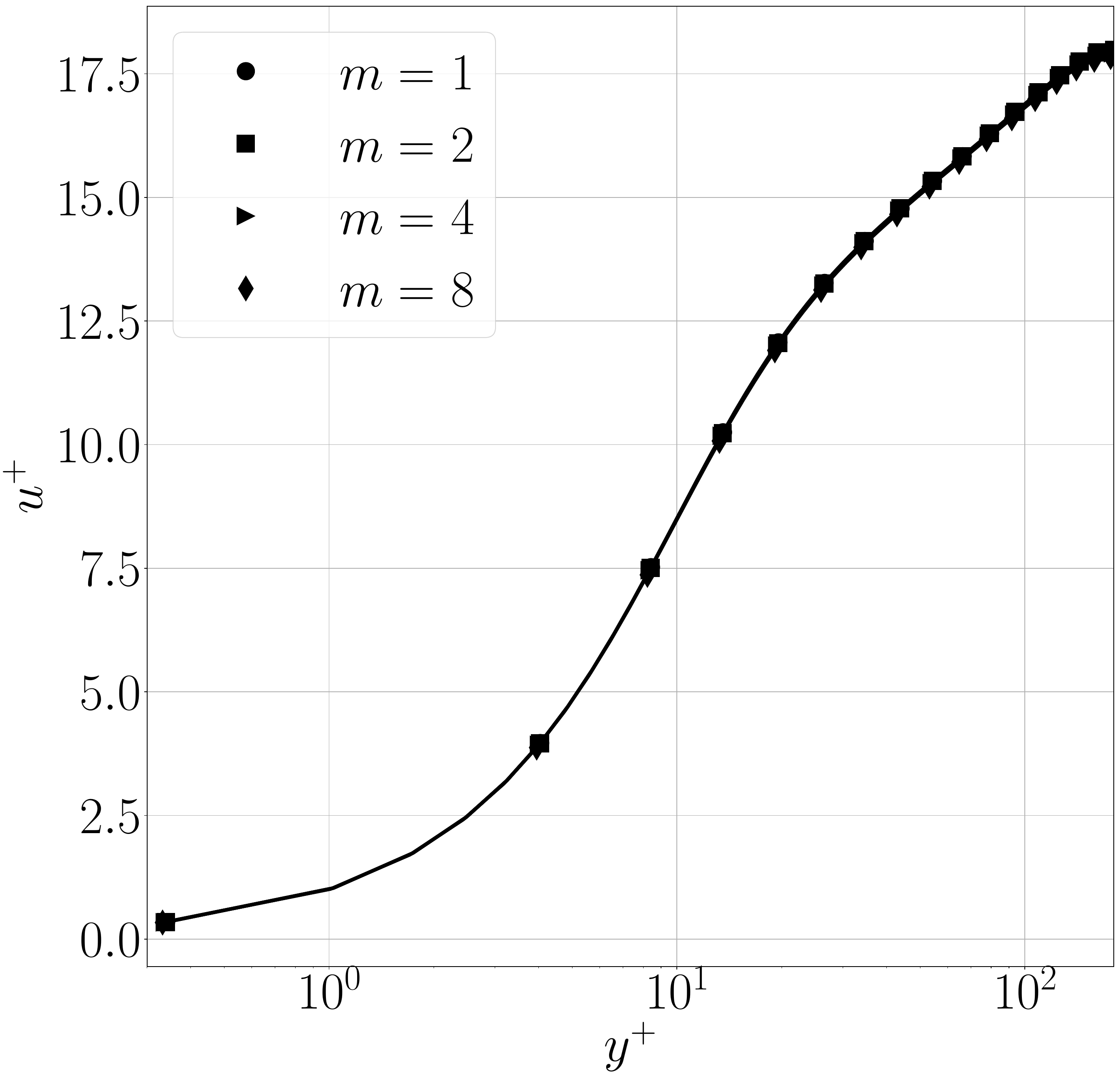}
  \includegraphics[height=0.3\textwidth]{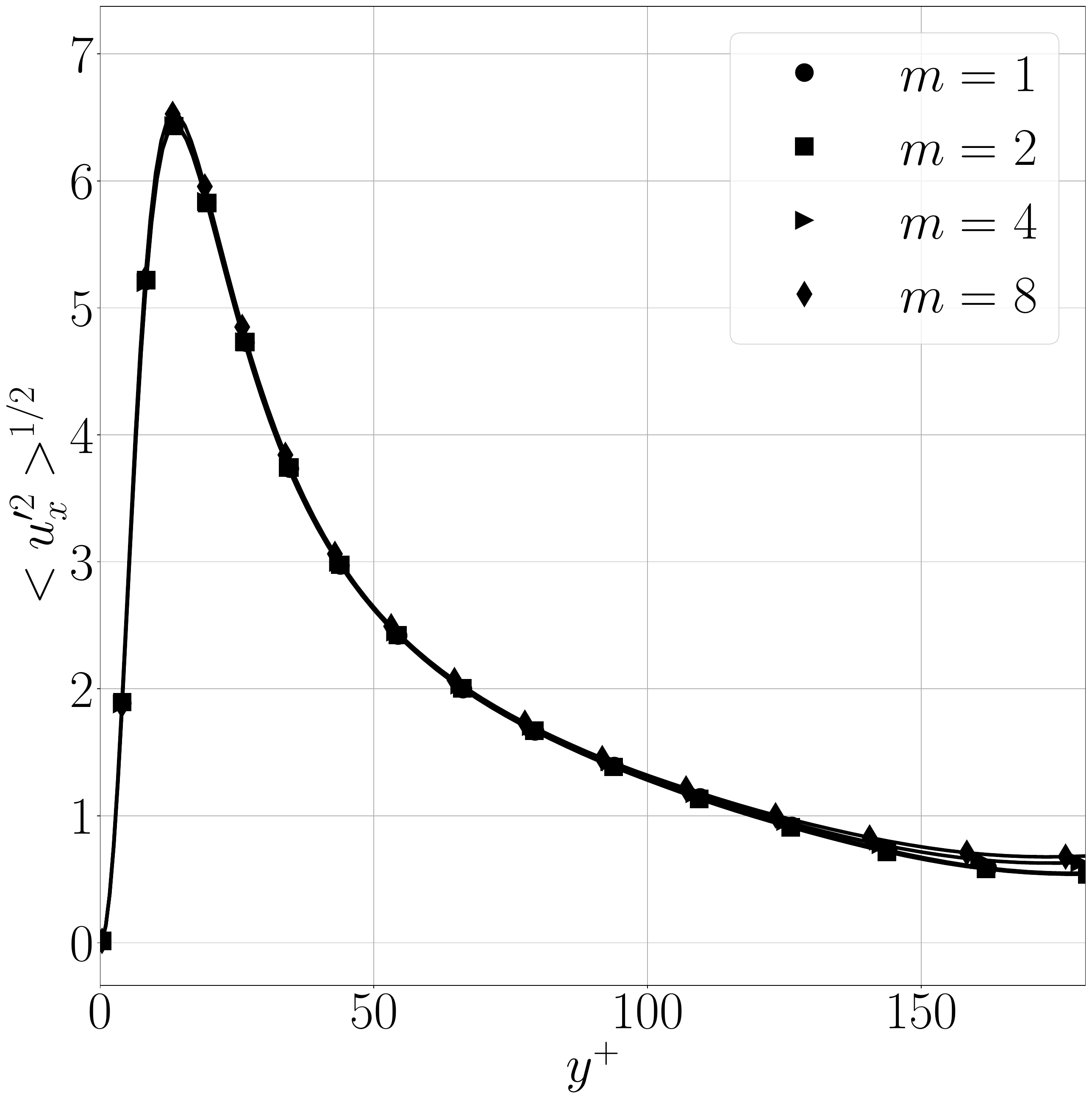}
  \caption{Average velocity in wall units (left) and rms streamwise
  velocity (right) profiles for 1, 2, 4, and 8 rhs in a turbulent planar
  channel flow of $\text{Re}_\tau=180$.}
  \label{fig:cf_results}
\end{figure}

Figure \ref{fig:cf_results} shows the results for 1, 2, 4 and 8 rhs in the
aforementioned set-up. It can be seen that making use of the ensemble averaging
technique with parallel-in-time simulations provide the equivalent results to the
classical time average simulation ($m=1$) for both first and second order averages,
for all used number of flow states, which validates the method in such a way that the 
obtained average solutions are equivalent, with minor differences for the second-order
averages at the centre of the channel.

The tested conditions in terms of $T_D,T_T,T_A$ give $\beta=10$ and $\gamma=15$ which will be used to
verify Eq. \eqref{eq:fullpm}. Table \ref{tab:dt-wct} shows the average wall-clock time per iteration and 
time-step for both meshes used and for all the flow states tested in this section. Note that the values presented 
for $\delta_m$ do not vary much compared to the expected value of $1/r_f^4$, which in this case would be 0.0625,
apart from the 4 rhs case, which differs slightly more than the other three cases.

\begin{table}[h]
  \centering
  \caption{Average time-steps, wall-clock times per iteration [s] and $\delta_m$ for both coarse ($80^3$) and fine ($160^3$) meshes and
  all the simulated flow states for a turbulent planar channel flow of $\text{Re}_\tau=180$.}
  \begin{tabular}{cccccc}
    $m$ & $\overline{\Delta t}_m^C$ & $\bar{t}_m^C$ & $\overline{\Delta t}_m^F$ & $\bar{t}_m^F$ & $\delta_m$ ($\Pi_m$) \\ \toprule
    1 & $2.66\times10^{-3}$ & 0.1167 & $1.30\times10^{-3}$ & 0.9152 & 0.0624 (16.018) \\
    2 & $2.538\times10^{-3}$ & 0.1651 & $1.286\times10^{-3}$ & 1.3267 & 0.0630 (15.858) \\
    4 & $2.438\times10^{-3}$ & 0.2554 & $1.739\times10^{-3}$ & 2.2454 & 0.0812 (12.320) \\
    8 & $2.340\times10^{-3}$ & 0.4795 & $1.092\times10^{-3}$ & 4.2377 & 0.0528 (18.943) 
  \end{tabular}
  \label{tab:dt-wct}
\end{table}

\begin{figure}[h]
  \centering
  \includegraphics[height=0.3\textwidth]{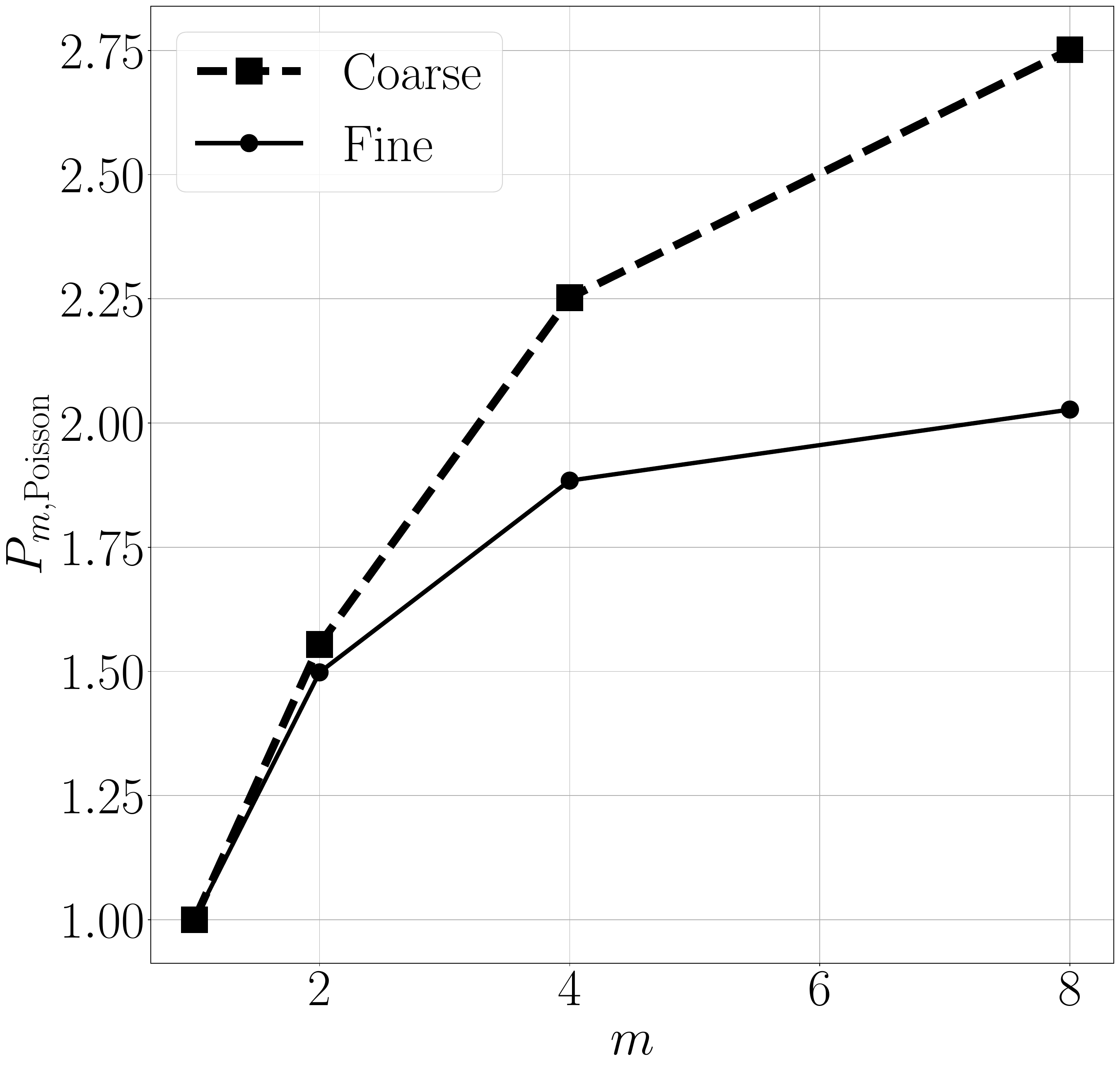}
  \includegraphics[height=0.3\textwidth]{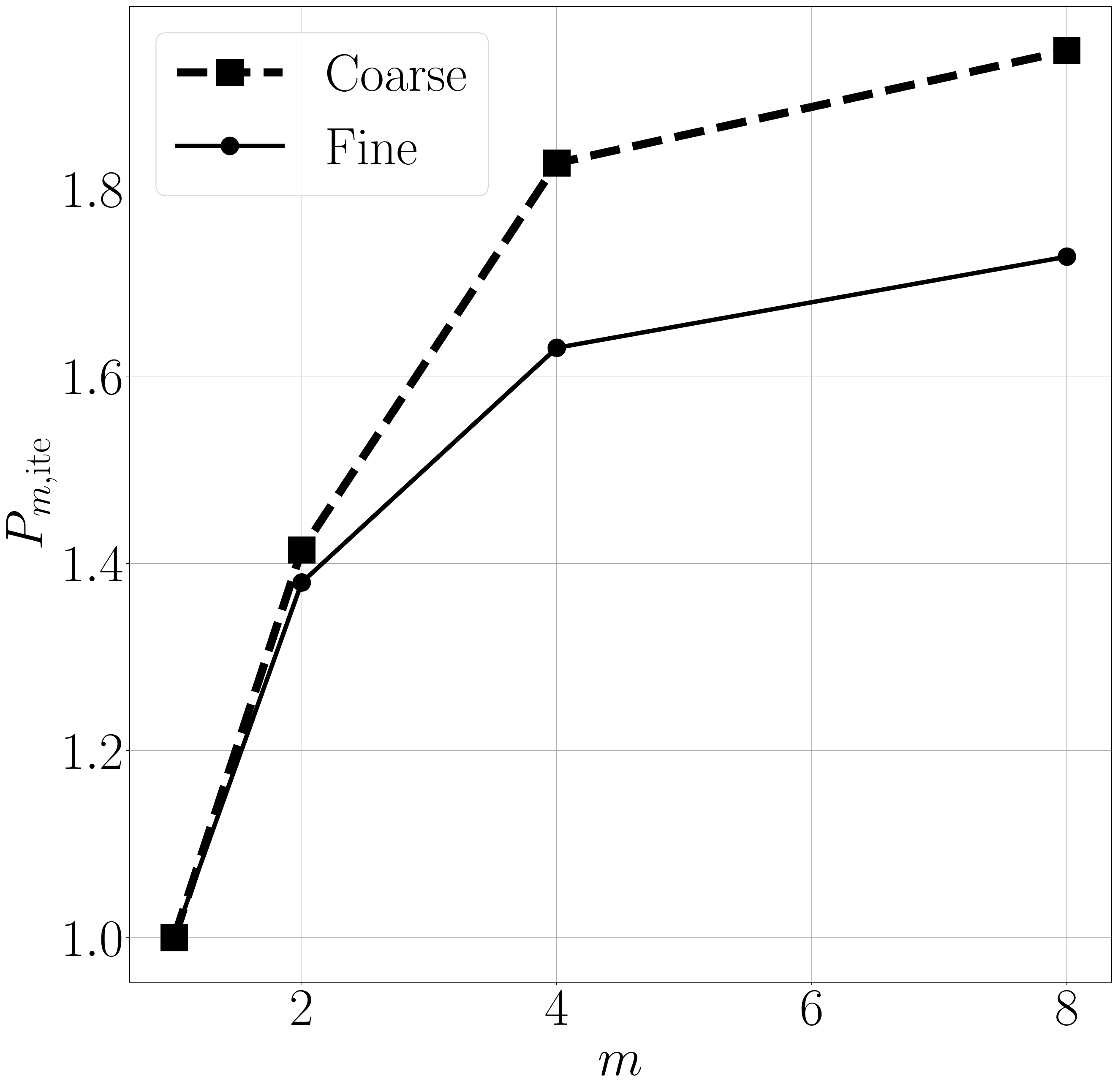}
  \includegraphics[height=0.3\textwidth]{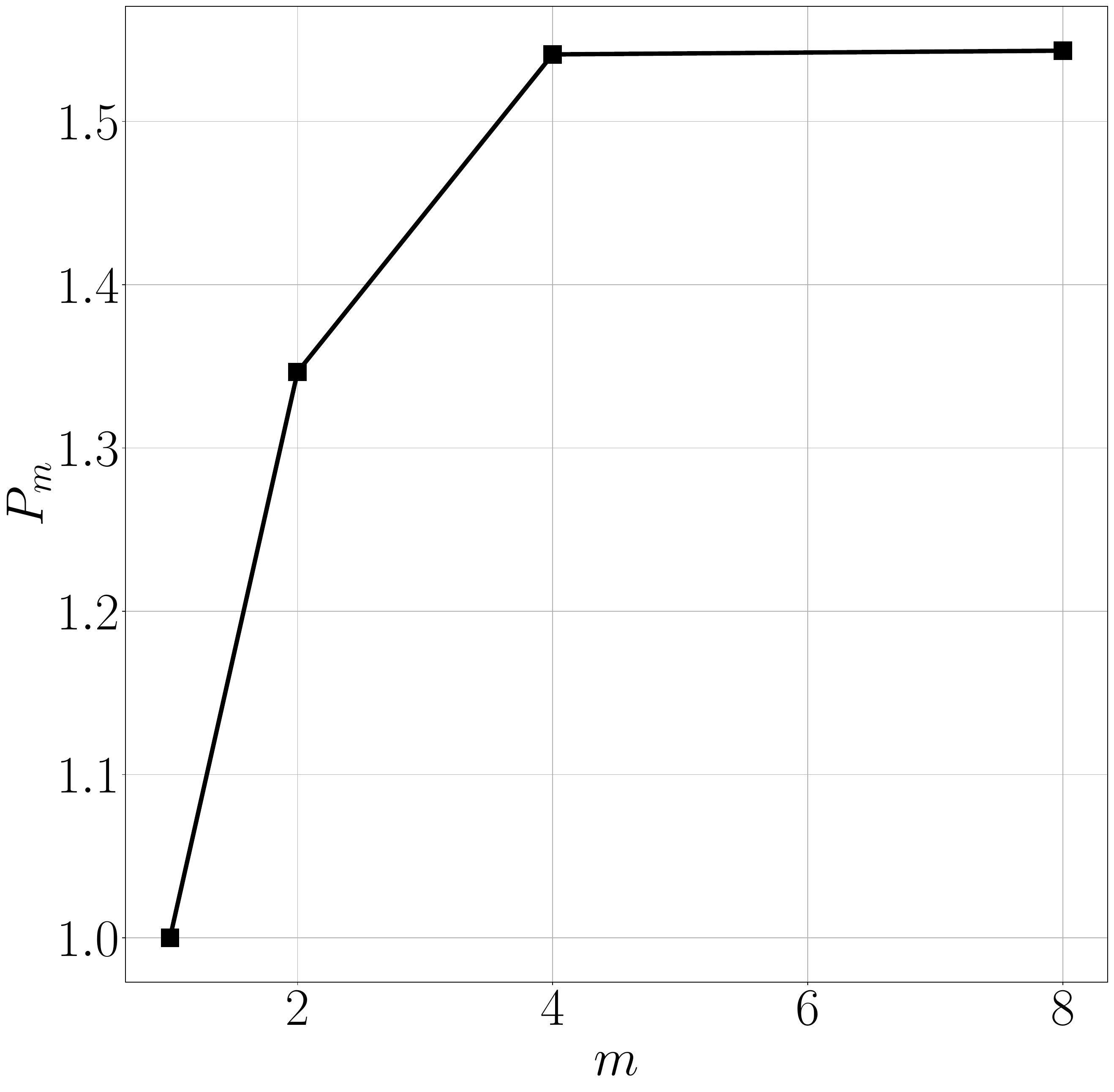}
  \caption{Speed-up values obtained in the numerical solution of the Poisson equation (left), a whole projection method iteration (center), 
  and the whole simulation (right) for both coarse ($80^3$) and fine ($160^3$) meshes and all simulated flow states for a turbulent planar
  channel flow of $\text{Re}_\tau=180$.}
  \label{fig:pm_cf}
\end{figure}

Thus, the speed-ups for the whole simulation given the aforementioned conditions can be obtained, and are displayed in Table \ref{tab:pm_cf}.
These results indicate that there is a speed-up ranging from 30\% with 2 flow states, to 54\% with both 4 and 8, with this speed-up value
obtained without the need of using any additional computational resources. Figure \ref{fig:pm_cf} shows the trend of these values, where it
can be observed that for 2 flow states, the obtained speed-ups both in Poisson and iteration are similar for both coarse and fine grids, yet for both 4 and 8 rhs, the difference between both meshes increases.

\begin{table}[h]
  \centering
  \caption{Speed-up values obtained in the numerical solution of the Poisson equation and the whole projection method iteration for 
  both coarse ($80^3$) and fine ($160^3$) meshes and all simulated flow states in a turbulent planar channel flow of $\text{Re}_\tau=180$.}
  \begin{tabular}{cccccc}
    $m$ & $P_{m,\text{Poisson}}^C$ & $P_{m,ite}^C$ & $P_{m,\text{Poisson}}^F$ & $P_{m,ite}^F$ & $P_{m,sim}$ \\ \toprule
    1 & 1 & 1 & 1 & 1 & 1 \\
    2 & 1.553 & 1.414 & 1.498 & 1.379 & 1.346 \\
    4 & 2.252 & 1.828 & 1.884 & 1.630 & 1.541 \\
    8 & 2.751 & 1.948 & 2.027 & 1.728 & 1.543
  \end{tabular}
  \label{tab:pm_cf}
\end{table}

Given the iteration speed-up values for the fine grid, Eq. \eqref{eq:pm_full_v2} can be verified, as well as the hypothesis of $\Phi=1$.
First of all, both estimation methods have been compared using an $L_2$ norm of the relative error vector $\varepsilon_i = |a_i-b_i|/a_i$,
as in any case these vectors will have a zero-value, yielding a difference of 1.3\%, which makes the assumption of $\Phi=1$ valid in this
case. Both methods using the actual $\Phi$ values and $\Phi=1$ yield a 9.5\% error and 8.2\% error when compared to the actual speed-ups,
as shown in Table \ref{tab:results_pm}.

\begin{table}[h]
  \centering
  \caption{Comparison of estimated simulation speed-up methods with the actual simulation speed-up given $\gamma=15$ and $\beta=10$.}
  \begin{tabular}{cccc}
    $m$ & $P_{m,\Phi}$ & $P_{m,\Phi=1}$ & $P_{m,sim}$ \\ \toprule
    1 & 1 & 1 & 1 \\
    2 & 1.313 & 1.313 & 1.346 \\
    4 & 1.491 & 1.501 & 1.541 \\
    8 & 1.602 & 1.591 & 1.543
    \end{tabular}
    \label{tab:results_pm}
\end{table}

On the other hand, given the actual iteration speed-ups presented in Table \ref{tab:pm_cf}, and $\beta=10$, the speed-ups
in the whole simulation considering a single mesh set-up can be obtained with Eq. \eqref{eq:pm_full_v2} by setting 
$\tilde{\beta} = \beta$ and $\Phi=1$. In this case, the
obtained speed-ups with equivalent $T_T, T_A$ would be 1.233, 1.210 and 0.943 for 2, 4, and 8 rhs, showing an improvement of an
11.22\%, 33.02\% and 60.01\%, respectively for the tested flow states.
 
\begin{figure}[h]
    \centering
  \raisebox{-0.5\height}{\includegraphics{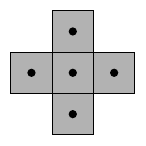}}%
  \hspace{1em}
  \raisebox{-0.5\height}{\includegraphics{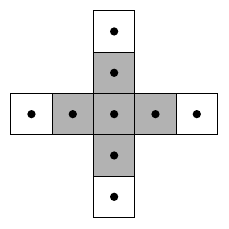}}%
  \hspace{1em}
  \raisebox{-0.5\height}{\includegraphics{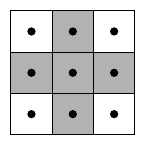}}
    \caption{Two-dimensional representations of the stencil for \texttt{7p}
    (left), \texttt{13p} (center), \texttt{27p} (right). The shaded cells
    represent the first neighbors in all representations.}
    \label{fig:discretization}
  \end{figure}

Moreover, the test case has been run for different number of non-zeros per row
in order to test the method for higher-order space discretizations, which would
make it appropriate for other high-order methods, which generally
have denser sparsity patterns compared to FVM just using the first neighbours. 
Under these circumstances, the method has
been tested in a framework considering only the first neighbour cells, which
will have 7 non-zeros per row (\texttt{7p}); considering the first and second
neighbour cells, which will have 13 non-zeros per row (\texttt{13p}), and
considering all neighbours of the cell, including the diagonal neighbours,
which makes it 27 non-zeros per row (\texttt{27p}). Figure
\ref{fig:discretization} shows a two-dimensional representation of the used
cells in each Laplacian discretization.

These tests have been performed by running a single time unit in the fine
160$^3$ mesh so that the presented time for the different elements to run has
had enough iterations so that its average value is representative. Moreover,
a timer has been added to monitor the time devoted to performing \texttt{SpMM}
operations within Chronos so that not only the speed-up present in the Poisson
solution and the whole iteration can be shown, but the speed-up obtained in the
kernel itself can also be studied. Moreover, this will allow setting proper
values for $\chi$ for Eq. \eqref{eq:poipm}.

\begin{figure}[h]
  \centering
  \includegraphics[height=0.3\textwidth]{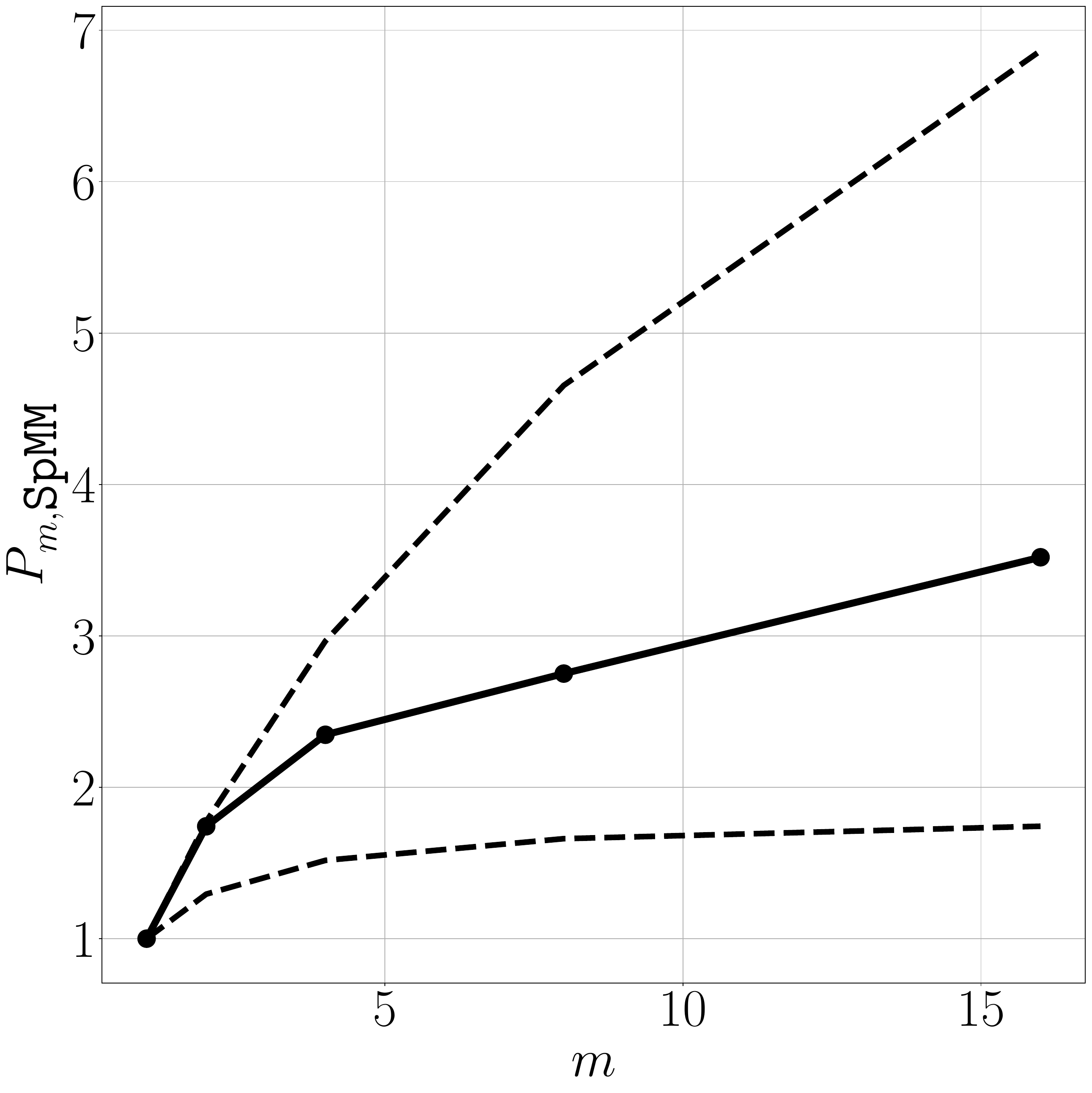}
  \includegraphics[height=0.3\textwidth]{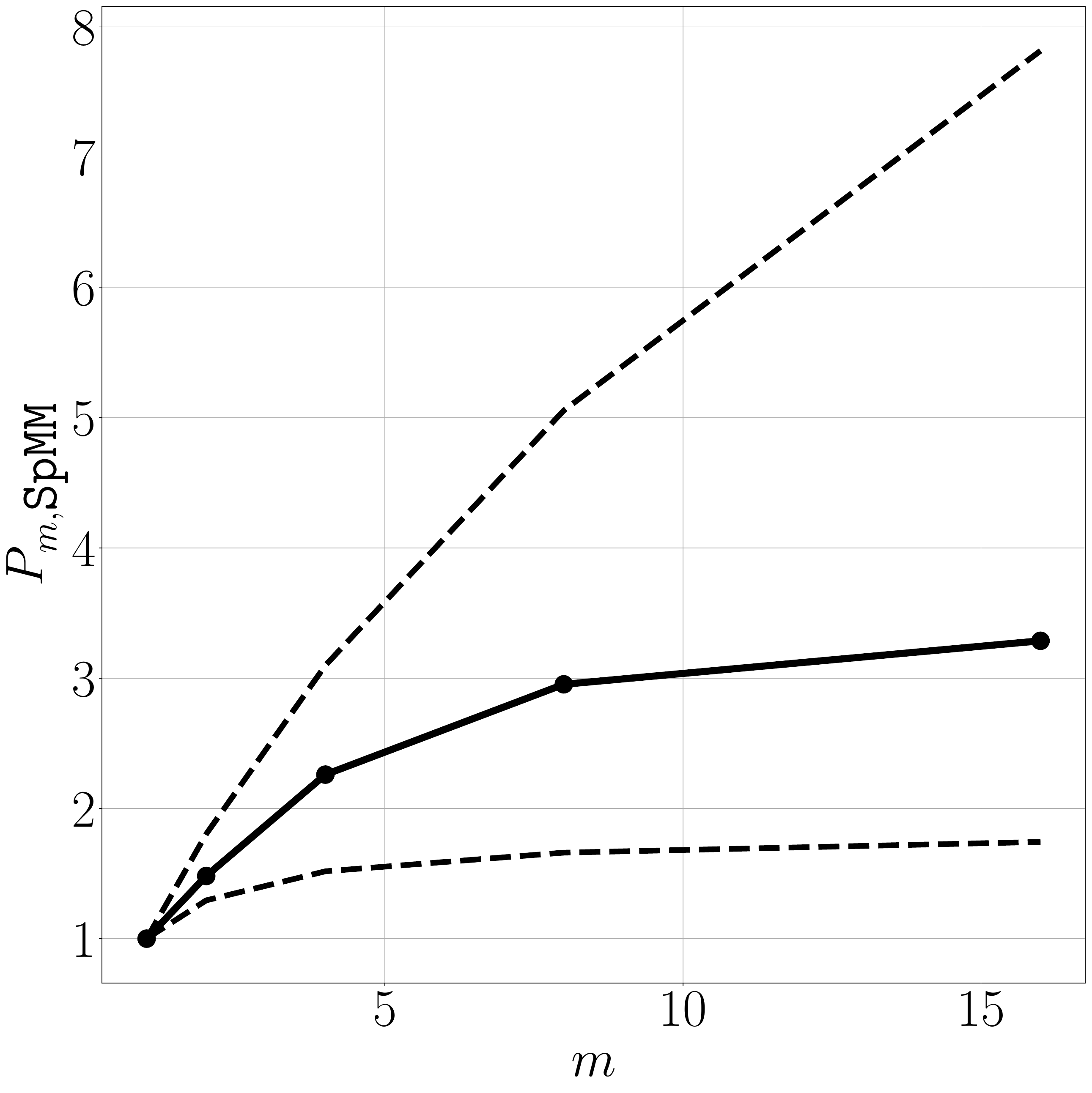}
  \includegraphics[height=0.3\textwidth]{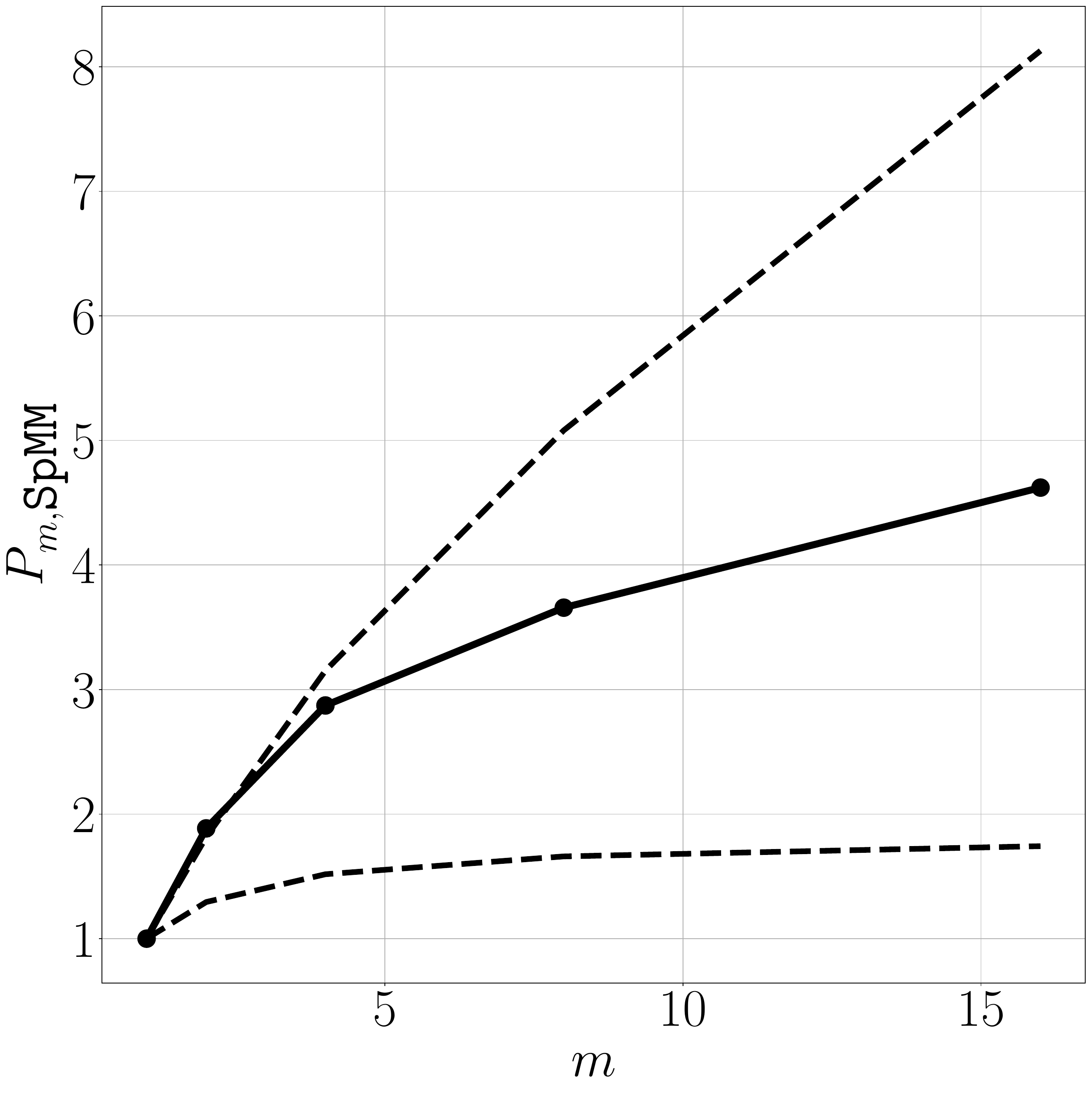}
  \caption{Speed-up in the execution of the \texttt{SpMM} operation in a turbulent planar channel flow
  of $\text{Re}_\tau=180$, compared to the theoretical upper and lower bounds. Left: \texttt{7p}, Center: \texttt{13p},
  Right: \texttt{27p}.}
  \label{fig:pm_spmm_cf}
\end{figure}

\begin{figure}[h]
  \centering
  \includegraphics[height=0.3\textwidth]{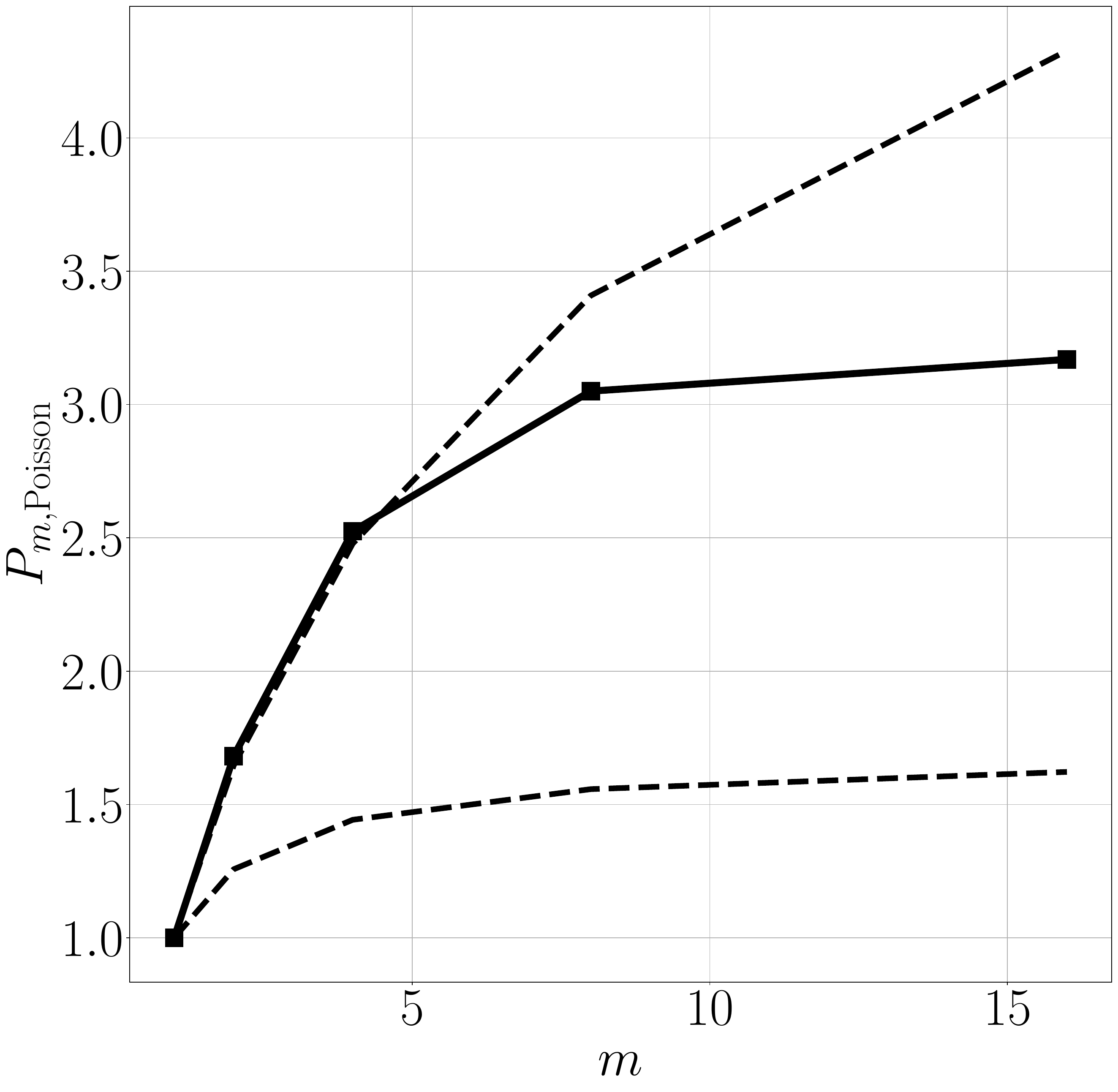}
  \includegraphics[height=0.3\textwidth]{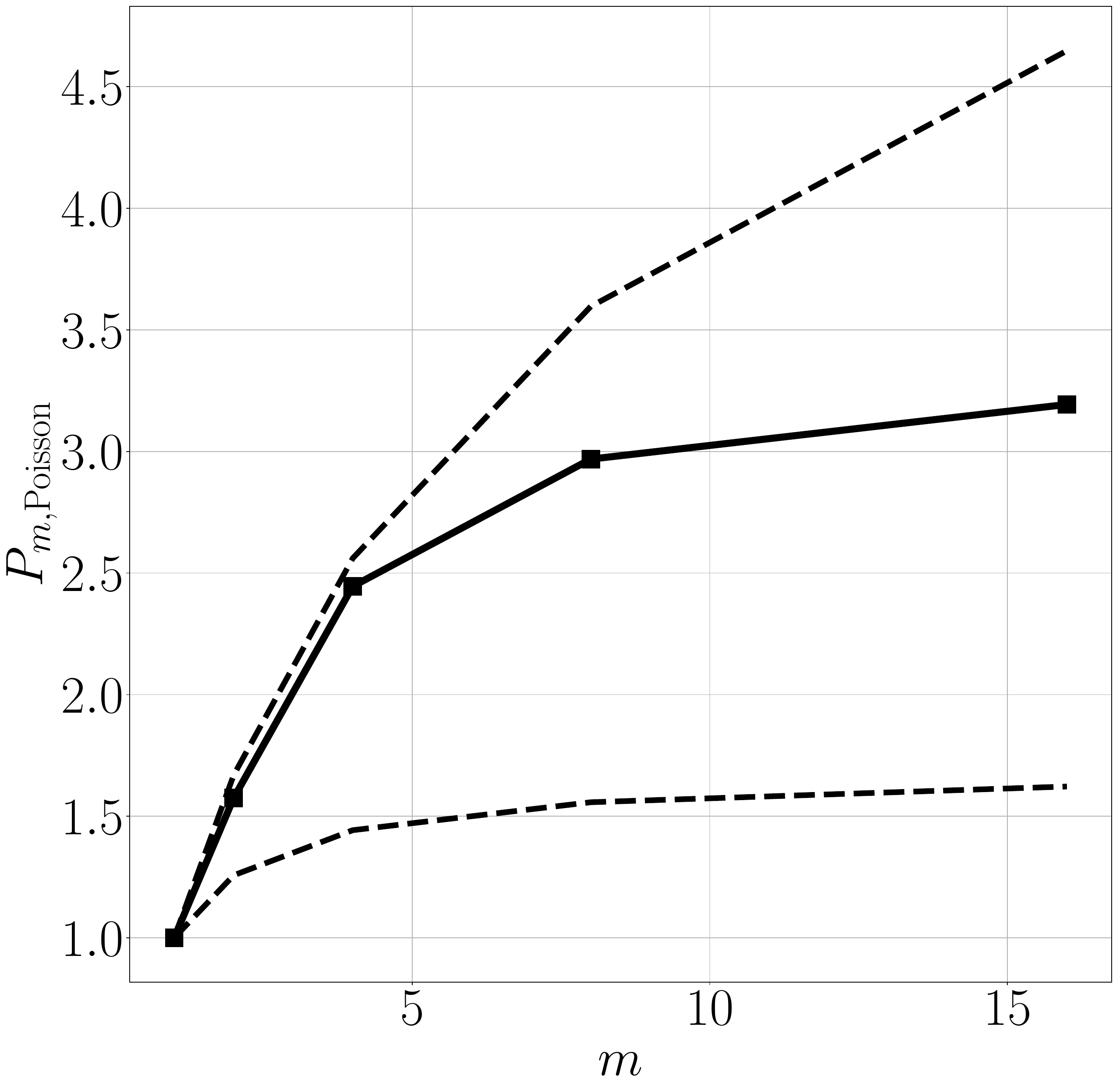}
  \includegraphics[height=0.3\textwidth]{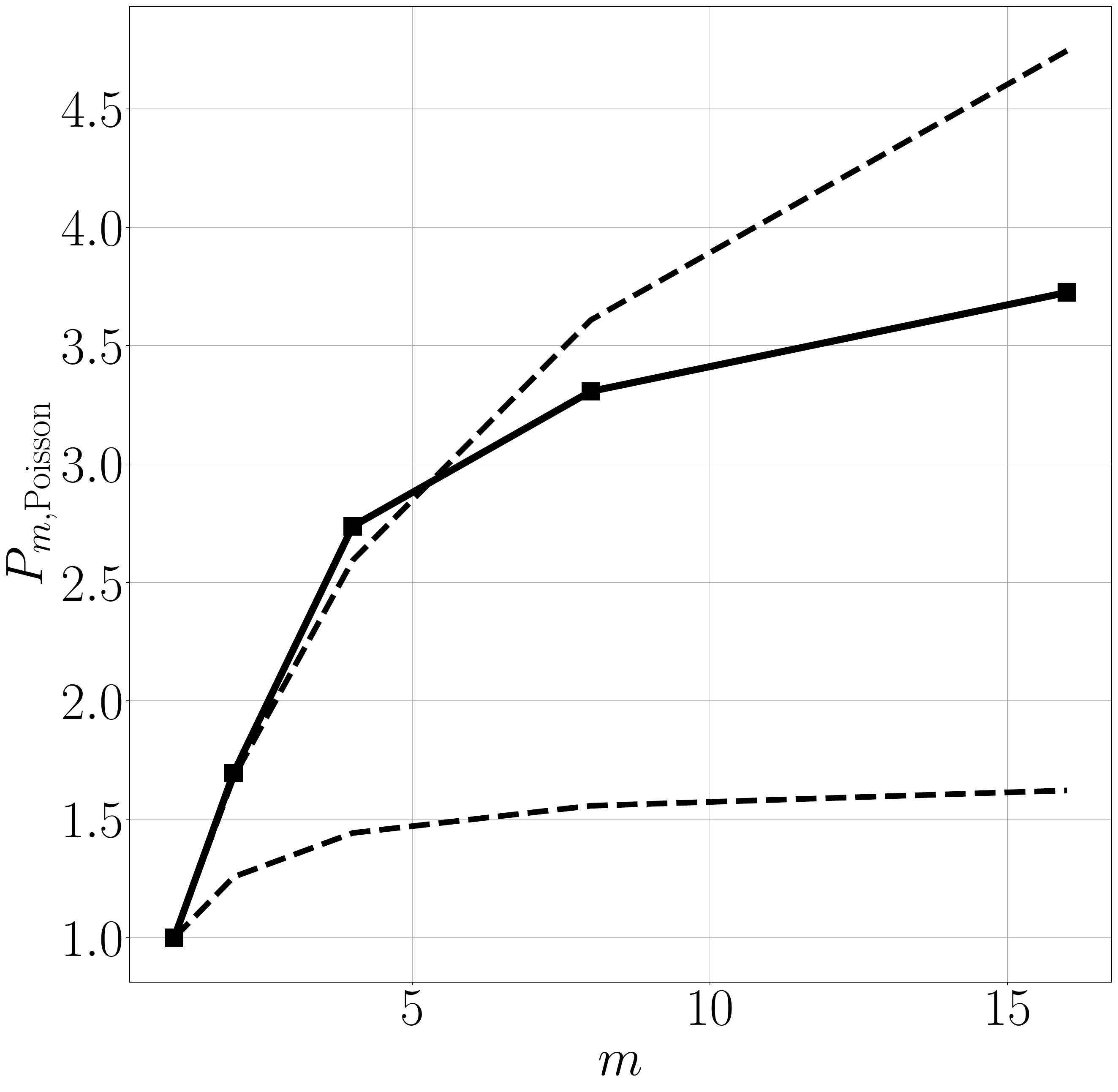}
  \caption{Speed-up in the numerical solution of the Poisson equation in a turbulent planar channel flow of $\text{Re}_\tau=180$ using
  a CG+aAMG solver from Chronos compared to theoretical bounds considering
  a weight factor of \texttt{SpMV} of 90\%. Left: \texttt{7p}, Center: \texttt{13p},
  Right: \texttt{27p}.}
  \label{fig:pm_poi_cf}
\end{figure}

Figure \ref{fig:pm_spmm_cf} shows that the behaviour for all three cases run,
\texttt{7p, 13p}, and \texttt{27p}, lies between the expected bounds of
speed-up, yet the case using only the first neighbours has the
closer-to-optimal behaviour, compared to the \texttt{13p} and \texttt{27p}
cases, which are still close to the upper-bound, yet further than the
\texttt{7p} case. With regards to the extension to the whole solution of the
Poisson equation using the aAMG preconditioned Conjugate Gradient implemented
in Chronos, speed-up figures are also obtained in Figure
\ref{fig:pm_poi_cf}. In this case, the bounds for both \texttt{SpMM} and Poisson
equation have not been calculated using Eq. \eqref{eq:pmspmm} as it assumes that
only a single combination of number of rows, columns and number of non-zeros is
present, which is different from the set of operations present in the CG+aAMG
framework. Thus, following \ref{sec:wgt_spmm}, the bounds for these operations
have been computed using a weighted average considering all possible
combinations of rows, columns and number of non-zeros.

\begin{figure}[h]
  \centering
  \includegraphics[height=0.3\textwidth]{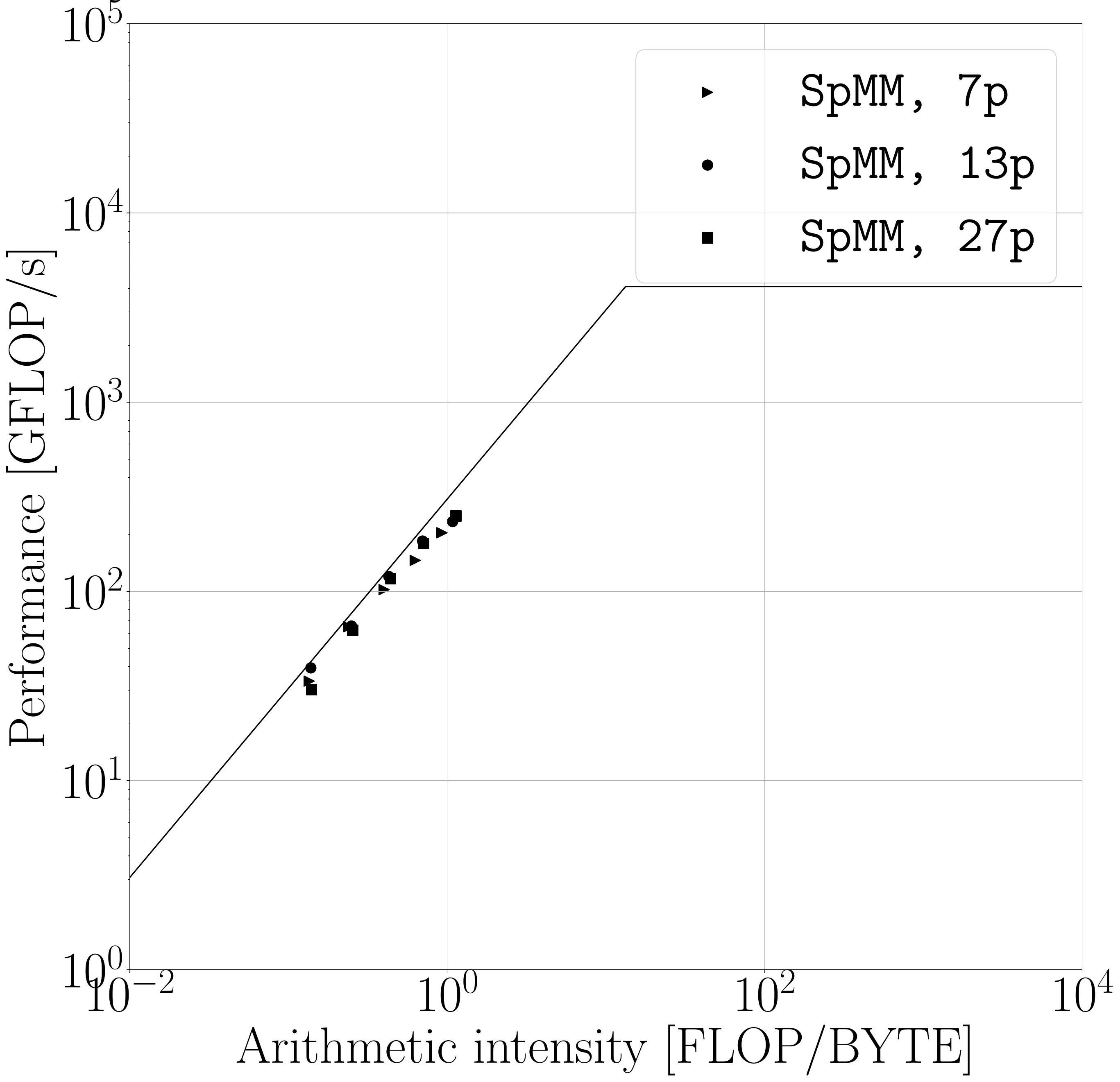}
  \caption{Performance evaluation of the \texttt{SpMM} operations within the
  CG+aAMG solver framework run in a single MN5-GPP node in the numerical simulation of a turbulent planar 
  channel flow with $\text{Re}_\tau=180$.}
  \label{fig:spmm_roofline}
\end{figure}

In order to test if not only the speed-up is good but the performance is as
expected, the roofline of these operations has also been extracted by using
a weighted average of the arithmetic intensities and number of floating-point
operations per second of all combinations of rows, columns and number of
non-zeros so that a representative value is present. Figure
\ref{fig:spmm_roofline} shows that the obtained performance is close to the
theoretical limit given by the manufacturer and the performance grows with the increased arithmetic
intensity. 

\begin{figure}[h]
  \centering
  \includegraphics[height=0.3\textwidth]{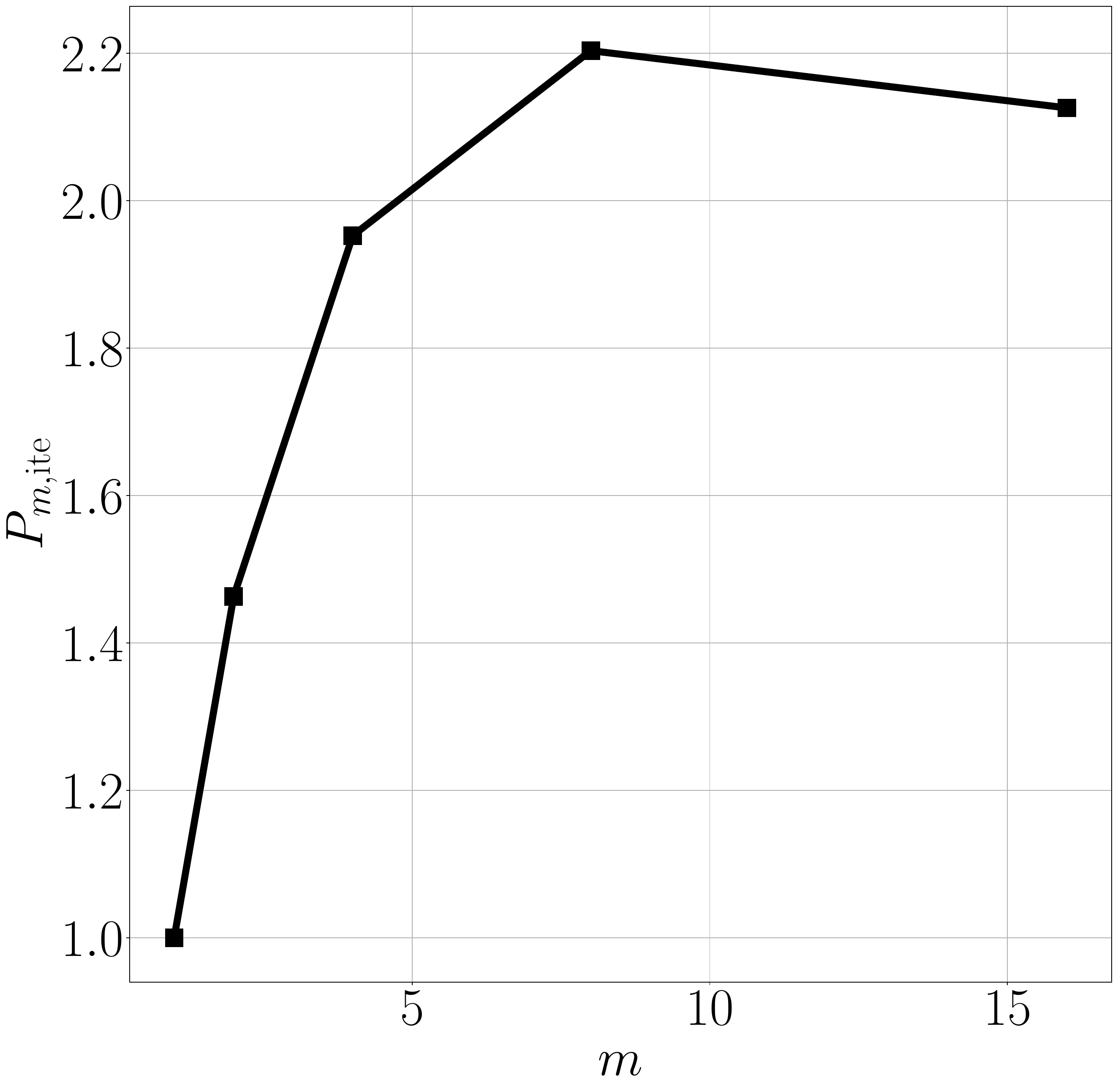}
  \includegraphics[height=0.3\textwidth]{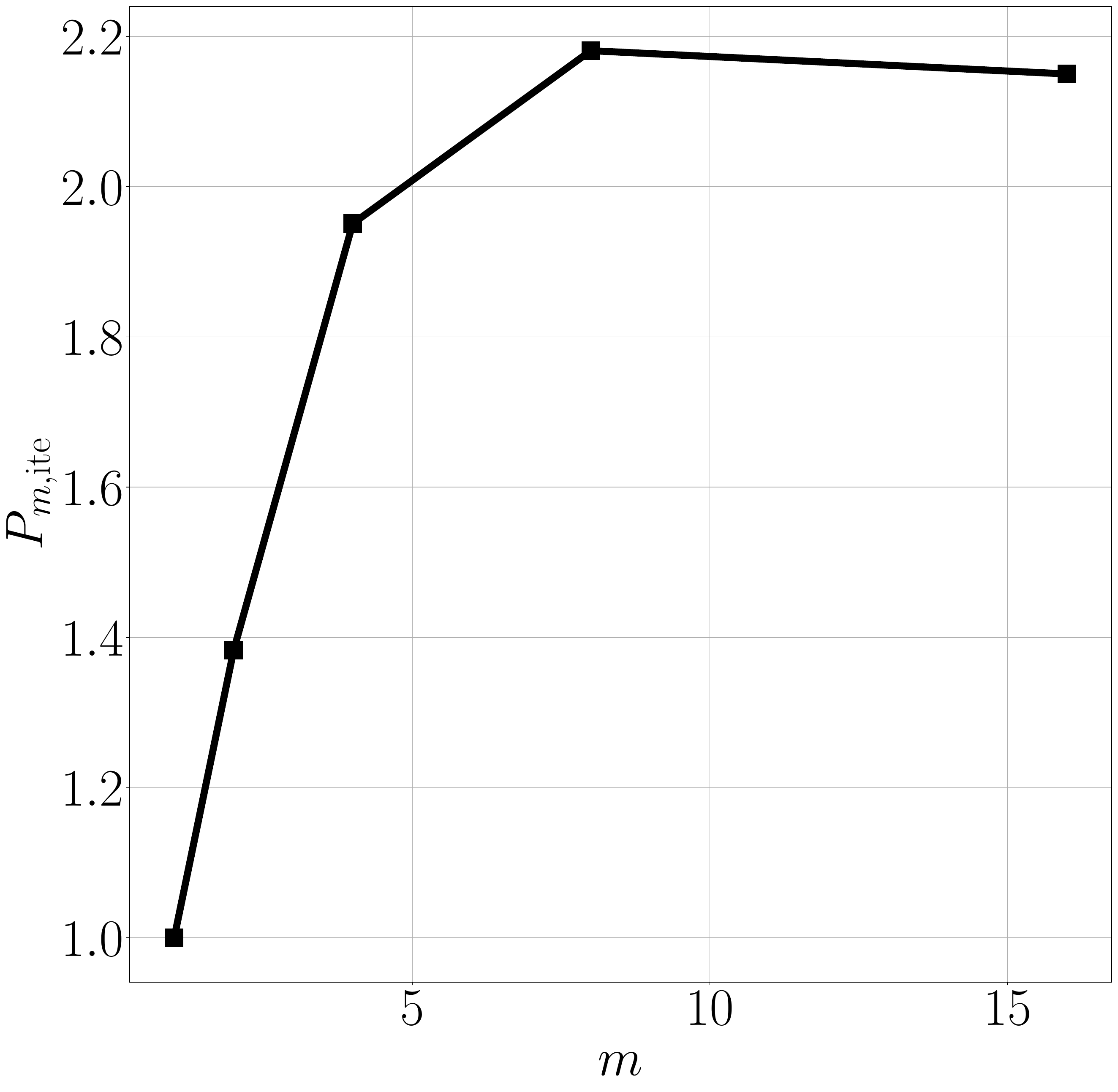}
  \includegraphics[height=0.3\textwidth]{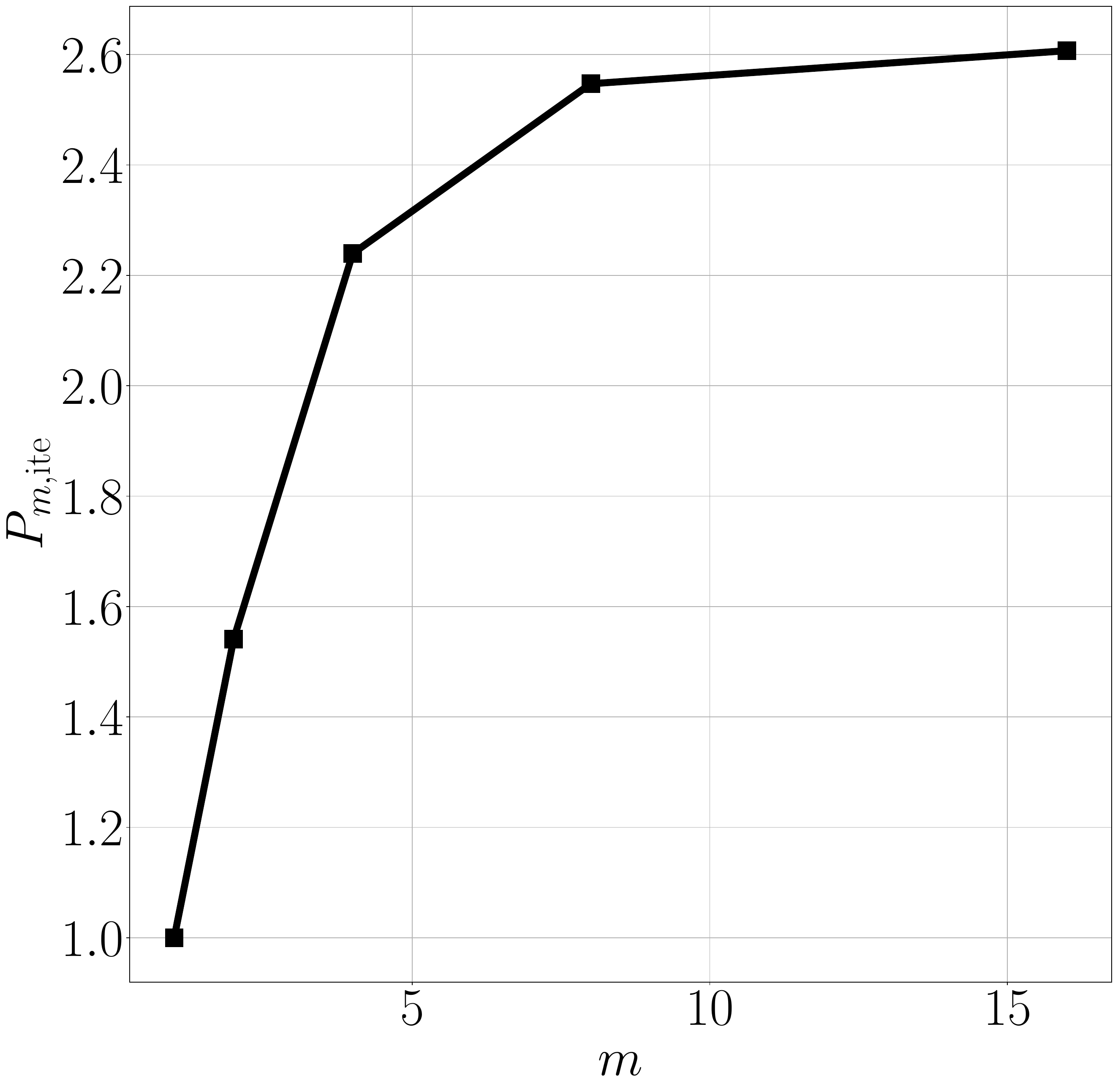}
  \caption{Speed-up in a whole iteration in a turbulent planar channel flow with $\text{Re}_\tau=180$ using a third-order Heun Runge-Kutta
  scheme. Left: \texttt{7p}, Center: \texttt{13p},
  Right: \texttt{27p}.}
  \label{fig:pm_ite}
\end{figure}

Moreover, the benefit obtained throughout the whole iteration has also been
tested for the three test cases previously mentioned (Figure \ref{fig:pm_ite}). By doing so, these
results may be extrapolated using Eq. \eqref{eq:pm_full_v2} given the $\tilde{\beta}$
of the case, either having mesh refinement or not. The extrapolated results for the \texttt{7p} case
can be observed in Table \ref{tab:extrap_7p}, in which the performance obtained using a mesh refinement
or not is compared. The improvements are the most remarkable for smaller values of $\beta$ and a bigger
number of flow states, as then the influence of a bigger effective times ratio is higher than for already
big times ratios or a smaller number of flow states.

\begin{table}[h]
  \centering
  \caption{Estimated simulation speed-ups considering the iteration speed-ups obtained with \texttt{7p}, for times ratio
  $\beta$ values ranging from 1 to 40 and 1 to 8 simultaneous flow states. Left: without mesh refinement. Right: $\Pi=16,\gamma=15$.}
\begin{tabular}{@{}ccccccc@{}}
  \cmidrule(l){3-7}
  \multirow{4}{*}{\rotatebox{90}{$m$}}      & \multicolumn{1}{c|}{8} & 0.44 & 0.91 & 1.21 & 1.49 & 1.69   \\
                                            & \multicolumn{1}{c|}{4} & 0.74 & 1.24 & 1.46 & 1.62 & 1.73   \\
                                            & \multicolumn{1}{c|}{2} & 0.97 & 1.25 & 1.33 & 1.38 & 1.42   \\
                                            & \multicolumn{1}{c|}{1} & 1.00 & 1.00 & 1.00  & 1.00  & 1.00 \\ \cmidrule(l){3-7} 
                                            &                         & 1    & 2    & 10     & 20    & 40  \\
                                            &                         & \multicolumn{5}{c}{$\beta$}    
\end{tabular}
  \hspace{0.4in}
\begin{tabular}{@{}ccccccc@{}}
  \cmidrule(l){3-7}
  \multirow{4}{*}{\rotatebox{90}{$m$}}      & \multicolumn{1}{c|}{8} & 1.12 & 1.70 & 1.82 & 1.90 & 1.94   \\
                                            & \multicolumn{1}{c|}{4} & 1.40 & 1.73 & 1.79 & 1.82 & 1.84   \\
                                            & \multicolumn{1}{c|}{2} & 1.31 & 1.41 & 1.43 & 1.44 & 1.45   \\
                                            & \multicolumn{1}{c|}{1} & 1.00 & 1.00 & 1.00  & 1.00  & 1.00 \\ \cmidrule(l){3-7} 
                                            &                         & 1    & 2    & 10     & 20    & 40  \\
                                            &                         & \multicolumn{5}{c}{$\beta$}    
\end{tabular}

  \label{tab:extrap_7p}
\end{table}

\subsection{Rayleigh-Bénard convection}

On the other hand, to test the robustness of the method when adding additional transport equations such as the energy
equation, the method has been tested in a Rayleigh-Bénard convection with $\text{Ra}=10^9$ and $\text{Pr}=0.71$, using
a 160x320x64 mesh with only being the $z$ direction periodic. In the $y$ direction the mesh has been refined with a
hyperbolic tangent function with $\gamma=1.5$. 

In order to let the flow develop to a steady-state, which will make the relevant data more uniform throughout the iterations,
the simulation has been run for 50 time units. Note that this is not enough to
obtain a statisically steady state \cite{trias_direct_2010} and thus the presented
results only aim at wall-clock times and speed-ups. For the \texttt{7p} case, a mesh refinement from a 80x160x32 mesh to the 160x320x64
mesh has been performed after 30 time units, in order to test the values of $\delta$ obtained by adding this additional transport
equation.

The simulations here presented have been run in 64 cores of a node of MN5-GPP supercomputer, which sets the load per CPU to 51k cells,
which is a load close to the strong scaling limit of the code
\citep{mosqueda-otero_portable_2025}, which has been set at around 34k cells per CPU.
As for the planar channel flow, the parallelization method used has been MPI-only.

As a full simulation letting the flow develop has not been performed, only actual figures for Poisson and iteration speed-up
will be presented for a multi-mesh set-up. On the other hand, estimations for the whole simulation can be extrapolated by setting a transition time $T_T$ of 50 time-units
as it shows to be sufficient time to develop the flow completely, and an averaging time of 400 time-units, set by \citet{demou_direct_2019} in previous
simulations of equivalent Rayleigh-Bénard convection simulations. This sets a times ratio of $\beta=8$ for the estimations. Moreover, $T_D$ will
be set to 40 so that the flow can redevelop in the finer mesh. By doing so, the estimations will be performed with $\beta=8$ and $\gamma=4$,
which gives smaller benefits in terms of speed-up compared to the $\beta=10$ and $\gamma=15$ used in the previous section.

\begin{figure}[h]
   \centering
   \includegraphics[height=0.3\textwidth]{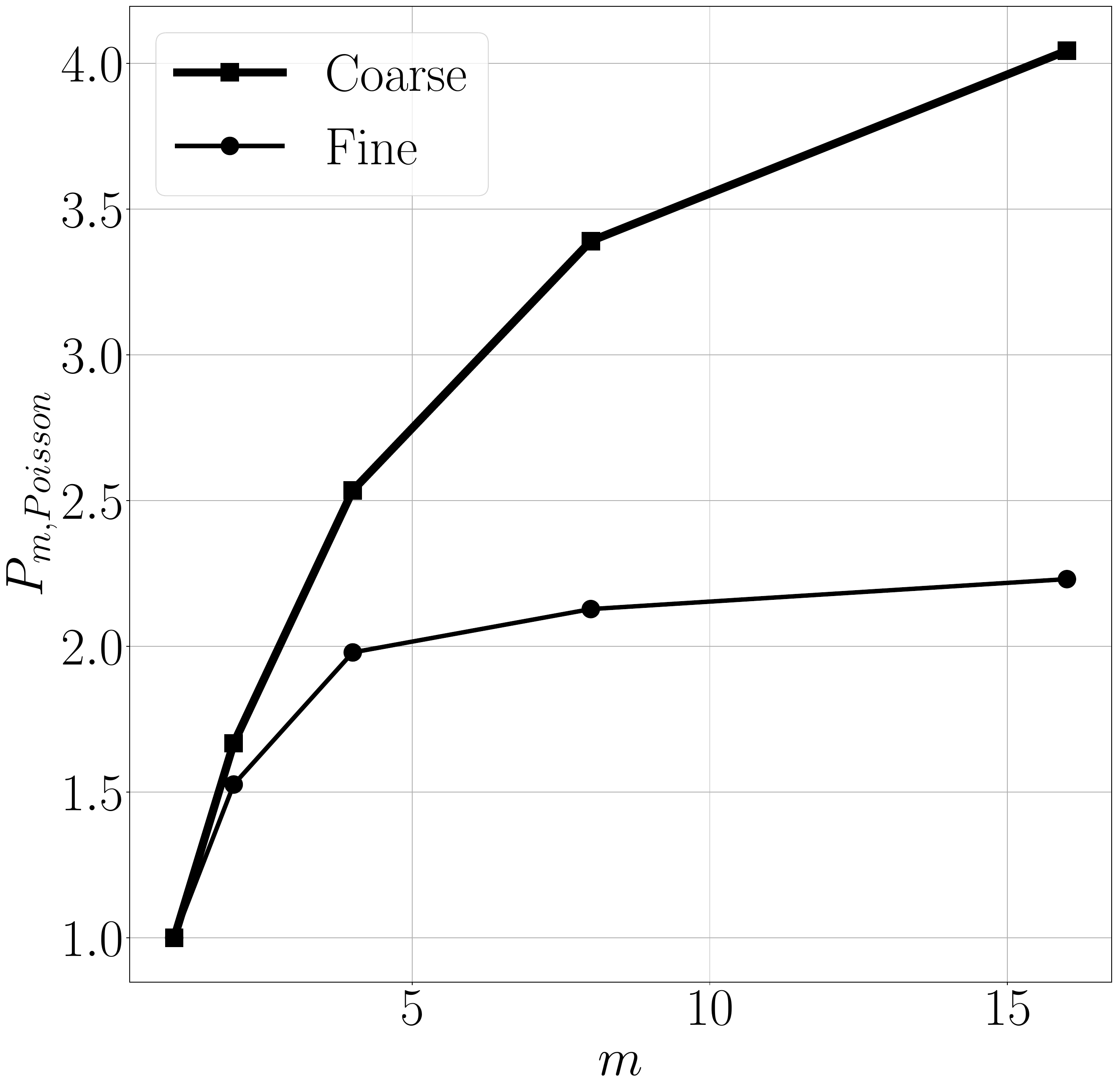}
   \includegraphics[height=0.3\textwidth]{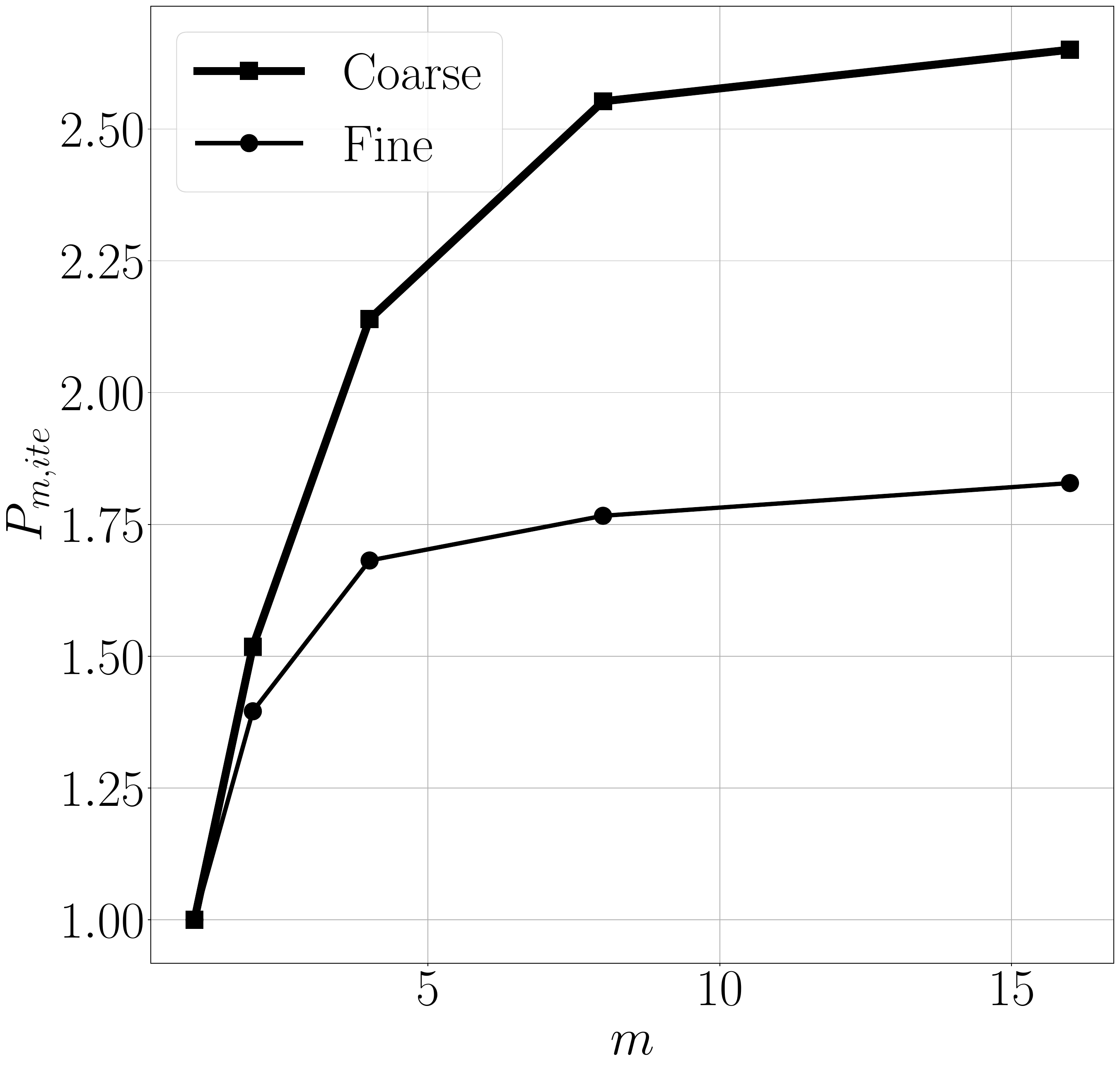}
   \caption{Speed-ups for the numerical solution of the Poisson equation using a CG+aAMG solver from Chronos (left),
   and for the whole iteration (right) for both fine and coarse meshes and up to 16 flow states.}
   \label{fig:mm_rbc}
\end{figure}

Figure \ref{fig:mm_rbc} shows the speed-ups for both Poisson equation and whole iteration in both fine and coarse meshes in the aforementioned
multi-mesh set-up in a symmetry-preserving second-order discretization for the Navier-Stokes equations which leads to sparse matrices with 7 non-zeros
per row (equivalent to \texttt{7p} case). It may be observed that in terms of the Poisson equation, the obtained speed-ups do not vary much compared to
the obtained results for the planar channel flow. This is as expected, given that the numerical solution of the Poisson equation does not change regardless
of the presence of an additional transport equation as the energy equation is in this case. Nonetheless, the obtained speed-ups for the whole iteration
in this case show a noticeable reduction when compared to the results from Fig. \ref{fig:pm_ite}, as the heavier \texttt{SpMM} operations, which
are present in the numerical solution of the Poisson equation, are diluted with more lighter \texttt{SpMM} operations, thus reducing the overall speed-up,
given the presence of an additional transport equation that dilutes the bigger impact of the \texttt{SpMM} operations in the whole iteration.

Moreover, the average \texttt{SpMM} times have been computed for the 20 time units in which the fine mesh has been running, so that the speed-ups for the kernel
itself can be presented. Similar results to the ones obtained in Fig. \ref{fig:pm_spmm_cf} are obtained, where the differences obtained, which max at around 15\% with 8 rhs, 
can be explained by the smaller load present in the current analysis compared to the channel flow simulations.

\begin{figure}[h]
  \centering
  \includegraphics[height=0.3\textwidth]{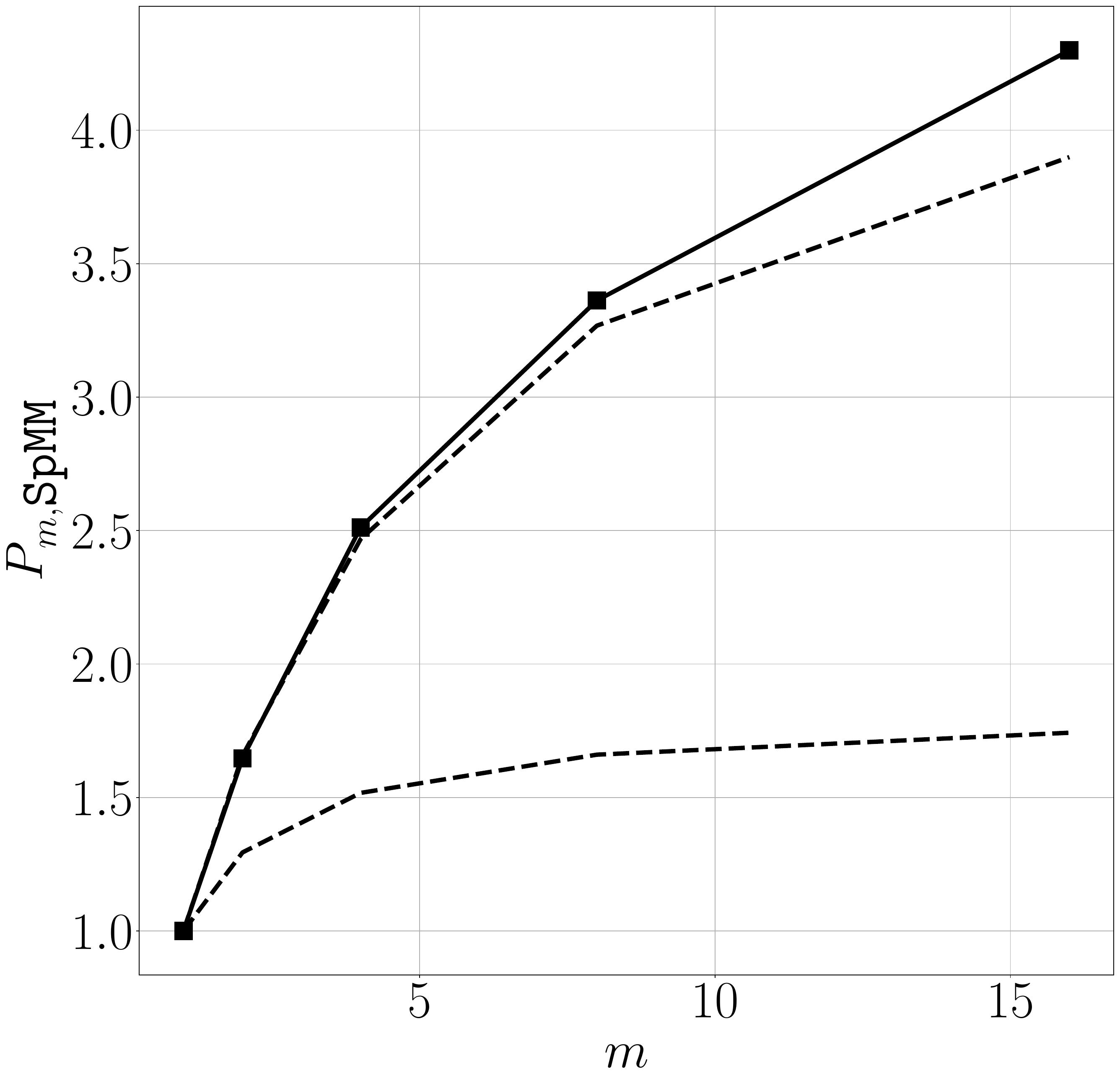}
  \includegraphics[height=0.3\textwidth]{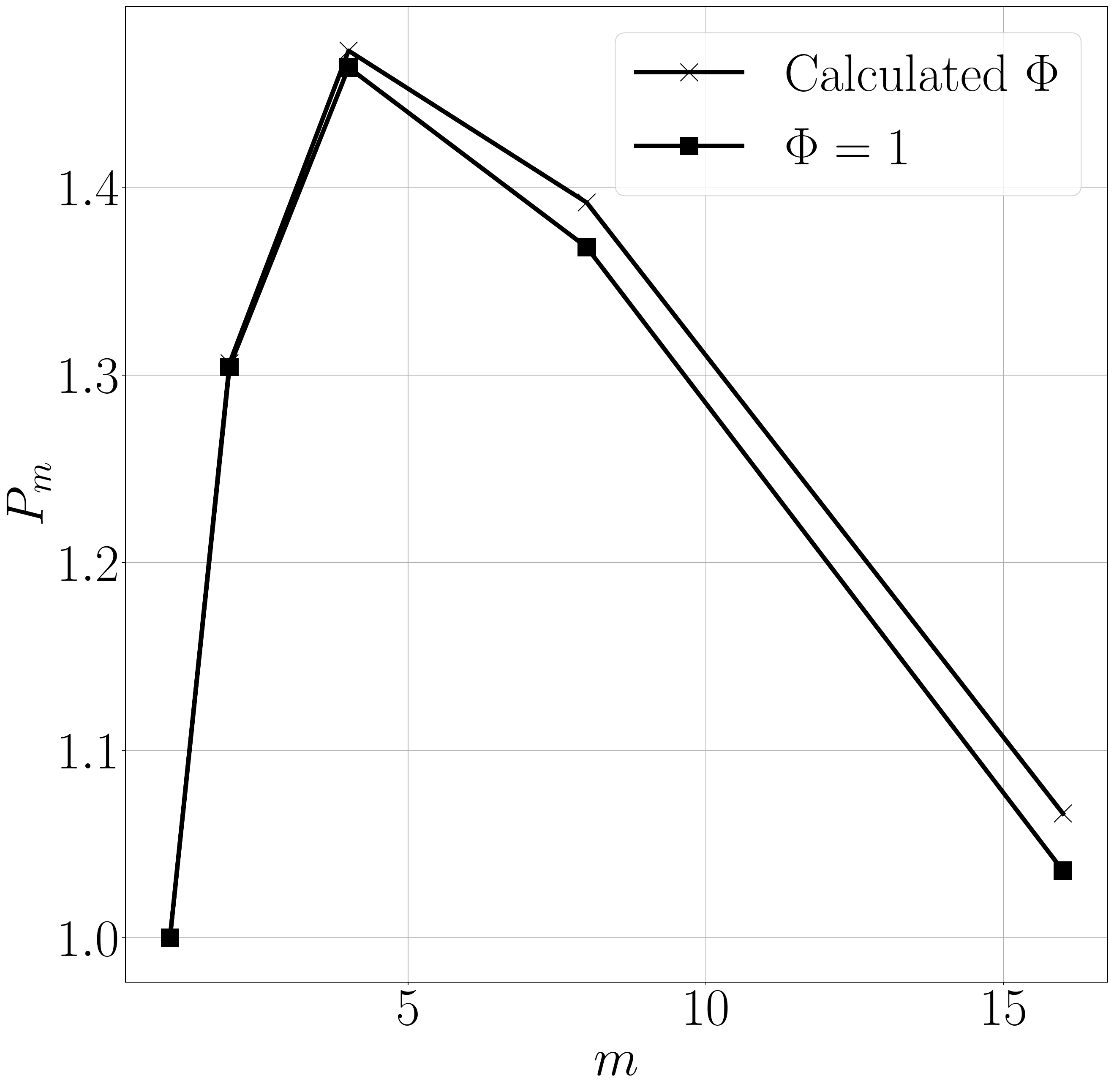}
  \caption{Speed-ups in the execution of the \texttt{SpMM} operation in the Rayleigh-Bénard convection, compared to the theoretical upper and lower bounds (left), and estimation of the simulation speed-up given $\gamma=4$ and $\beta=8$ for both methods, calculating $\Phi$ or setting $\Phi=1$ (right).}
  \label{fig:spmm_rbc}
\end{figure}

In terms of simulation speed-up estimation, Table \ref{tab:dt-wct-rbc} presents the boost in the coarse mesh, which has a wider range than for the channel flow test (Tab. \ref{tab:dt-wct}),
with $\Pi_m$ ranging from 11 to 15, growing with $m$. Nonetheless, the same analysis for the channel flow has been performed, where both methods, calculating $\Phi$ and estimating $\Phi=1$ have been tested and shown in Fig.\ref{fig:spmm_rbc} (right). The estimation shows that in this case, with the presented conditions of $\beta = 8$, $\Pi = 11$ and $\gamma = 4$, the simulation speed-up will have its optimum value at $m=4$, with the benefits of the method reducing notably for a greater number of flow states. In the comparison between the methods, the both perform similarly, with a difference in between the methods of 2.85\% using the same method as for the turbulent channel flow. Hence, the difference between both methods is slightly larger than for the case without an additional transport eqequation (1.3\%) yet it is still acceptable with a difference below 3\%.

\begin{table}[h]
  \centering
  \caption{Average time-steps, wall-clock times per iteration [s] and $\delta_m$ for both coarse ($80\times160\times32$) and fine ($160\times320\times24$) meshes and
  all the simulated flow states.}
  \begin{tabular}{cccccc}
    $m$ & $\overline{\Delta t}_m^C$ & $\bar{t}_m^C$ & $\overline{\Delta t}_m^F$ & $\bar{t}_m^F$ & $\delta_m$ ($\Pi_m$) \\ \toprule
    1 & $1.450\times10^{-2}$ & 0.1439 & $8.000\times10^{-3}$ & 0.8391 & 0.0944 (11.000) \\
    2 & $1.353\times10^{-2}$ & 0.1896 & $7.440\times10^{-3}$ & 1.2026 & 0.0866 (11.530) \\
    4 & $1.248\times10^{-2}$ & 0.2691 & $7.140\times10^{-3}$ & 1.9961 & 0.0771 (12.964) \\
    8 & $1.193\times10^{-2}$ & 0.4511 & $6.749\times10^{-3}$ & 3.8002 & 0.0671 (14.899) \\
    16 & $1.152\times10^{-2}$ & 0.8688 & $6.485\times10^{-3}$ & 7.3423 & 0.0667 (15.006) 
  \end{tabular}
  \label{tab:dt-wct-rbc}
\end{table}

\subsection{Industrial case: 30P30N}

The method has been tested as well in an unstructured mesh to validate the
performance of the method in an industrially relevant case. In order to do so,
the numerical simulation of a Re=$10^7$ around a 30P30N airfoil is presented
for a single mesh. Instead of performing an ensemble average, in this case
the multiple right-hand sides are devoted to multiple flow states concerning
the angle of attack $\alpha$ of the airfoil. Thus, by performing $n$
simulations simultaneously, multiple studies are performed at the same time,
which allows the obtention of curves such as the polar curve of the airfoil
running all simulations all at once. This framework, opposite to an ensemble averaging
situation, will preserve the whole iteration speed-up for the entirety of the simulation,
as the benefit is just obtained by running multiple full simulations, instead of $m$ short
versions of the simulations which ensemble averaged provide the same results.

\begin{figure}[h]
  \centering
  \includegraphics[width=0.4\textwidth]{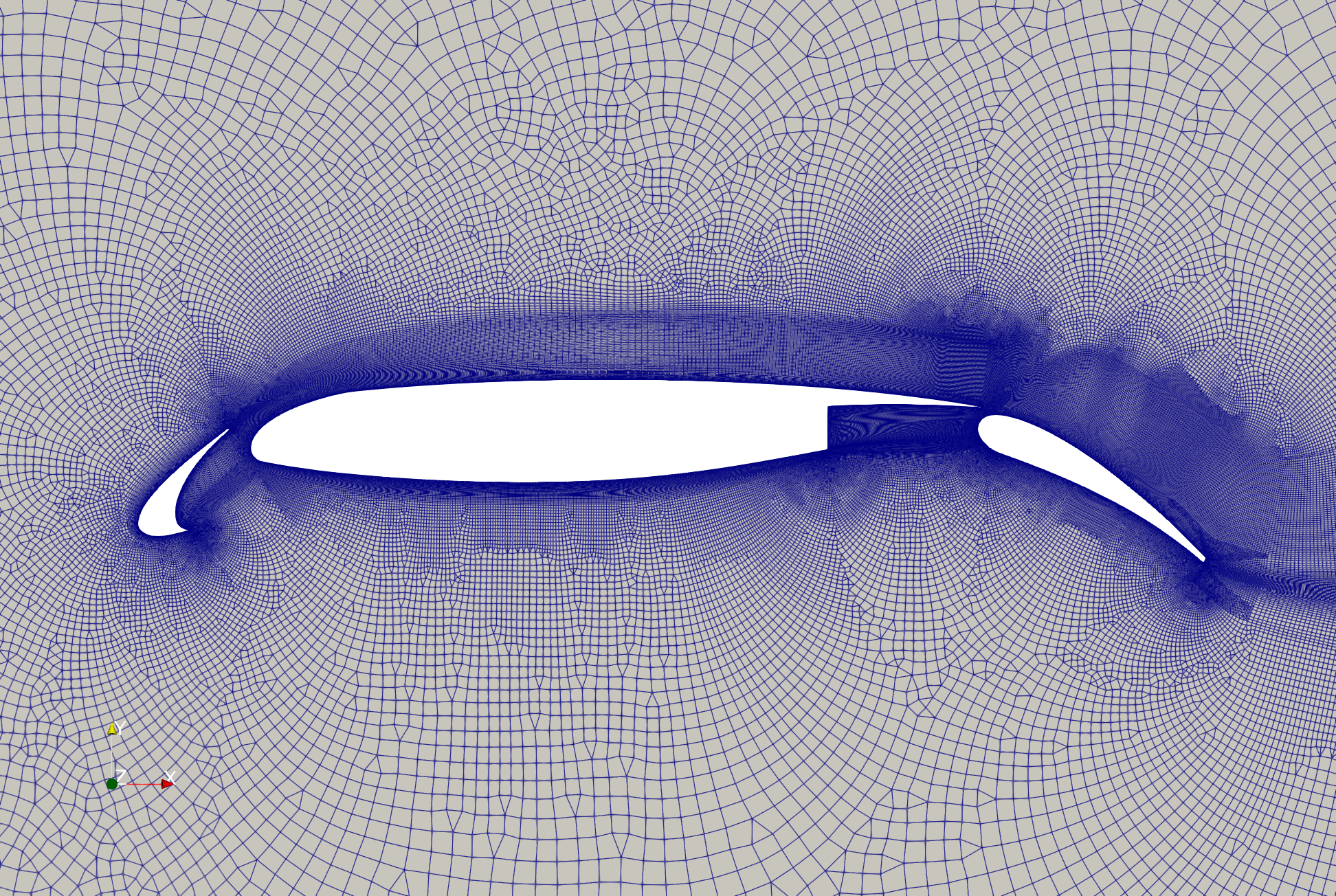}
  \caption{Close-up view of the mesh used for the simulation of the 30P30N airfoil.}
  \label{fig:30p30n_mesh}
\end{figure}

The simulated domain is $150c\times150c\times2c$, where $c$ stands for the airfoil's chord,
and the used mesh (Fig. \ref{fig:30p30n_mesh}) has 14M cells. The test case has been run for 1, 2, 4, and 8 right-hand sides,
which in this framework would correspond to simulating multiple angles-of-attack $\alpha$ simultaneously,
so that relevant figures for an airfoil such as the $C_L-\alpha$ relation, or the polar curve, can be obtained
faster opposite to running each simulation one after the other (or simultaneously, using more computational
resources). The method has been run in a full node of the MN5-GPP supercomputer (112 CPUs) using an MPI-only
framework, leading to a higher load of 115k cells per CPU compared to the previous cases, which had loads close to half this
value.

The method has been run for 100 iterations in a third-order RK scheme in order to test the performance
of all \texttt{SpMM} kernel, Poisson solution and projection method iteration, so that the application of the method not
only in a canonical flow situation such as the ones previously depicted but also in a more industrial case, with
a highly unstructured mesh, is presented.

\begin{figure}[h]
  \centering
  \includegraphics[height=0.3\textwidth]{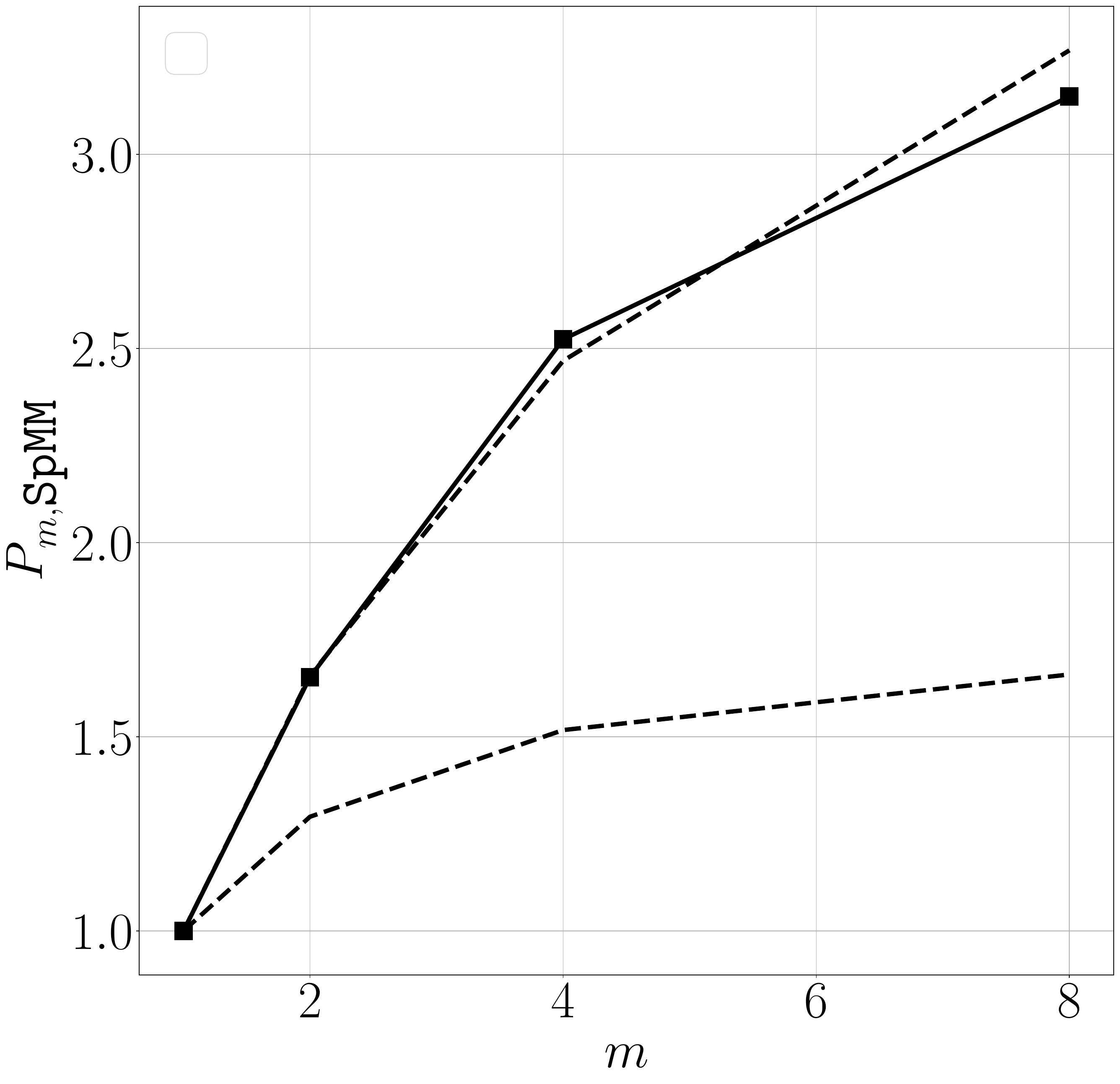}
  \includegraphics[height=0.3\textwidth]{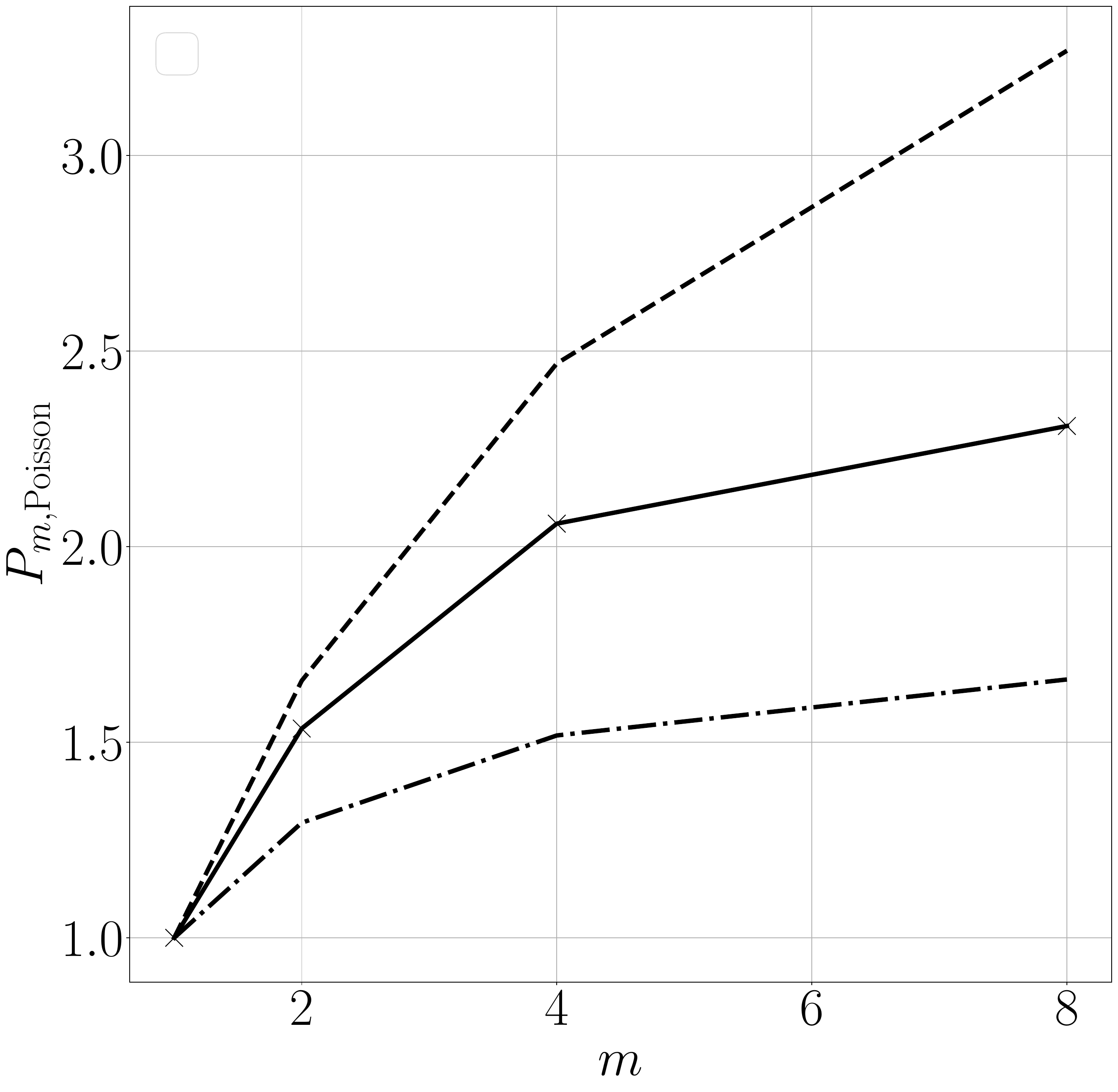}
  \includegraphics[height=0.3\textwidth]{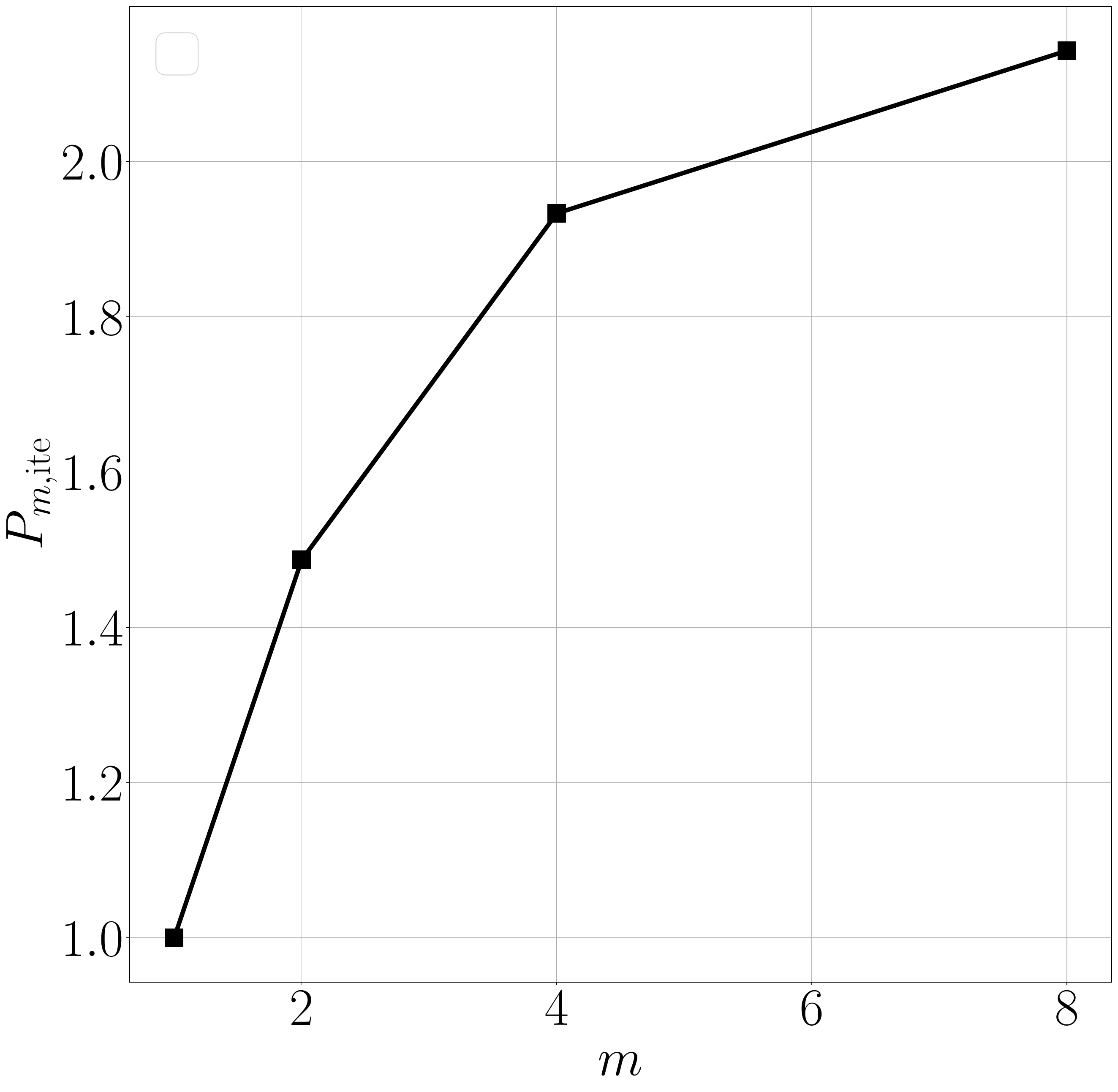}
  \caption{Speed-up values obtained in the \texttt{SpMM} kernel (left), the numerical solution of the Poisson equation (center), and a whole projection method iteration (right), for the presented mesh and all simulated flow states. The first two figures show the speed-ups compared to the theoretical bounds (dashed lines).}
  \label{fig:30p30n_results}
\end{figure}

The obtained speed-ups in the application of the \texttt{SpMM} kernel compared to the classical \texttt{SpMV} (Figure \ref{fig:30p30n_results}) 
show equivalent
performance compared to the results presented both for the turbulent planar channel flow (Fig. \ref{fig:pm_spmm_cf}) and the Rayleigh-Bénard
convection (Fig. \ref{fig:spmm_rbc}), making the obtained speed-up equivalent both for structured and unstructured meshes. With regards to the
numerical solution of the Poisson equation, the obtained results are slightly below the results found for the planar channel flow
(Fig. \ref{fig:pm_poi_cf}) and the Rayleigh-Bénard convection (Fig. \ref{fig:mm_rbc}). Finally, as the
speed-ups obtained in the Poisson equation differ from structured to unstructured meshes, the overall iteration performance is reduced in around a 20\% compared to the fine mesh in both of the structured cases. Nonetheless, the overall
speed-up in these simulations will indeed, in case of a multiple parameter run, correspond to the speed-up on each iteration, thus
leading to greater simulation speed-ups in this case compared ot the other two cases in which an ensemble averaging parallel-in-time
technique was applied, with iteration (and thus simulation) speed-ups of up to 80\% compared to the 55\% obtained in the complete
channel flow simulation (Fig. \ref{fig:pm_cf}).

\section{Discussion} \label{sec:disc}

In this work a novel approach aimed at increasing the arithmetic intensity of CFD simulations has been proposed. This was done
following the works presented by \citet{krasnopolsky_approach_2018}, \citet{tosi_use_2022}, and \citet{alsalti-baldellou_lighter_2024}, 
where the arithmetic intensity of the sparse matrix-vector product (\texttt{SpMV}) kernel was incremented by replacing them with 
sparse matrix-matrix products (\texttt{SpMM}) when possible, making use of running multiple flow states simultaneously, 
multiple parameters, or exploiting symmetries or repeated geometries, respectively. The novelty of the proposed method is two-fold: 
(i) extending the method proposed by \citep{krasnopolsky_approach_2018} from the \texttt{SpMV} operations in the Poisson
equation to all the operators which are represented with a \texttt{SpMV}: gradient, divergence and Laplacian, as well as any other
\texttt{SpMV}s present in the simulation, and (ii) a novel mesh-refinement strategy to reduce the wall-cloclk time of the
transition phase in parallel-in-time simulations.

The overall simulation performance is thus affected by this new mesh-refinement strategy. By initiating the simulation on a coarser mesh and running it for
a fraction of the transition time, until the refinement time $T_D$ is reached (where the results are interpolated to the target mesh), the wall-clock time
spent to reach the transition time $T_T$ is effectively reduced. This will thus reduce the computational cost of the simulations, thus offering a practical
balance between efficiency and accuracy, which is eventually preserved, as the
relevant part of the simulation, the averaging, is performed in the finer grid
which was to be used in the first place.

Considering the trade-offs between convergence, numerical accuracy and performance also becomes a critical factor. While this mesh-refinement strategy allows
reducing the wall-clock time spent to reach $T_T$, it is of uppermost importance to ensure that at $T_D$ the obtained results are statistically independent, as
otherwise the ergodicity principle may not be applied. Hence, selecting $T_D$ and $T_T$ involves balancing accuracy and performance. Greater
values of $T_D$ compared to $T_T$, i.e. greater values of $\gamma$, will lead to bigger effective times ratios, $\tilde{\beta}$, and thus bigger overall simulation
speed-ups. Nonetheless, if the flow has not become statistically independent again at $T_T$, e.g. $T_{DT}$ is not large enough, the accuracy of the simulation will
be affected when ensemble averaging is applied, thus leading to incorrect results obtained by pushing the performance of the method.
Moreover, the transformation of \texttt{SpMV} to \texttt{SpMM} serves to boost the $I$ of these kernels. This is allowed by the presence of
repeated matrix block structures (explicitly such as in \citet{alsalti-baldellou_lighter_2024}, where the proper ordering of the mesh allowed these structures to appear; or implicitly,
as presented in Eq. \eqref{eq:ns-nrhs}) which let the matrix be loaded only once for all the multiple right-hand sides that are run, which slightly mitigates
the bottleneck caused by the limitations present in memory bandwidth, which makes most of the current CFD implementations memory bound \citep{williams_roofline_2009}. This
transformation leads to significant speed-ups in the numerical solution of the Poisson equation. Moreover, other operators such as the gradient, the divergence and
the Laplacian benefits from this new implementation as all of them are \texttt{SpMV}-based.

Hence, the evaluation of the speed-up metrics further validates our approach. In this paper, the upper and lower bounds both for the \texttt{SpMM} kernel as well as
for the solution of the Poisson equation were derived and compared against the experimental results obtained in canonical cases such as a turbulent channel flow, which
showed the behaviour of the method in a pure momentum case, and a Rayleigh-Bénard convection, which proved the method to perform nicely with the addition of transport equations,
which eventually diluted the heavier \texttt{SpMM} operations, which are the core of the overall speed-up. The obtained results show the potential of the method when
running ensemble averaging simulations in parallel-in-time frameworks.

Moreover, the paper also addresses the differences between the performance both in structured and unstructured meshes. Nonetheless, the tests performed with a 30P30N airfoil
show promising results and thus the method should extend nicely to unstructured industrial cases as well, considering the need of having different optimization parameters. 

Therefore, the use of the techniques aforementioned in this paper (Sec. \ref{sec:theory})
generates a benefit in wall-clock times in the given conditions for both
parallel-in-time as well as multiple parameter frameworks. {Section \ref{sec:appl} further discusses the general applicability of the method across different CFD frameworks, highlighting its versatility and potential for widespread adoption, as it may be beneficial for a wide range of CFD solvers that rely on sparse linear algebra operations, regardless of the specific numerical methods used for spatial discretization.}
However, this
benefit would become of even greater relevance if the expected wall-clock time
of CFD simulations would increase in the following years. Section \ref{sec:wct}
performs a global study of this scenario.

\subsection{{General applicability of the method across different CFD frameworks}} \label{sec:appl}

{While the present work demonstrates the efficacy of the proposed method in the context of a projection method for incompressible flows within a finite-volume method framework, the underlying principles of increasing arithmetic intensity through sparse matrix-matrix products (\texttt{SpMM}) can be extended to other CFD frameworks as well, as the method is grounded in the algebraic abstraction of the discrete operators.}

{Regardless of whether the underlying spatial discretization employs cell-centered finite volumes, or high-order finite elements, the resultant linear opreators (e.g., gradient, divergence, and Laplacian) are universally represented in memory as sparse matrices. Thus, the \texttt{SpMM} kernel can be applied to any CFD solver that relies on sparse linear algebra operations, entirely decoupling the high-performance computing optimizations from the specific numerical methods used for spatial discretization. This means that the method can be implemented in a wide range of CFD solvers, including those based on finite element methods, spectral methods, or discontinuous Galerkin methods, as long as the linear operators can be expressed in a sparse matrix format. Moreover, other time-advancing schemes, such as SIMPLE-like methods, can also benefit from the proposed method as they can also be expressed in terms of \texttt{SpMM} operations, thus extending the applicability of the method beyond projection methods.}

{Furthermore, in most general-purpose CFD solvers, the most computationally intensive operations are often those arising from the numerical solution of the Poisson equation for pressure correction. The proposed method's focus on optimizing the \texttt{SpMM} operations naturally extends to block-Krylov subspace methods, which are commonly used in the solution of linear systems in CFD. Therefore, the method's applicability is not limited to a specific solver architecture but can be broadly applied across various CFD frameworks, making it a versatile tool for enhancing performance in a wide range of applications.}

{On the other hand, the proposed in-flight mesh refinement strategy is, by construction, independent of the spatial discretization method, as well as the kind of grid used. The interpolation of the flow state from the coarse to the fine mesh does not depend on whether a structure or an unstructured grid is used, nor on the numerical method used for the spatial discretization. This means that the proposed mesh refinement strategy can be implemented in a wide variety of CFD solvers without being constrained by the choice of spatial discretization or grid type.}

{In summary, the method's reliance on the algebraic structure of the linear operators rather than the specifics of the spatial discretization ensures its general applicability across different CFD frameworks, enabling performance improvements in a variety of solver architectures and numerical methods.}

\subsection{Estimation of the growth of the wall clock time} \label{sec:wct}

The use of parallel-in-time techniques, as presented in Sec.
\ref{sec:theory}, should become a must if the wall clock time of simulations
increases over the years, $Y$, for a load and cost independent of the Reynolds number.
Hence, let us consider the following key parameters, assuming ideal speed-up:

\begin{subequations}
  \begin{align}
    \text{WCT}&\propto\frac{(N_xN_yN_z)N_t}{P}\beta_2^{-Y}, \label{eq:wct} \\
    \text{LOAD}&\propto\frac{N_xN_yN_z}{P}, \label{eq:load} \\
    \text{COST}&\propto \text{WCT}~P\beta_1^{-Y}, \label{eq:cost}
     \end{align}
\end{subequations}

where WCT is the wall-clock time of a generic DNS/LES simulation of a turbulent
flow, LOAD represents the load per computing unit, $P$ is the number of
computing units. This cost is here
represented as COST and can be associated with either energy or money.
Furthermore, $N_x,N_y,N_z$ represent the grid size, and $N_t$ is the number of
time-steps. Finally, $\beta_1,\beta_2\geq1$ are introduced to model the
improvements over time of the performance and cost of the above-mentioned
computing devices, e.g., CPUs, GPUs.

Firstly, given a certain Reynolds number, Re, the total number of control volumes
can be easily estimated using the classical Kolmogorov theory of turbulence \cite{moin_direct_1998},

\begin{equation}
  N_xN_yN_z\propto(\text{Re}^{3/4})^3=\text{Re}^{9/4}.
  \label{eq:grid_re}
\end{equation}

The timestep $\Delta t$ can either scale as $\Delta t\propto\text{Re}^{-1/2}$
according to K41 theory from
\citet{kolmogorov_local_1941,kolmogorov_degeneration_1941}, or $\Delta t\propto\text{Re}^{-3/4}$ according
to the CFL condition for the convective term \citep{courant_uber_1928}. Moreover, $N_t\propto\Delta
t^{-1}$. Hence, in general,

\begin{equation}
  N_t\propto\text{Re}^\alpha,~~\alpha=\left\{\frac{1}{2},\frac{3}{4}\right\}.
\end{equation}

Assuming that load is independent of the Reynolds number, i.e.
$\text{LOAD}\propto\text{Re}^0$, Eq.\eqref{eq:load} together with Eq.
\eqref{eq:grid_re} imply that $P\propto\text{Re}^{9/4}$.

As stated, the simulation cost is hypothesized independent
of Reynolds, i.e., COST$=K\propto\text{Re}^0$. Then,
\begin{equation}
  \text{Re}^{9/4+\alpha}\beta_1^{-Y}\beta_2^{-Y} = K,
\end{equation}
\begin{equation}
  \text{Re}=K^\frac{1}{9/4+\alpha}\tilde{\beta}^\frac{Y}{9/4+\alpha},
  \label{eq:re_b1b2}
\end{equation}

where $\tilde{\beta}=\beta_1\beta_2$. Introducing Eq. \eqref{eq:re_b1b2} into
Eq. \eqref{eq:wct}. Let $\beta_2=\tilde{\beta}^\gamma$. Then,

\begin{equation}
  \text{WCT}\propto
  K^\frac{\alpha}{9/4+\alpha}\tilde{\beta}^{\left(\frac{\alpha}{9/4+\alpha}-\gamma\right)Y}
  \label{eq:wct_end}
\end{equation}

Hence, if the exponent $\frac{\alpha}{9/4+\alpha}-\gamma$ is positive, the
wall-clock time will increase with $Y$, making parallel-in-time a must. If the
exponent becomes negative, the presented method remains relevant as 
wall-clock times are notably reduced, as shown in Table. \ref{tab:ub_7nnz},\ref{tab:ub_27nnz}.

Operating, $\gamma=\log\beta_2/\log(\beta_1\beta_2)$. Hence, introducing this
the the exponent, and operating, if the condition

\begin{equation}
  \beta_1>\beta_2^\frac{9/4}{\alpha}
  \label{eq:condition}
\end{equation}

is held, then the exponent will be positive, thus leading to an ever-increasing
wall-clock time.

\begin{figure}[h]
  \centering
  \includegraphics[width=0.4\textwidth]{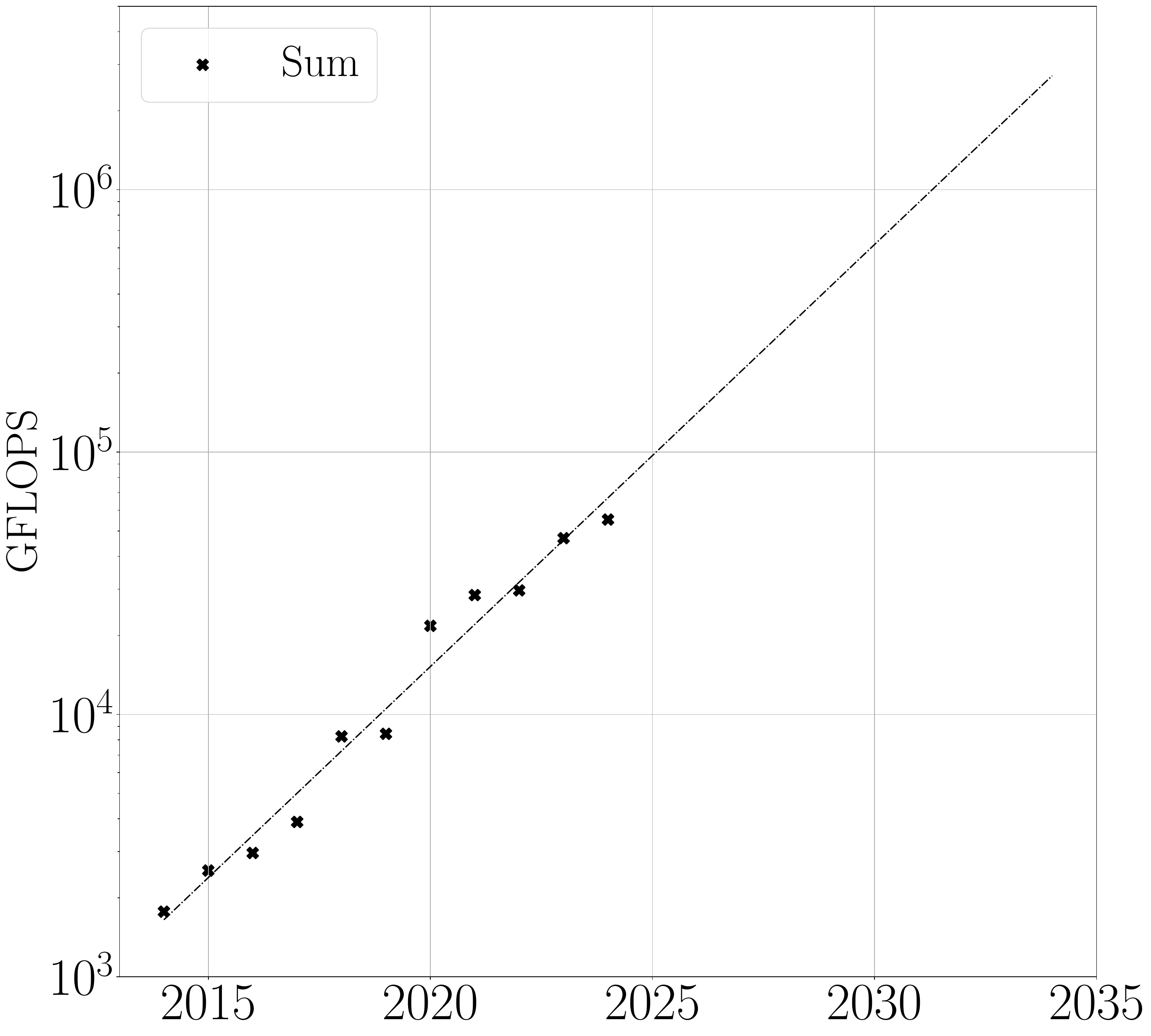}
  \includegraphics[width=0.4\textwidth]{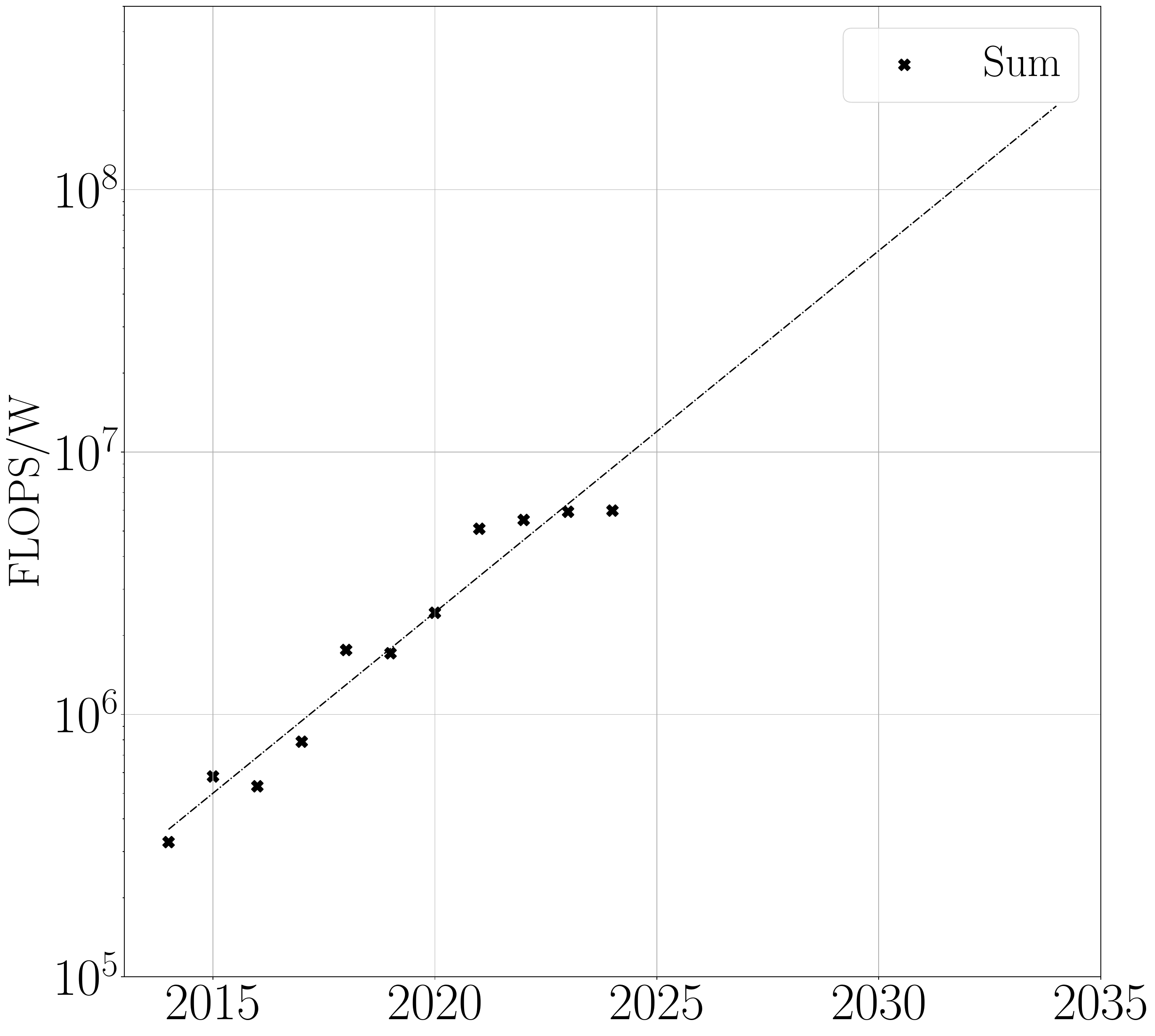}
  \caption{Data from the sum of the Top10 of the HPCG (left)
  and the performance-per-watt (right), and trendline for the sum of the Top10
  values for the next decade.}
  \label{fig:trends}
\end{figure}

In order to estimate the values of $\beta_1,\beta_2$, these can be approximated
with the performance and power trends of current and former supercomputers to
estimate both the sparse algebra performance evolution as well as the
efficiency of these
systems. The chosen benchmark has been HPCG \cite{dongarra_high-performance_2016}, as its set of tests is strongly related
to computational physics applications as it consists of sparse algebra operations, which
aligns better with the CFD framework than the dense algebra from TOP500.
The sum of the 10
best performers is presented in Fig. \ref{fig:trends}, in which the trend of
the sum is estimated and presented for the next decade.

Hence, the slope of the HPCG evolution can be directly related to $\beta_2$, as
an increase in the performance in sparse algebra can be directly related to
reductions in wall-clock time. Hence, $\beta_2\approx1.45$. On the other hand, the
slope of the performance-per-watt evolution is related to $\beta_1$ as
an improvement on the flops generated per watt relates to a cost reduction of the simulation. 
Hence, the data provides $\beta_1\approx1.37$. This shows
that the performance of the supercomputers increases faster than their
efficiency, while Eq. \eqref{eq:condition} required $\beta_1$ to be bigger than $\beta_2$
for $\alpha<9/4$ which is the case for both K41 and CFL values. This result
implies that the wall-clock time should decrease in the next few decades,
assuming ideal speed-up and a cost and load independent of the Reynolds
number.

\begin{figure}[h]
  \centering
  \includegraphics[width=.4\textwidth]{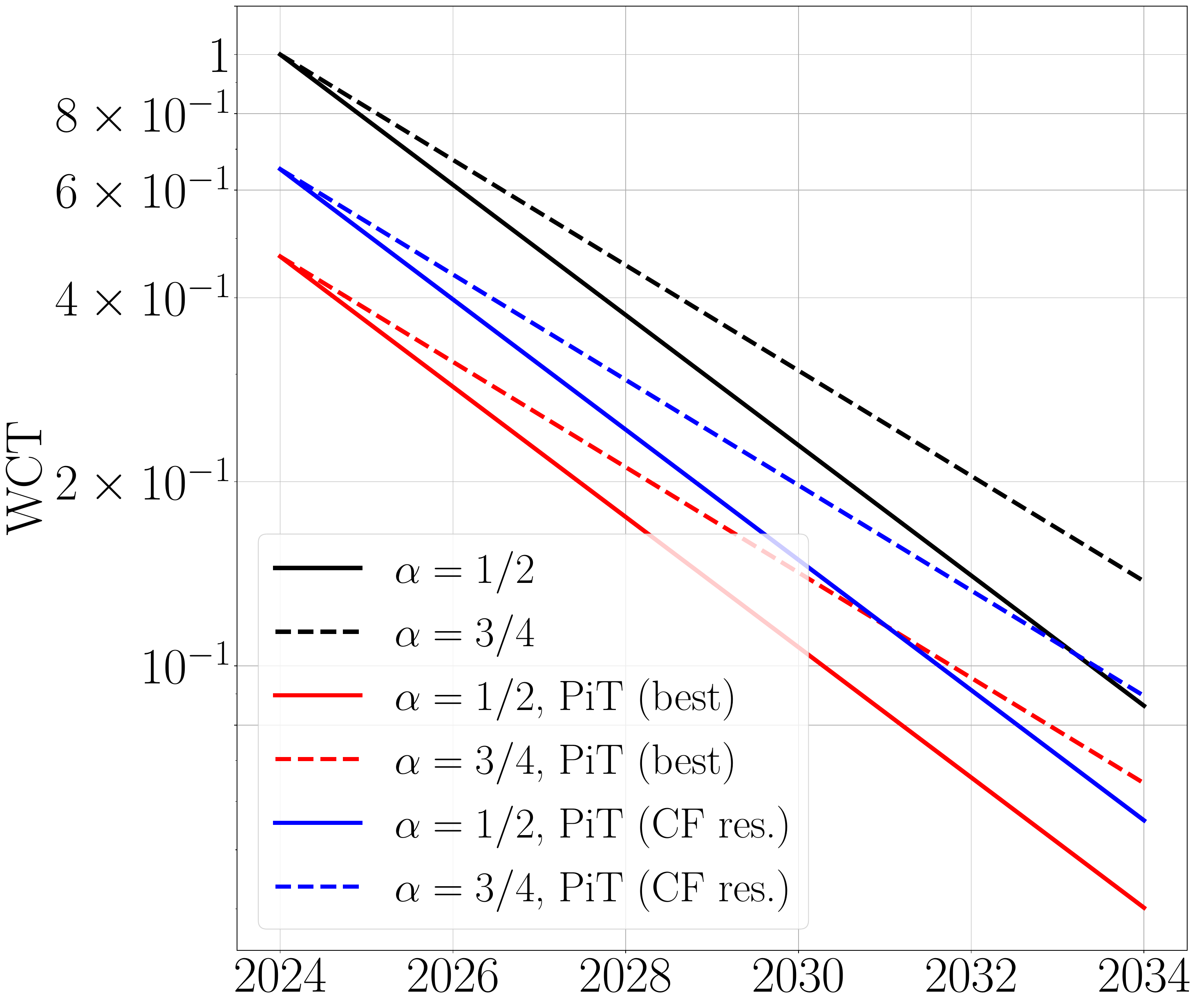}
  \caption{Evolution of wall-clock time (WCT) for the next decade assuming the
  hypotheses of ideal speed-up and constant cost and load, together with
  the values of $\beta_1,\beta_2$ extracted from HPCG (black),
  considering the best-case scenario from Table \ref{tab:extrap_7p} (red), and
  the actual speed-ups from Table \ref{tab:results_pm} (blue).}
  \label{fig:wct}
\end{figure}

Hence, the evolution of the wall-clock time can be estimated by Eq.
\eqref{eq:wct_end}. As $K$ is a constant, it is assumed to be 1 in this derivation. Then, for both
$\alpha=1/2$ and $\alpha=3/4$, in the next decade, the wall-clock time, under
the hypotheses previously mentioned, is expected to evolve as in Figure
\ref{fig:wct}. Moreover, the best-case scenario from Table \ref{tab:extrap_7p}
is depicted in red, while the speed-up results from the full turbulent channel flow presented 
in this paper are shown in blue.

It shows that for convective-dominated flows, i.e., $\alpha=3/4$,
wall-clock time is expected to reduce slower than diffusion-dominated
flows such as the ones dominated by K41 theory. As the main goal of DNS and LES
is the study of turbulence, those flows are generally convective-dominated in the
bulk of the flow. Nonetheless, the parameter $\alpha$ here depends on the effect
on the $\Delta t$. In highly turbulent flows, $\Delta t$ is generally limited by
the smallest scales, which are diffusion-dominated. Therefore, $\alpha=1/2$ should
be more representative than $\alpha=3/4$.

\section{Conclusions} \label{sec:conc}

Some computational physics applications, when run in modern supercomputers, are strongly limited by their memory bandwidths instead
of the peak performance of those machines. This is due to the low arithmetic intensity of the basic kernels, which are usually
based on sparse linear algebra, that makes them memory-bound, and thus the full potential of the modern supercomputers cannot 
be extracted in this field.

This work presented a methodology to mitigate this reduced arithmetic intensity by replacing the sparse matrix-vector products (\texttt{SpMV}) by
the more compute-intensive sparse matrix-matrix products (\texttt{SpMM}), which essentially enhances the performance of those
multiplications, thus making them faster. This may happen only in situations in which repeated matrix block structures appear in
the operators, i.e., these can be defined as the Kronecker product of this smaller matrix. In cases where off-diagonal terms appear, such as in
\citep{alsalti-baldellou_lighter_2024}, alternatives have already been proposed. In case of CFD simulations, this fully-diagonal matrix block structure appears in
a multiple parameter setup, as well as in an ensemble averaging parallel-in-time simulation, in which these repeated block structures are implicit.

It is in the latter, ensemble averaging parallel-in-time simulations, where a novel mesh-refinement method has been proposed to reduce the wall-clock time spent
in the transition to a developed flow by developing the flow in a coarser mesh to then map those results into the target mesh, thus exploiting both faster iterations
in the coarser mesh as well as the allowance of having greater timesteps as the stability of the simulations allowed so.

The results presented in this work show the benefit of replacing \texttt{SpMV} by \texttt{SpMM} both in structured and unstructured frameworks. In the former,
full simulation speed-ups reach to up to 1.55x, without the use of additional computational power, by incrementing the arithmetic intensity of \texttt{SpMV}, which
become indeed up to 3.0x faster. This speed-up translates to both the Poisson equation and the whole iteration. Given the dependance on the sparsity pattern of the
matrices, together with the number of non-zeros per row of these, using other
higher-order discretization methods, may become even more benefitial, as shown
with the tests with matrices that have used a greater number of non-zeros
per row (13 and 27), in which the \texttt{SpMM} speed-up grew to 3.3x and 4.1x, respectively. Estimates in the simulation speed-up considering the 27 non-zeros per row
discretization reach to values closer to 1.8x under the same conditions of the tested turbulent channel flow. The method has proven to be robust and applicable both to
additional transport equations as well as to unstructured meshes, providing results similar in magnitude compared to the turbulent channel flow.

Future lines of work will include applying the underlying implementation in GPU-accelerated supercomputers, together with extending the application to full industrial
unstructured simulations in order to verify the estimated speed-ups presented in this work, thus proving the actual method to work in full simulations for both structured
and unstructured domains. Moreover, high-order symmetry-preserving schemes will also be applied, in order to verify the benefits that can be induced by the increment in
performance obtained in the matrices which had a higher value of non-zeros per row.


\section*{Acknowledgements}

This work is supported by the \textit{Ministerio de Economía y Competitividad},
Spain, SIMEX project (PID2022-142174OB-I00). J.P.-R. is also supported by the
Catalan Agency for Management of University and Research Grants (AGAUR), 2022
FI\_B 00810. A.A.-B. was financially supported by the EU's
  Horizon Europe programme under the Marie Sk\l odowska‑Curie grant
  agreement No.~101208388.  Views and opinions expressed are however
  those of the authors only and do not necessarily reflect those of
  the European Union or the European Research Executive
  Agency. Neither the European Union nor the granting authority can be
  held responsible for them. 
Calculations were performed on the MareNostrum 5 supercomputer
at the BSC. The authors thankfully acknowledge the support of these organizations.
The authors would also like to acknowledge the team of M3E s.r.l. for providing access
to the Chronos package and for their technical support and collaboration.


\appendix
\section{Computation of the estimates for aAMG-preconditioned \texttt{SpMM} operations} \label{sec:wgt_spmm}

The computation of the upper-bound is usually performed by using Eq. \eqref{eq:pmspmm}, which arises from
the ratio of the operational intensity of the $m$-rhs \texttt{SpMM} and the equivalent \texttt{SpMV} for
the given sparse matrix, i.e. number of rows, columns, number of non-zeros and sparsity pattern.

Nonetheless, the presence of an adaptive algebraic multigrid (aAMG) as a preconditioner for the conjugate
gradient solver generates a variety of different sparse matrices within the solution. This comes from the
hierarchy generated by the preconditioner based of purely algebraic information in which the size of the
problem is reduced in a series of levels. Each of these levels will indeed have a different number of rows,
columns, non-zeros and sparsity pattern which will make unique bounds for each of them, given the traditional
definition from Eq. \eqref{eq:pmspmm}. Moreover, in a non-preconditioned conjugate gradient, all the matrices
used in \texttt{SpMM} operations are square, yet in this aAMG framework, the restriction and prolongation
matrices, i.e. those used to jump in between levels, will indeed be rectangular.

Given the algebraic nature of the procedure, these intermediate sparse matrices will be completely case
dependant and thus an exhausive analytical study for each case would not be applicable in any other case. Thus,
in order to estimate the upper-bound for these \texttt{SpMM} operations, a ponderated arithmetic intensity for
all number of right-hand sides is calculated to then estimate the upper-bound by computing the ratios.

In the case of the $80^3$ mesh for the turbulent channel flow, the \texttt{SpMM} distribution in the solution
of the Poisson equation in the \texttt{7p} case for a time unit is shown in
Table \ref{tab:sparse-matrices}.

\begin{table}[h]
  \centering
  \caption{Sparse matrices used in the CG+aAMG Poisson solver in the 1-rhs
  case, where $N$ is the number of cells of the simulation.}
  \begin{tabular}{ccccc}
    $n_c/N$ & $n_r/N$ & $\text{nnz}(A)/N$ & N.calls & Weight,
    $\omega$ \\ \toprule
    0.1265 & 0.1265 & 6.87 & 64333 & 14.85\% \\
    0.0515 & 0.0515 & 16.43 & 40983 & 9.46\% \\
    0.0154 & 0.0154 & 25.35 & 40983 & 9.46\% \\
    0.0034 & 0.0034 & 48.89 & 40983 & 9.46\% \\
    0.0006 & 0.0006 & 102.35 & 40983 & 9.46\% \\
    1.0000 & 1.0000 & 6.94 & 16 & 0.004\% \\
    0.4039 & 0.4039 & 16.65 & 15 & 0.003\% \\
    0.1177 & 0.1177 & 25.74 & 15 & 0.003\% \\
    0.0252 & 0.0252 & 47.81 & 15 & 0.003\% \\
    0.0045 & 0.0045 & 83.97 & 15 & 0.003\% \\
    0.0006 & 0.0006 & 181.45 & 15 & 0.003\% \\
    0.1265 & 0.0515 & 4.38 & 20484 & 4.73\% \\
    0.0515 & 0.0154 & 6.26 & 20484 & 4.73\% \\
    0.0154 & 0.0034 & 12.22 & 20484 & 4.73\% \\
    0.0034 & 0.0006 & 19.61 & 20484 & 4.73\%\\
    0.0006 & 0.00009 & 22.03 & 20484 & 4.73\%\\
    0.00009 & 0.0006 & 19.61 & 20484 & 4.73\%\\
    0.0006 & 0.0034 & 12.22 & 20484 & 4.73\%\\
    0.0034 & 0.0154 & 6.26 & 20484 & 4.73\%\\
    0.0154 & 0.0515 & 4.34 & 20484 & 4.73\%\\
    0.0515 & 0.1265 & 1.78 & 20484 & 4.73\%\\\bottomrule
  \end{tabular}
  \label{tab:sparse-matrices}
\end{table}

Hence, this upper-bound will be computed as

\begin{equation}
  P_{m,\text{\texttt{SpMM}, aAMG}}^{ub}
  = \sum_{i=1}^{n_\text{\texttt{SpMM}}}
  {\omega_iP_{m,\text{\texttt{SpMM}}}(\text{nnz}(A_i),n_{r,i},n_{c,i})}
\end{equation}

This same methodology has been used to compute the performance of the
\texttt{SpMM} when building the roofline plot. The individual wall-clock times
for every different \texttt{SpMM} operation within the CG+aAMG Poisson solution are
accumulated, $t_i$, while the number of floating-point operations ($OC$, from
operation count) that these products require can be extracted from Eq. \eqref{eq:ai} as

\begin{equation}
  \mathrm{OC}_i(\text{nnz}(A_i),m) = (2\text{nnz}(A_i)+1)m.
\end{equation}

Hence, the average performance $\pi_\text{\texttt{SpMM}}$ of \texttt{SpMM} within the CG+aAMG framework
can be estimated as

\begin{equation}
  \pi_\text{\texttt{SpMM}}
  = \sum_{i=1}^{n_\text{\texttt{SpMM}}}{\omega_i\frac{\mathrm{OC}_i}{t_i}}.
\end{equation}


\bibliography{export}

\begin{thebibliography}{42}
\expandafter\ifx\csname natexlab\endcsname\relax\def\natexlab#1{#1}\fi
\providecommand{\url}[1]{\texttt{#1}}
\providecommand{\href}[2]{#2}
\providecommand{\path}[1]{#1}
\providecommand{\DOIprefix}{doi:}
\providecommand{\ArXivprefix}{arXiv:}
\providecommand{\URLprefix}{URL: }
\providecommand{\Pubmedprefix}{pmid:}
\providecommand{\doi}[1]{\href{http://dx.doi.org/#1}{\path{#1}}}
\providecommand{\Pubmed}[1]{\href{pmid:#1}{\path{#1}}}
\providecommand{\bibinfo}[2]{#2}
\ifx\xfnm\relax \def\xfnm[#1]{\unskip,\space#1}\fi
\bibitem[{Williams et~al.(2009)Williams, Waterman, and
  Patterson}]{williams_roofline_2009}
\bibinfo{author}{S.~Williams}, \bibinfo{author}{A.~Waterman},
  \bibinfo{author}{D.~Patterson},
\newblock \bibinfo{title}{Roofline: {An} insightful visual performance model
  for multicore architectures},
\newblock \bibinfo{journal}{Communications of the ACM} \bibinfo{volume}{52}
  (\bibinfo{year}{2009}) \bibinfo{pages}{65--76}.
  \DOIprefix\doi{10.1145/1498765.1498785}.
\bibitem[{Greathouse and Daga(2014)}]{greathouse_efficient_2014}
\bibinfo{author}{J.~L. Greathouse}, \bibinfo{author}{M.~Daga},
\newblock \bibinfo{title}{Efficient {Sparse} {Matrix}-{Vector} {Multiplication}
  on {GPUs} {Using} the {CSR} {Storage} {Format}},
\newblock in: \bibinfo{booktitle}{International {Conference} for {High}
  {Performance} {Computing}, {Networking}, {Storage} and {Analysis}, {SC}},
  volume \bibinfo{volume}{2015-January}, \bibinfo{publisher}{IEEE Computer
  Society}, \bibinfo{year}{2014}, pp. \bibinfo{pages}{769--780}.
  \DOIprefix\doi{10.1109/SC.2014.68}, \bibinfo{note}{issue: January ISSN:
  21674337}.
\bibitem[{Kreutzer et~al.(2014)Kreutzer, Hager, Wellein, Fehske, and
  Bishop}]{Kreutzer2014}
\bibinfo{author}{M.~Kreutzer}, \bibinfo{author}{G.~Hager},
  \bibinfo{author}{G.~Wellein}, \bibinfo{author}{H.~Fehske},
  \bibinfo{author}{A.~R. Bishop},
\newblock \bibinfo{title}{A unified sparse matrix data format for efficient
  general sparse matrix-vector multiplication on modern processors with wide
  simd units},
\newblock \bibinfo{journal}{SIAM Journal on Scientific Computing}
  \bibinfo{volume}{36} (\bibinfo{year}{2014}) \bibinfo{pages}{C401--C423}.
  \DOIprefix\doi{10.1137/130930352}.
\bibitem[{Bell and Garland(2009)}]{Bell2009}
\bibinfo{author}{N.~Bell}, \bibinfo{author}{M.~Garland},
\newblock \bibinfo{title}{Implementing sparse matrix-vector multiplication on
  throughput-oriented processors},
\newblock in: \bibinfo{booktitle}{Proceedings of the Conference on High
  Performance Computing Networking, Storage and Analysis}, SC '09,
  \bibinfo{publisher}{ACM}, \bibinfo{year}{2009}, pp. \bibinfo{pages}{1--11}.
  \DOIprefix\doi{10.1145/1654059.1654078}.
\bibitem[{Markidis et~al.(2018)Markidis, Chien, Laure, Peng, and
  Kestor}]{Markidis2018}
\bibinfo{author}{S.~Markidis}, \bibinfo{author}{S.~W.~D. Chien},
  \bibinfo{author}{E.~Laure}, \bibinfo{author}{I.~B. Peng},
  \bibinfo{author}{G.~Kestor},
\newblock \bibinfo{title}{Nvidia tensor core programmability, precision and
  performance},
\newblock in: \bibinfo{booktitle}{2018 IEEE/ACM International Workshop on
  Performance, Portability and Productivity in HPC (P3HPC)},
  \bibinfo{publisher}{IEEE}, \bibinfo{year}{2018}, pp. \bibinfo{pages}{13--23}.
  \DOIprefix\doi{10.1109/P3HPC.2018.8647431}.
\bibitem[{Anzt et~al.(2019)Anzt, Cojean, Yen, and Dongarra}]{Anzt2019}
\bibinfo{author}{H.~Anzt}, \bibinfo{author}{T.~Cojean}, \bibinfo{author}{C.~D.
  Yen}, \bibinfo{author}{J.~Dongarra},
\newblock \bibinfo{title}{Mixed-precision iterative refinement leveraging
  low-precision gpu tensor cores},
\newblock \bibinfo{journal}{Parallel Computing} \bibinfo{volume}{81}
  (\bibinfo{year}{2019}) \bibinfo{pages}{102--117}.
  \DOIprefix\doi{10.1016/j.parco.2018.11.006}.
\bibitem[{Makarashvili et~al.(2017)Makarashvili, Merzari, Obabko, Siegel, and
  Fischer}]{makarashvili_performance_2017}
\bibinfo{author}{V.~Makarashvili}, \bibinfo{author}{E.~Merzari},
  \bibinfo{author}{A.~Obabko}, \bibinfo{author}{A.~Siegel},
  \bibinfo{author}{P.~Fischer},
\newblock \bibinfo{title}{A performance analysis of ensemble averaging for high
  fidelity turbulence simulations at the strong scaling limit},
\newblock \bibinfo{journal}{Computer Physics Communications}
  \bibinfo{volume}{219} (\bibinfo{year}{2017}) \bibinfo{pages}{236--245}.
  \DOIprefix\doi{10.1016/j.cpc.2017.05.023}, \bibinfo{note}{publisher: Elsevier
  B.V.}
\bibitem[{Nastac et~al.(2017)Nastac, Labahn, Magri, and
  Ihme}]{nastac_lyapunov_2017}
\bibinfo{author}{G.~Nastac}, \bibinfo{author}{J.~W. Labahn},
  \bibinfo{author}{L.~Magri}, \bibinfo{author}{M.~Ihme},
\newblock \bibinfo{title}{Lyapunov exponent as a metric for assessing the
  dynamic content and predictability of large-eddy simulations},
\newblock \bibinfo{journal}{Physical Review Fluids} \bibinfo{volume}{2}
  (\bibinfo{year}{2017}). \DOIprefix\doi{10.1103/PhysRevFluids.2.094606},
  \bibinfo{note}{publisher: American Physical Society}.
\bibitem[{Tosi et~al.(2022)Tosi, N{\'u}{\~n}ez, Pons-Prats, Principe, and
  Rossi}]{tosi_use_2022}
\bibinfo{author}{R.~Tosi}, \bibinfo{author}{M.~N{\'u}{\~n}ez},
  \bibinfo{author}{J.~Pons-Prats}, \bibinfo{author}{J.~Principe},
  \bibinfo{author}{R.~Rossi},
\newblock \bibinfo{title}{On the use of ensemble averaging techniques to
  accelerate the {Uncertainty} {Quantification} of {CFD} predictions in wind
  engineering},
\newblock \bibinfo{journal}{Journal of Wind Engineering and Industrial
  Aerodynamics} \bibinfo{volume}{228} (\bibinfo{year}{2022}).
  \DOIprefix\doi{10.1016/j.jweia.2022.105105}, \bibinfo{note}{publisher:
  Elsevier B.V.}
\bibitem[{Krasnopolsky(2018)}]{krasnopolsky_approach_2018}
\bibinfo{author}{B.~I. Krasnopolsky},
\newblock \bibinfo{title}{An approach for accelerating incompressible turbulent
  flow simulations based on simultaneous modelling of multiple ensembles},
\newblock \bibinfo{journal}{Computer Physics Communications}
  \bibinfo{volume}{229} (\bibinfo{year}{2018}) \bibinfo{pages}{8--19}.
  \DOIprefix\doi{10.1016/j.cpc.2018.03.023}, \bibinfo{note}{publisher: Elsevier
  B.V.}
\bibitem[{Alsalti-Baldellou et~al.(2024{\natexlab{a}})Alsalti-Baldellou,
  {\'A}lvarez-Farr{\'e}, Colomer, Gorobets, P{\'e}rez-Segarra, Oliva, and
  Trias}]{alsalti-baldellou_lighter_2024}
\bibinfo{author}{{\`A}.~Alsalti-Baldellou},
  \bibinfo{author}{X.~{\'A}lvarez-Farr{\'e}}, \bibinfo{author}{G.~Colomer},
  \bibinfo{author}{A.~Gorobets}, \bibinfo{author}{C.~D. P{\'e}rez-Segarra},
  \bibinfo{author}{A.~Oliva}, \bibinfo{author}{F.~X. Trias},
\newblock \bibinfo{title}{Lighter and faster simulations on domains with
  symmetries},
\newblock \bibinfo{journal}{Computers and Fluids} \bibinfo{volume}{275}
  (\bibinfo{year}{2024}{\natexlab{a}}).
  \DOIprefix\doi{10.1016/j.compfluid.2024.106247}, \bibinfo{note}{publisher:
  Elsevier Ltd}.
\bibitem[{Alsalti-Baldellou et~al.(2024{\natexlab{b}})Alsalti-Baldellou, Janna,
  {\'A}lvarez-Farr{\'e}, and Trias}]{alsalti-baldellou_multigrid_2024}
\bibinfo{author}{{\`A}.~Alsalti-Baldellou}, \bibinfo{author}{C.~Janna},
  \bibinfo{author}{X.~{\'A}lvarez-Farr{\'e}}, \bibinfo{author}{F.~X. Trias},
\newblock \bibinfo{title}{A {Multigrid} {Reduction} {Framework} for {Domains}
  with {Symmetries}},
\newblock \bibinfo{journal}{SIAM Journal on Scientific Computing}
  \bibinfo{volume}{46} (\bibinfo{year}{2024}{\natexlab{b}})
  \bibinfo{pages}{B860--B883}. \URLprefix
  \url{https://epubs.siam.org/doi/10.1137/24M1638513}.
  \DOIprefix\doi{10.1137/24M1638513}.
\bibitem[{{\'A}lvarez et~al.(2018){\'A}lvarez, Gorobets, Trias, Borrell, and
  Oyarzun}]{alvarez_hpc2fully-portable_2018}
\bibinfo{author}{X.~{\'A}lvarez}, \bibinfo{author}{A.~Gorobets},
  \bibinfo{author}{F.~X. Trias}, \bibinfo{author}{R.~Borrell},
  \bibinfo{author}{G.~Oyarzun},
\newblock \bibinfo{title}{{HPC2}{\textemdash}{A} fully-portable, algebra-based
  framework for heterogeneous computing. {Application} to {CFD}},
\newblock \bibinfo{journal}{Computers and Fluids} \bibinfo{volume}{173}
  (\bibinfo{year}{2018}) \bibinfo{pages}{285--292}.
  \DOIprefix\doi{10.1016/j.compfluid.2018.01.034}, \bibinfo{note}{publisher:
  Elsevier Ltd}.
\bibitem[{Krasnopolsky(2018)}]{krasnopolsky_optimal_2018}
\bibinfo{author}{B.~I. Krasnopolsky},
\newblock \bibinfo{title}{Optimal {Strategy} for {Modelling} {Turbulent}
  {Flows} with {Ensemble} {Averaging} on {High} {Performance} {Computing}
  {Systems}},
\newblock \bibinfo{journal}{Lobachevskii Journal of Mathematics}
  \bibinfo{volume}{39} (\bibinfo{year}{2018}) \bibinfo{pages}{533--542}.
  \DOIprefix\doi{10.1134/S199508021804008X}, \bibinfo{note}{publisher: Pleiades
  Publishing}.
\bibitem[{Alsalti-Baldellou et~al.(2023)Alsalti-Baldellou,
  {\'A}lvarez-Farr{\'e}, Trias, and Oliva}]{alsalti-baldellou_exploiting_2023}
\bibinfo{author}{A.~Alsalti-Baldellou},
  \bibinfo{author}{X.~{\'A}lvarez-Farr{\'e}}, \bibinfo{author}{F.~X. Trias},
  \bibinfo{author}{A.~Oliva},
\newblock \bibinfo{title}{Exploiting spatial symmetries for solving {Poisson}'s
  equation},
\newblock \bibinfo{journal}{Journal of Computational Physics}
  \bibinfo{volume}{486} (\bibinfo{year}{2023}).
  \DOIprefix\doi{10.1016/j.jcp.2023.112133}, \bibinfo{note}{publisher: Academic
  Press Inc.}
\bibitem[{Trias et~al.(2014)Trias, Lehmkuhl, Oliva, P{\'e}rez-Segarra, and
  Verstappen}]{Trias2014}
\bibinfo{author}{F.~X. Trias}, \bibinfo{author}{O.~Lehmkuhl},
  \bibinfo{author}{A.~Oliva}, \bibinfo{author}{C.~D. P{\'e}rez-Segarra},
  \bibinfo{author}{R.~W. Verstappen},
\newblock \bibinfo{title}{Symmetry-preserving discretization of
  {Navier}-{Stokes} equations on collocated unstructured grids},
\newblock \bibinfo{journal}{Journal of Computational Physics}
  \bibinfo{volume}{258} (\bibinfo{year}{2014}) \bibinfo{pages}{246--267}.
  \DOIprefix\doi{10.1016/j.jcp.2013.10.031}, \bibinfo{note}{publisher: Academic
  Press Inc.}
\bibitem[{Nikitin(2009)}]{nikitin_disturbance_2009}
\bibinfo{author}{N.~V. Nikitin},
\newblock \bibinfo{title}{Disturbance growth rate in turbulent wall flows},
\newblock \bibinfo{journal}{Fluid Dynamics} \bibinfo{volume}{44}
  (\bibinfo{year}{2009}) \bibinfo{pages}{652--657}.
  \DOIprefix\doi{10.1134/S0015462809050032}.
\bibitem[{Keefe et~al.(1992)Keefe, Moin, and Kim}]{Keefe1992}
\bibinfo{author}{L.~Keefe}, \bibinfo{author}{P.~Moin},
  \bibinfo{author}{J.~Kim},
\newblock \bibinfo{title}{The dimension of attractors underlying periodic
  turbulent {Poiseuille} flow},
\newblock \bibinfo{journal}{Journal of Fluid Mechanics} \bibinfo{volume}{242}
  (\bibinfo{year}{1992}) \bibinfo{pages}{1--29}.
  \DOIprefix\doi{10.1017/S0022112092002258}.
\bibitem[{Badii et~al.(1988)Badii, Heinzelmann, Meier, and
  Politi}]{badii_correlation_1988}
\bibinfo{author}{R.~Badii}, \bibinfo{author}{K.~Heinzelmann},
  \bibinfo{author}{P.~F. Meier}, \bibinfo{author}{A.~Politi},
\newblock \bibinfo{title}{Correlation functions and generalized {Lyapunov}
  exponents},
\newblock \bibinfo{journal}{Physical Review A} \bibinfo{volume}{37}
  (\bibinfo{year}{1988}) \bibinfo{pages}{1323--1328}. \URLprefix
  \url{https://link.aps.org/doi/10.1103/PhysRevA.37.1323}.
  \DOIprefix\doi{10.1103/PhysRevA.37.1323}.
\bibitem[{Mendes et~al.(2019)Mendes, Da~Silva, and Beims}]{mendes_decay_2019}
\bibinfo{author}{C.~F.~O. Mendes}, \bibinfo{author}{R.~M. Da~Silva},
  \bibinfo{author}{M.~W. Beims},
\newblock \bibinfo{title}{Decay of the distance autocorrelation and {Lyapunov}
  exponents},
\newblock \bibinfo{journal}{Physical Review E} \bibinfo{volume}{99}
  (\bibinfo{year}{2019}) \bibinfo{pages}{062206}. \URLprefix
  \url{https://link.aps.org/doi/10.1103/PhysRevE.99.062206}.
  \DOIprefix\doi{10.1103/PhysRevE.99.062206}.
\bibitem[{Cheskidov and Foias(2006)}]{cheskidov_global_2006}
\bibinfo{author}{A.~Cheskidov}, \bibinfo{author}{C.~Foias},
\newblock \bibinfo{title}{On global attractors of the {3D}
  {Navier}{\textendash}{Stokes} equations},
\newblock \bibinfo{journal}{Journal of Differential Equations}
  \bibinfo{volume}{231} (\bibinfo{year}{2006}) \bibinfo{pages}{714--754}.
  \URLprefix
  \url{https://linkinghub.elsevier.com/retrieve/pii/S002203960600338X}.
  \DOIprefix\doi{10.1016/j.jde.2006.08.021}.
\bibitem[{Bortolan et~al.(2024)Bortolan, de~Carvalho, Mar{\'i}n-Rubio, and
  Valero}]{Bortolan2024}
\bibinfo{author}{M.~C. Bortolan}, \bibinfo{author}{A.~N. de~Carvalho},
  \bibinfo{author}{P.~Mar{\'i}n-Rubio}, \bibinfo{author}{J.~Valero},
\newblock \bibinfo{title}{Weak global attractor for the {3D}-{Navier}-{Stokes}
  equations via the globally modified {Navier}-{Stokes} equations}
  (\bibinfo{year}{2024}). \URLprefix \url{http://arxiv.org/abs/2402.06435}.
  \DOIprefix\doi{10.48550/arXiv.2402.06435}, \bibinfo{note}{arXiv: 2402.06435}.
\bibitem[{Burstedde et~al.(2011)Burstedde, Wilcox, and
  Ghattas}]{burstedde_p4est_2011}
\bibinfo{author}{C.~Burstedde}, \bibinfo{author}{L.~C. Wilcox},
  \bibinfo{author}{O.~Ghattas},
\newblock \bibinfo{title}{p4est: Scalable algorithms for parallel adaptive mesh
  refinement on forests of octrees},
\newblock \bibinfo{journal}{SIAM Journal on Scientific Computing}
  \bibinfo{volume}{33} (\bibinfo{year}{2011}) \bibinfo{pages}{1103--1133}.
  \URLprefix \url{https://doi.org/10.1137/100791634}.
  \DOIprefix\doi{10.1137/100791634}.
  \href{http://arxiv.org/abs/https://doi.org/10.1137/100791634}{{\tt
  arXiv:https://doi.org/10.1137/100791634}}.
\bibitem[{Zhang et~al.(2019)Zhang, Almgren, Beckner, Bell, Blaschke, Chan, Day,
  Friesen, Gott, Graves, Katz, Myers, Nguyen, Nonaka, Rosso, Williams, and
  Zingale}]{Zhang2019}
\bibinfo{author}{W.~Zhang}, \bibinfo{author}{A.~Almgren},
  \bibinfo{author}{V.~Beckner}, \bibinfo{author}{J.~Bell},
  \bibinfo{author}{J.~Blaschke}, \bibinfo{author}{C.~Chan},
  \bibinfo{author}{M.~Day}, \bibinfo{author}{B.~Friesen},
  \bibinfo{author}{K.~Gott}, \bibinfo{author}{D.~Graves},
  \bibinfo{author}{M.~P. Katz}, \bibinfo{author}{A.~Myers},
  \bibinfo{author}{T.~Nguyen}, \bibinfo{author}{A.~Nonaka},
  \bibinfo{author}{M.~Rosso}, \bibinfo{author}{S.~Williams},
  \bibinfo{author}{M.~Zingale},
\newblock \bibinfo{title}{Amrex: a framework for block-structured adaptive mesh
  refinement},
\newblock \bibinfo{journal}{Journal of Open Source Software}
  \bibinfo{volume}{4} (\bibinfo{year}{2019}) \bibinfo{pages}{1370}. \URLprefix
  \url{https://doi.org/10.21105/joss.01370}.
  \DOIprefix\doi{10.21105/joss.01370}.
\bibitem[{Martin et~al.(2025)Martin, Adams, Colella, Graves, Johansen, Johnson,
  Keen, Ligocki, McCorquodale, Modiano, Schwartz, Sternberg, and
  Straalen}]{Martin2025}
\bibinfo{author}{D.~Martin}, \bibinfo{author}{M.~Adams},
  \bibinfo{author}{P.~Colella}, \bibinfo{author}{D.~Graves},
  \bibinfo{author}{H.~Johansen}, \bibinfo{author}{J.~Johnson},
  \bibinfo{author}{N.~Keen}, \bibinfo{author}{T.~Ligocki},
  \bibinfo{author}{P.~McCorquodale}, \bibinfo{author}{D.~Modiano},
  \bibinfo{author}{P.~Schwartz}, \bibinfo{author}{T.~Sternberg},
  \bibinfo{author}{B.~V. Straalen},
\newblock \bibinfo{title}{Chombo design document}  (\bibinfo{year}{2025}).
  \DOIprefix\doi{10.6084/m9.figshare.28599755.v1}.
\bibitem[{Verstappen and Veldman(2003)}]{Verstappen2003}
\bibinfo{author}{R.~W. C.~P. Verstappen}, \bibinfo{author}{A.~E.~P. Veldman},
\newblock \bibinfo{title}{Symmetry-preserving discretization of turbulent
  flow},
\newblock \bibinfo{journal}{Journal of Computational Physics}
  \bibinfo{volume}{187} (\bibinfo{year}{2003}) \bibinfo{pages}{343--368}.
  \DOIprefix\doi{10.1016/S0021-9991(03)00126-8}, \bibinfo{note}{publisher:
  Academic Press Inc.}
\bibitem[{Sanderse and Koren(2012)}]{sanderse_accuracy_2012}
\bibinfo{author}{B.~Sanderse}, \bibinfo{author}{B.~Koren},
\newblock \bibinfo{title}{Accuracy analysis of explicit {Runge}-{Kutta} methods
  applied to the incompressible {Navier}-{Stokes} equations},
\newblock \bibinfo{journal}{Journal of Computational Physics}
  \bibinfo{volume}{231} (\bibinfo{year}{2012}) \bibinfo{pages}{3041--3063}.
  \DOIprefix\doi{10.1016/j.jcp.2011.11.028}, \bibinfo{note}{publisher: Academic
  Press Inc.}
\bibitem[{Capuano et~al.(2017)Capuano, Coppola, R{\'a}ndez, and
  de~Luca}]{capuano_explicit_2017}
\bibinfo{author}{F.~Capuano}, \bibinfo{author}{G.~Coppola},
  \bibinfo{author}{L.~R{\'a}ndez}, \bibinfo{author}{L.~de~Luca},
\newblock \bibinfo{title}{Explicit {Runge}{\textendash}{Kutta} schemes for
  incompressible flow with improved energy-conservation properties},
\newblock \bibinfo{journal}{Journal of Computational Physics}
  \bibinfo{volume}{328} (\bibinfo{year}{2017}) \bibinfo{pages}{86--94}.
  \DOIprefix\doi{10.1016/j.jcp.2016.10.040}, \bibinfo{note}{publisher: Academic
  Press Inc.}
\bibitem[{Trias et~al.(2024)Trias, {\'A}lvarez-Farr{\'e}, Alsalti-Baldellou,
  Gorobets, and Oliva}]{trias_efficient_2024}
\bibinfo{author}{F.~X. Trias}, \bibinfo{author}{X.~{\'A}lvarez-Farr{\'e}},
  \bibinfo{author}{A.~Alsalti-Baldellou}, \bibinfo{author}{A.~Gorobets},
  \bibinfo{author}{A.~Oliva},
\newblock \bibinfo{title}{An efficient eigenvalue bounding method: {CFL}
  condition revisited},
\newblock \bibinfo{journal}{Computer Physics Communications}
  \bibinfo{volume}{305} (\bibinfo{year}{2024}).
  \DOIprefix\doi{10.1016/j.cpc.2024.109351}, \bibinfo{note}{publisher: Elsevier
  B.V.}
\bibitem[{Falgout et~al.(2006)Falgout, Jones, and Yang}]{falgout_design_2006}
\bibinfo{author}{R.~D. Falgout}, \bibinfo{author}{J.~E. Jones},
  \bibinfo{author}{U.~M. Yang},
\newblock \bibinfo{title}{The {Design} and {Implementation} of hypre, a
  {Library} of {Parallel} {High} {Performance} {Preconditioners}},
\newblock in: \bibinfo{editor}{A.~Bruaset}, \bibinfo{editor}{A.~Tveito} (Eds.),
  \bibinfo{booktitle}{Numerical {Solution} of {Partial} {Differential}
  {Equations} on {Parallel} {Computers}. {Lecture} {Notes} in {Computational}
  {Science} and {Engineering}}, volume~\bibinfo{volume}{51},
  \bibinfo{publisher}{Springer}, \bibinfo{address}{Berlin, Heidelberg},
  \bibinfo{year}{2006}, pp. \bibinfo{pages}{267--294}.
\bibitem[{Isotton et~al.(2021)Isotton, Frigo, Spiezia, and
  Janna}]{isotton_chronos_2021}
\bibinfo{author}{G.~Isotton}, \bibinfo{author}{M.~Frigo},
  \bibinfo{author}{N.~Spiezia}, \bibinfo{author}{C.~Janna},
\newblock \bibinfo{title}{Chronos: {A} {General} {Purpose} {Classical} {AMG}
  {Solver} for {High} {Performance} {Computing}},
\newblock \bibinfo{journal}{SIAM Journal on Scientific Computing}
  \bibinfo{volume}{43} (\bibinfo{year}{2021}) \bibinfo{pages}{C335--C357}.
  \DOIprefix\doi{10.1137/21M1398586}.
\bibitem[{Paludetto~Magri et~al.(2019)Paludetto~Magri, Franceschini, and
  Janna}]{paludetto_magri_novel_2019}
\bibinfo{author}{V.~A. Paludetto~Magri}, \bibinfo{author}{A.~Franceschini},
  \bibinfo{author}{C.~Janna},
\newblock \bibinfo{title}{A {Novel} {Algebraic} {Multigrid} {Approach} {Based}
  on {Adaptive} {Smoothing} and {Prolongation} for {Ill}-{Conditioned}
  {Systems}},
\newblock \bibinfo{journal}{SIAM Journal on Scientific Computing}
  \bibinfo{volume}{41} (\bibinfo{year}{2019}) \bibinfo{pages}{A190--A219}.
  \URLprefix \url{https://epubs.siam.org/doi/10.1137/17M1161178}.
  \DOIprefix\doi{10.1137/17M1161178}.
\bibitem[{De~Sterck et~al.(2006)De~Sterck, Yang, and
  Heys}]{de_sterck_reducing_2006}
\bibinfo{author}{H.~De~Sterck}, \bibinfo{author}{U.~M. Yang},
  \bibinfo{author}{J.~J. Heys},
\newblock \bibinfo{title}{Reducing {Complexity} in {Parallel} {Algebraic}
  {Multigrid} {Preconditioners}},
\newblock \bibinfo{journal}{SIAM Journal on Matrix Analysis and Applications}
  \bibinfo{volume}{27} (\bibinfo{year}{2006}) \bibinfo{pages}{1019--1039}.
  \URLprefix \url{http://epubs.siam.org/doi/10.1137/040615729}.
  \DOIprefix\doi{10.1137/040615729}.
\bibitem[{De~Sterck et~al.(2008)De~Sterck, Falgout, Nolting, and
  Yang}]{de_sterck_distancetwo_2008}
\bibinfo{author}{H.~De~Sterck}, \bibinfo{author}{R.~D. Falgout},
  \bibinfo{author}{J.~W. Nolting}, \bibinfo{author}{U.~M. Yang},
\newblock \bibinfo{title}{Distance-two interpolation for parallel algebraic
  multigrid},
\newblock \bibinfo{journal}{Numerical Linear Algebra with Applications}
  \bibinfo{volume}{15} (\bibinfo{year}{2008}) \bibinfo{pages}{115--139}.
  \URLprefix \url{https://onlinelibrary.wiley.com/doi/10.1002/nla.559}.
  \DOIprefix\doi{10.1002/nla.559}.
\bibitem[{Trias et~al.(2010)Trias, Gorobets, Soria, and
  Oliva}]{trias_direct_2010}
\bibinfo{author}{F.~X. Trias}, \bibinfo{author}{A.~Gorobets},
  \bibinfo{author}{M.~Soria}, \bibinfo{author}{A.~Oliva},
\newblock \bibinfo{title}{Direct numerical simulation of a differentially
  heated cavity of aspect ratio 4 with {Rayleigh} numbers up to 1011 - {Part}
  {I}: {Numerical} methods and time-averaged flow},
\newblock \bibinfo{journal}{International Journal of Heat and Mass Transfer}
  \bibinfo{volume}{53} (\bibinfo{year}{2010}) \bibinfo{pages}{665--673}.
  \DOIprefix\doi{10.1016/j.ijheatmasstransfer.2009.10.026}.
\bibitem[{Mosqueda-Otero et~al.(2025)Mosqueda-Otero, Alsalti-Baldellou,
  {\'A}lvarez-Farr{\'e}, Plana-Riu, Colomer, Trias, and
  Oliva}]{mosqueda-otero_portable_2025}
\bibinfo{author}{M.~Mosqueda-Otero},
  \bibinfo{author}{{\`A}.~Alsalti-Baldellou},
  \bibinfo{author}{X.~{\'A}lvarez-Farr{\'e}}, \bibinfo{author}{J.~Plana-Riu},
  \bibinfo{author}{G.~Colomer}, \bibinfo{author}{F.~X. Trias},
  \bibinfo{author}{A.~Oliva},
\newblock \bibinfo{title}{A {Portable} {Algebraic} {Implementation} for
  {Reliable} {Overnight} {Industrial} {LES}},
\newblock \bibinfo{journal}{Proceedings of the 35th Parallel CFD International
  Conference 2024 35th Parallel CFD International Conference 2024}
  \bibinfo{volume}{ParCFD 2024} (\bibinfo{year}{2025}) \bibinfo{pages}{pages
  119 -- 126}. \URLprefix \url{https://juser.fz-juelich.de/record/1041837}.
  \DOIprefix\doi{10.34734/FZJ-2025-02469}, \bibinfo{note}{artwork Size: pages
  119 - 126 Publisher: Forschungzentrum J{\"u}lich}.
\bibitem[{Demou and Grigoriadis(2019)}]{demou_direct_2019}
\bibinfo{author}{A.~D. Demou}, \bibinfo{author}{D.~G.~E. Grigoriadis},
\newblock \bibinfo{title}{Direct numerical simulations of
  {Rayleigh}{\textendash}{B{\'e}nard} convection in water with
  non-{Oberbeck}{\textendash}{Boussinesq} effects},
\newblock \bibinfo{journal}{Journal of Fluid Mechanics} \bibinfo{volume}{881}
  (\bibinfo{year}{2019}) \bibinfo{pages}{1073--1096}.
  \DOIprefix\doi{10.1017/jfm.2019.787}.
\bibitem[{Moin and Mahesh(1998)}]{moin_direct_1998}
\bibinfo{author}{P.~Moin}, \bibinfo{author}{K.~Mahesh},
\newblock \bibinfo{title}{{DIRECT} {NUMERICAL} {SIMULATION}: {A} {Tool} in
  {Turbulence} {Research}},
\newblock \bibinfo{journal}{Annual Review of Fluid Mechanics}
  \bibinfo{volume}{30} (\bibinfo{year}{1998}) \bibinfo{pages}{539--578}.
  \URLprefix
  \url{https://www.annualreviews.org/doi/10.1146/annurev.fluid.30.1.539}.
  \DOIprefix\doi{10.1146/annurev.fluid.30.1.539}.
\bibitem[{Kolmogorov(1941{\natexlab{a}})}]{kolmogorov_local_1941}
\bibinfo{author}{A.~Kolmogorov},
\newblock \bibinfo{title}{The local structure of turbulence in incompressible
  viscous fluid for very large {Reynolds} numbers},
\newblock \bibinfo{journal}{Doklady Akademii Nauk SSSR} \bibinfo{volume}{30}
  (\bibinfo{year}{1941}{\natexlab{a}}) \bibinfo{pages}{9--13}.
  \DOIprefix\doi{10.1098/rspa.1991.0075}.
\bibitem[{Kolmogorov(1941{\natexlab{b}})}]{kolmogorov_degeneration_1941}
\bibinfo{author}{A.~Kolmogorov},
\newblock \bibinfo{title}{On degeneration (decay) of isotropic turbulence in an
  incompressible viscous liquid},
\newblock \bibinfo{journal}{Doklady Akademii Nauk SSSR} \bibinfo{volume}{31}
  (\bibinfo{year}{1941}{\natexlab{b}}) \bibinfo{pages}{538--540}.
\bibitem[{Courant et~al.(1928)Courant, Friedrichs, and
  Lewy}]{courant_uber_1928}
\bibinfo{author}{R.~Courant}, \bibinfo{author}{K.~Friedrichs},
  \bibinfo{author}{H.~Lewy},
\newblock \bibinfo{title}{{\"U}ber die partiellen {Differenzengleichungen} der
  mathematischen {Physik}},
\newblock \bibinfo{journal}{Mathematische Annalen} \bibinfo{volume}{100}
  (\bibinfo{year}{1928}) \bibinfo{pages}{32--74}.
  \DOIprefix\doi{10.1007/BF01448839}.
\bibitem[{Dongarra et~al.(2016)Dongarra, Heroux, and
  Luszczek}]{dongarra_high-performance_2016}
\bibinfo{author}{J.~Dongarra}, \bibinfo{author}{M.~A. Heroux},
  \bibinfo{author}{P.~Luszczek},
\newblock \bibinfo{title}{High-performance conjugate-gradient benchmark: {A}
  new metric for ranking high-performance computing systems},
\newblock \bibinfo{journal}{The International Journal of High Performance
  Computing Applications} \bibinfo{volume}{30} (\bibinfo{year}{2016})
  \bibinfo{pages}{3--10}. \URLprefix
  \url{https://journals.sagepub.com/doi/10.1177/1094342015593158}.
  \DOIprefix\doi{10.1177/1094342015593158}.

\end{thebibliography}

\end{document}